\documentclass[superscriptaddress,prc,secnumarabic,preprintnumbers,amsmath,amssymb,aps,showpacs,bibnotes,altaffilletter,10pt]{revtex4-2}
\usepackage{orcidlink}
\usepackage{lineno}
\usepackage{hyperref}
\usepackage{geometry}
\usepackage{bm}
\usepackage{nicefrac}
\usepackage{graphicx}
\usepackage{tabularx}
\usepackage[title]{appendix}
\usepackage{url}
\usepackage{textcomp}
\usepackage[T1]{fontenc}

\def\nuc#1#2{${}^{#1}$#2}

\def\BBz{$\beta\beta(0\nu)$}

\def\MJ{\textsc{Majorana}}             
\def\mj{\textsc{Majorana}}             
\def\MJD{\textsc{Majorana Demonstrator}}
\def\Demo{\textsc{Demonstrator}}
\def\DEM{\textsc{Demonstrator}}

\newcommand{\LM}{LEGEND-1000}

\def\ppc{P-PC}                          
\def\ICPC{ICPC}
\def\cofs{$^{56}\mathrm{Co}$}
\def\cosixty{$^{60}\mathrm{Co}$}
\def\gese{$^{68}\mathrm{Ge}$}
\def\gess{$^{76}\mathrm{Ge}$}
\def\thtte{$^{228}\mathrm{Th}$}
\def\thttt{$^{232}\mathrm{Th}$}
\def\utte{$^{238}\mathrm{U}$}

\def\geoxide{$\mathrm{GeO}_2$}

\begin{document}

\title{The {\sc Majorana Demonstrator} experiment's construction, commissioning, and performance}


\newcommand{\blhill}{Department of Physics, Black Hills State University, Spearfish, SD 57799, USA}
\newcommand{\ITEP}{National Research Center ``Kurchatov Institute'', Kurchatov Complex of Theoretical and Experimental Physics, Moscow, 117218 Russia}
\newcommand{\JINR}{Joint Institute for Nuclear Research, Dubna, 141980 Russia} 
\newcommand{\lbnl}{Nuclear Science Division, Lawrence Berkeley National Laboratory, Berkeley, CA 94720, USA}
\newcommand{\lbnle}{Engineering Division, Lawrence Berkeley National Laboratory, Berkeley, CA 94720, USA}
\newcommand{\lanl}{Los Alamos National Laboratory, Los Alamos, NM 87545, USA}
\newcommand{\queens}{Department of Physics, Engineering Physics and Astronomy, Queen's University, Kingston, ON K7L 3N6, Canada}
\newcommand{\uw}{Center for Experimental Nuclear Physics and Astrophysics, and Department of Physics, University of Washington, Seattle, WA 98195, USA}
\newcommand{\unc}{Department of Physics and Astronomy, University of North Carolina, Chapel Hill, NC 27514, USA}
\newcommand{\duke}{Department of Physics, Duke University, Durham, NC 27708, USA}
\newcommand{\ncsu}{Department of Physics, North Carolina State University, Raleigh, NC 27695, USA}	
\newcommand{\ornl}{Oak Ridge National Laboratory, Oak Ridge, TN 37830, USA}
\newcommand{\ou}{Research Center for Nuclear Physics, Osaka University, Ibaraki, Osaka 567-0047, Japan}
\newcommand{\pnnl}{Pacific Northwest National Laboratory, Richland, WA 99354, USA}
\newcommand{\ttu}{Tennessee Tech University, Cookeville, TN 38505, USA}
\newcommand{\sdsmt}{South Dakota Mines, Rapid City, SD 57701, USA}
\newcommand{\usc}{Department of Physics and Astronomy, University of South Carolina, Columbia, SC 29208, USA}
\newcommand{\usd}{Department of Physics, University of South Dakota, Vermillion, SD 57069, USA}  
\newcommand{\ut}{Department of Physics and Astronomy, University of Tennessee, Knoxville, TN 37916, USA}
\newcommand{\tunl}{Triangle Universities Nuclear Laboratory, Durham, NC 27708, USA}
\newcommand{\mpi}{Max Planck Institute for Physics, 85748 Garching, Germany}
\newcommand{\tum}{Technical University of Munich, TUM School of Natural Sciences, Physics Department, 85748 Garching, Germany}
\newcommand{\williams}{Physics Department, Williams College, Williamstown, MA 01267, USA}
\newcommand{\ciemat}{Centro de Investigaciones Energ\'{e}ticas, Medioambientales y Tecnol\'{o}gicas, CIEMAT 28040, Madrid, Spain}
\newcommand{\iu}{IU Center for Exploration of Energy and Matter, and Department of Physics, Indiana University, Bloomington, IN 47405, USA}
\newcommand{\ucsd}{Hal\i c{\i}o\u{g}lu Data Science Institute, Department of Physics, University of California San Diego, CA 92093, USA}

\affiliation{\lbnl}
\affiliation{\pnnl}
\affiliation{\unc}
\affiliation{\tunl}
\affiliation{\usc}
\affiliation{\ornl}
\affiliation{\ITEP}
\affiliation{\usd}
\affiliation{\ncsu}
\affiliation{\lanl}
\affiliation{\JINR}
\affiliation{\uw}
\affiliation{\duke}
\affiliation{\sdsmt}
\affiliation{\ciemat}
\affiliation{\ut}
\affiliation{\ou}
\affiliation{\williams}
\affiliation{\blhill}
\affiliation{\ttu}
\affiliation{\ucsd}
\affiliation{\queens}
\affiliation{\mpi}
\affiliation{\tum}
\affiliation{\iu}

\author{N.~Abgrall\,\orcidlink{0009-0005-0777-8661}}\affiliation{\lbnl}
\author{E.~Aguayo}\affiliation{\pnnl} 
\author{I.J.~Arnquist}\affiliation{\pnnl} 
\author{F.T.~Avignone~III}\affiliation{\usc}\affiliation{\ornl}
\author{A.S.~Barabash\,\orcidlink{0000-0002-5130-0922}}\affiliation{\ITEP}
\author{C.J.~Barton\,\orcidlink{0000-0002-4698-3765}}\altaffiliation{Present address: Roma Tre University and INFN Roma Tre, Rome, Italy}\affiliation{\usd}
\author{P.J.~Barton\,\orcidlink{0000-0002-4894-0486}}\affiliation{\lbnl}		
\author{F.E.~Bertrand}\affiliation{\ornl} 
\author{E.~Blalock}\affiliation{\ncsu}\affiliation{\tunl} 
\author{B.~Bos}\affiliation{\unc}\affiliation{\tunl} 
% ...
			
\collaboration{{\sc{Majorana}} Collaboration}
\noaffiliation


\date{\today}

\begin{abstract}
\begin{description}
\item[Background] 
The \MJD , a modular array of  isotopically enriched high-purity germanium (HPGe) detectors, was constructed to demonstrate backgrounds low enough to justify building a tonne-scale experiment to search for the neutrinoless double-beta decay (\BBz ) of \gess .
\item[Purpose]
This paper presents a description of the instrument, its commissioning, and operations. 
It covers the electroforming, underground infrastructure, enrichment, detector fabrication, low-background and construction techniques, electronics, data acquisition, databases, and data processing of the \MJD .
\item[Method] The \MJD\ operated inside an ultra-low radioactivity passive shield at the 4850-foot~level of the Sanford Underground Research Facility (SURF) from 2015-2021. 
\item[Results and Conclusions]
The \MJD\ achieved the best energy resolution and second-best background level of any \BBz\ search. This enabled it to achieve an ultimate half-life limit on \BBz\ in \gess\ of $8.3\times 10^{25}$~yr (90\% C.L.) and perform a rich set of searches for other physics beyond the Standard Model.
\end{description}
\end{abstract}

\maketitle


\tableofcontents

\section{Introduction}
\label{sec:intro}

\subsection{Motivation}
\label{sec:motivation}

Neutrinoless double-beta decay (\BBz -decay) is a hypothetical nuclear decay, given as~\cite{PhysRev.56.1184}:
\begin{equation}
(A,Z) \rightarrow (A,Z+2) + 2e^-
\label{eqn:0nubb}
\end{equation} 
The observation of this lepton number violating process ($\Delta L =2$) is sufficient to show that the neutrino is a Majorana fermion~\cite{PhysRevD.25.2951}. 
Its discovery would also provide independent constraints on the masses of neutrinos and provide support for leptogenesis models that explain the matter-antimatter asymmetry in the universe. All of these are compelling reasons to search for this rare process, and there is a large international effort currently underway using a variety of isotopes and experimental techniques. 
For comprehensive experimental and theoretical reviews on \BBz -decay, see Refs.~\cite{avig08, rode11, elli12a, crem13, schwi13,bilenky15,delloro16,vergados16,henn16, engel17,barabash17,doli19,RevModPhys.95.025002,adams2022neutrinoless}. 
 
The \MJD , a modular array of  high-purity germanium (HPGe) detectors, was constructed to demonstrate backgrounds low enough to justify building a tonne-scale experiment to search for the \BBz -decay of \gess .
It operated inside an ultra-low radioactivity passive shield at the 4850-foot~level of the Sanford Underground Research Facility (SURF) in Lead, South Dakota, from 2015-2021. 
The \DEM\ achieved the best energy resolution of any \BBz -decay search and carried out a rich set of searches for physics beyond the Standard Model.

A subset of the \MJD\ collaboration have joined with a subset of the GERDA~\cite{agos17, GERDA:2023wbr} collaboration to pursue a tonne-scale \gess\ experiment called the Large Enriched Germanium Experiment for Neutrinoless $\beta\beta$ Decay (LEGEND)~\cite{abgr17b}. 
The results and experience from the \MJD\ have been incorporated into the design of LEGEND, which has two phases. 
LEGEND-200 is currently operating at INFN Gran Sasso National Laboratory (LNGS) in Italy and will deploy 200~kg of enriched detectors. 
\LM\  is the proposed next phase that will deploy 1000~kg of enriched detectors at LNGS with data-taking starting in~2030~\cite{L1000pcdr}. 
 
A detailed description of the \MJD\ experimental design was published in~2014, which was at the start of its construction~\cite{abgr14}. 
In this paper we provide an update, focusing on the construction, commissioning, upgrades, and operation of the \MJD .
It covers the enrichment, detector fabrication, low-background techniques, electronics, data acquisition, databases, and data processing of the \MJD .  
It also provides citations to all \MJD\ results and publications.  
Appendix ~\ref{se:vendors_appendix} provides more information about vendors and software suppliers mentioned in this paper.

\subsection{The \MJD\ Experiment Overview}
\label{se:experiment_overview}

\subsubsection{Design Considerations}
\label{se:design_considerations}

The kinematics of \BBz -decay (eqn.~\ref{eqn:0nubb}) require that the sum energy of the emitted electrons be equal to the $Q$-value of the decay.
The HPGe detectors of the \DEM\ serve as both source and detector of this decay and measure the sum energy directly; hence, the experimental signature of \BBz -decay is an excess of events with energy deposits in the region of interest (ROI) around the 2039.006(50)~keV $Q$-value of \gess ~\cite{PhysRevLett.86.4259}.
This signature can be mimicked by many other types of naturally occurring radioactivity from the \thttt\ and \utte\ decay chains, cosmic-rays, and cosmogenic isotopes.
Fortunately, the excellent energy resolution (0.13\%) of the HPGe detectors reduce the size of the ROI and therefore reduces backgrounds.
Further reduction requires careful design, material selection and special handling procedures, an underground location, and cleanliness, which are prominent considerations in the subsequent sections. 

\subsubsection{Design of the \MJD\ }
\label{se:experiment_design}

The \MJD\ relied on the well-known benefits of enriched HPGe detectors that use intrinsically low-background source material and have an understood isotopic enrichment process.
They also provide excellent energy resolution and sophisticated event reconstruction capabilities.
The \MJD 's specific detector design is called a p-type, point-contact (\ppc)~\cite{luke89,barb07} and had masses ranging from 0.6~-~1.1~kg. 
This design enables sensitive pulse-shape discrimination and provides excellent discrimination against multi-site background events.
Its low-energy threshold also enabled a variety of physics searches other than \BBz -decay. 

\begin{figure}[ht]
\begin{center}
\includegraphics[width=0.9\textwidth]{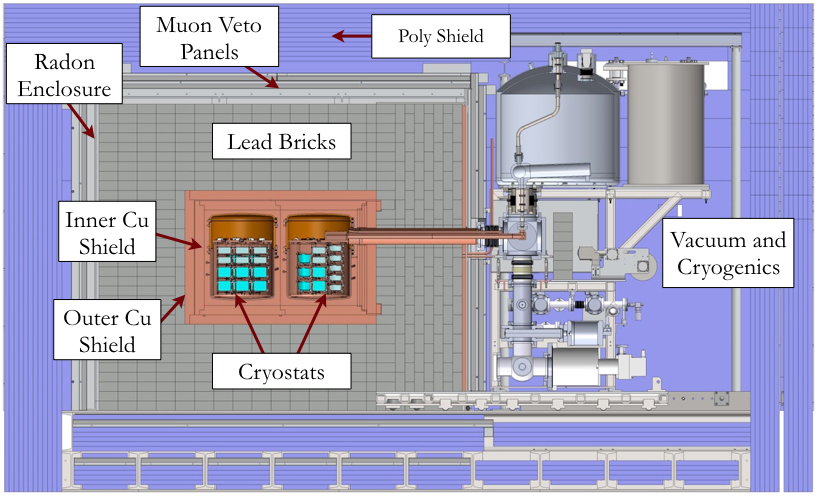}
\caption{The \MJD\ cross section, shown with shield layers, cryogenics, and both cryostats and their HPGe detectors installed. The entire assembly is~2.74 meters in height.}
\label{fig:ShieldOverview}
\end{center}
\end{figure}

The detectors were deployed in  two modular arrays housed inside two vacuum cryostats. 
The two cryostats combined contained 35~detectors with a total mass of 29.7~kg, fabricated with germanium enriched to 87.4$\pm$0.5\% in \nuc{76}{Ge}~\cite{arnq22}, and 23~detectors with a total mass of 14.4~kg, fabricated from natural Ge (7.8\% \nuc{76}{Ge}).
Module~1(2) housed 16.8~kg (12.9~kg) of enriched germanium detectors and 5.6~kg (8.8~kg) of natural germanium detectors. 
This was the configuration for most of the \DEM\ operations and physics data-taking.
It operated for brief periods in different configurations, as described in Sec.~\ref{sec:data sets} and below. 

The completed  \MJD\ layout is shown in Fig.~\ref{fig:ShieldOverview}.
Starting from the innermost cavity, the two cryostat modules were surrounded by an inner layer of electroformed copper, an outer layer of commercial C10100 oxygen-free high-conductivity copper, high-purity lead, an active muon veto, borated polyethylene, and polyethylene. 
The cryostats, copper, and lead shielding were all enclosed in a radon exclusion box that was purged with liquid-nitrogen boil-off gas. 
The experiment was located in a clean room at the 4850-foot level (4300 meters water equivalent or m.w.e.) of SURF ~\cite{heis15}. 
The radioassay program developed to ensure the \MJD\ met background goals is described in Ref.~\cite{abgr16}.

The array was cooled with a horizontally-oriented thermosyphon contained within the ``crossarm tube'' that was under vacuum. 
The crossarm tube also serves as the penetration through the shield for cabling and the pumping path for the cryostat vacuum.
The HPGe detectors were calibrated using radioactive line sources deployed through the shielding into a helical track surrounding each module~\cite{abgr17a}. 
Custom readout electronics provided amplified, differential pulses from the HPGe detector charge collection process~\cite{Majorana:2021mtz}.
These signals were recorded  by digitizers developed for the GRETINA experiment~\cite{lee04, zimm12} which were read out using the 
Object-oriented Real-time Control and Acquisition (ORCA) software package~\cite{howe04, orcaweb, orcagit}.

Assembling and operating the \MJD\ was a complex process that required accurate record keeping. 
The collaboration developed and deployed several databases to record material processing and reprocessing, parts fabrication histories, enriched detector cosmic-ray exposures, slow-controls and environmental conditions, and analysis parameters. 

The two modules were installed sequentially with data collected from Module~1 while Module~2 was assembled.   
In~2020 the collaboration performed a significant upgrade on the cables and connectors in Module~2 and also installed four enriched (to 88$\pm$1\%) ORTEC inverted-coaxial point-contact (\ICPC ) detectors~\cite{cooper201125}, which replaced 5~\ppc\ detectors.
ICPC detectors are the baseline design for \LM , and the \MJD\ provided an opportunity to study these novel detectors in a low-background setting. 
The experiment stopped its \BBz -decay search in March 2021 and was converted into a search for decays of isomeric \nuc{180}{Ta}, including possible stimulated decays from dark matter~\cite{lehnert17, lehnert20}. 
The experimental configuration and results of this phase are not topics of this paper and are discussed in~\cite{arnq24a}. 

\subsection{Physics Results Summary}
\label{se:physics_results}

The \DEM\ published three results for the search of \BBz-decay. 
The first, published in 2017, included 9.95~kg~yr of exposure. No counts were observed in the \BBz -decay region, resulting in a half-life limit of $1.9\times10^{25}$~yr at 90\% CL for \gess\ \BBz-decay. 
Since these data were taken during commissioning and early operations, no blindness scheme was employed~\cite{aals18}. 
After collecting additional enriched exposure, for a total of $26.0\pm0.5$~kg~yr, the collaboration performed a blinded analysis and observed one event in the region of interest (ROI) with 0.65~events expected from background, resulting in a lower limit on the  half-life of $2.7\times10^{25}$~yr (90\% CL) with a median sensitivity of $4.8\times 10^{25}$~yr (90\% CL)~\cite{alvi19b}. 

In 2022 the collaboration published its final \BBz -decay result, based on an accumulated 64.5~kg~yr of enriched active exposure~\cite{arnq22}. 
Four events were observed in the 10~keV ROI, which was consistent with the measured background rate of $16.6^{+0.14}_{-0.13}$~cts/(FWHM t yr).
With a world-leading energy resolution of 2.52~keV FWHM, it set a half-life limit of $8.3\times 10^{25}$~yr (90\% C.L.). 
This provided a range of upper limits on $m_{\beta\beta}$ of $(113-269)$~meV (90\% C.L.), depending on the choice of nuclear matrix elements. 
The \DEM\ measured a background rate higher than what was expected from an initial assay-based projection of \mbox{$<2.5$~cts/(FWHM t yr)~\cite{abgr16}}.
The collaboration is further analyzing possible background sources with a detailed model using a GEANT4~\cite{agost03} based simulation package MaGe~\cite{bosw11}.
A boosted decision tree analysis was also performed by the collaboration and achieved similar results, with potential future applications in background identification and rejection in LEGEND~\cite{arnq22b}.
An assay-based background projection for the \DEM\ using Monte Carlo Uncertainty Propagation was recently posted by the collaboration~\cite{arnquist2024assay}.

The high granularity of the \DEM\ also allowed it to set the world-leading limits in the search of the double beta-decay of $^{76}$Ge to excited states of $^{76}$Se with half-life limits in the range  of $(0.75-4.0)\times10^{24}$~years, depending on the decay mode~\cite{PhysRevC.103.015501}.
A recent update has been posted, further improving these limits~\cite{arnquist2024final}.

The \DEM\ published several limits on other Beyond the Standard Model (BSM) processes, starting with data acquired during 2015 commissioning runs ~\cite{abgr16b}.
Its limits on bosonic dark matter were an improvement over other germanium experiments, EDELWEISS~\cite{EDEL} and CDEX~\cite{CDEX16}, due to the lower cosmogenic activity in \mj\ enriched detectors.
Subsequently, the collaboration used additional exposure and more sophisticated analyses to perform more sensitive searches for bosonic dark matter and also expand the reach of the \DEM\ to perform competitive searches for keV-scale sterile neutrino and fermionic dark matter~\cite{arnq22a}.

The \DEM\ searched for solar axions via coherent Bragg conversion in the germanium crystals and set a world-leading experimental limit on the axion-photon coupling strength of $g_{a\gamma} < 1.45\times10^{-9}\,\mathrm{GeV}^{-1}$ (95\% CL) in the 1 to 100~eV/$c^2$ mass range~\cite{arnq22c}. 
This work used a temporal-energy analysis and a crystal axis averaging technique developed by members of the collaboration~\cite{xu17}.

The collaboration also published the first limits for 15~tri-nucleon decay-specific modes and invisible decay modes for germanium isotopes~\cite{alvis19c}, as well as world-leading limits on cosmic-ray Lightly-Ionizing Particles (LIPs)~\cite{alvis18}. 
The large path length due to thick detectors and the low thresholds allowed for a LIP sensitivity down to $1/1000$ of an electron charge ($e$). 
These were the first results for a non-accelerator experiment on the natural flux of LIPS with charges less than $e/200$ and a significant improvement of the existing limits for charges between $e/6$ and $e/200$~\cite{alvis18}.

The low backgrounds, good energy resolution, and long exposure allowed for sensitive tests of the foundations of quantum mechanics by searching for forbidden transitions of atomic electrons. 
In some scenarios, electrons that are newly produced in pair production could transition into Pauli-forbidden states in a germanium atom~\cite{elli12}. 
An improved limit on the parameter $\beta^2/2$, which quantifies the probability of such a Pauli Exclusion Principle violating (PEPv) process, was measured by the \DEM\ using calibration data to be $\beta^2/2<6.3\times10^{-4}$ at 95\% CL.
Another PEPv process could occur when an electron spontaneously de-excites into a forbidden level in an atom, and the \DEM\ set a model-dependent limit of $\beta^2/2 < 1.0 \times 10^{-48}$ (90\% CL)
The \DEM\ also set limits on the mean electron lifetime with its most recent value being $\tau_e>3.2\times10^{25}$~\cite{abgr16b, arnq22d}.

Quantum measurement and the associated wave function collapse is a long-standing problem in quantum mechanics, and the Continuous Spontaneous Localization (CSL) model provides a mathematically well-motivated model for this process~\cite{ghirardi86, bassi13}. 
The \DEM\ provided a factor 40-100 improved limits over comparable experiments on the predicted collapse rate for the white spectrum CSL model by searching for spontaneous x-ray emissions~\cite{arnq22e}.

The \DEM\ performed a measurement of in-situ cosmic-ray isotope production in the HPGe detectors, of which \nuc{77}{Ge} (11.3 hour half-life) is especially germane as a potential background in \LM\ ~\cite{PhysRevC.105.014617}. 
The collaboration also quantified background neutron production from $^{13}$C$(\alpha,n))^{16}$O  reactions induced by $\alpha$ particles emitted within the calibration sources~\cite{PhysRevC.105.064610}.
Finally, the collaboration studied the effect of charge trapping in HPGe PPC detectors on energy resolution and developed a modified digital pole-zero correction as mitigation strategy~\cite{arnq22f}.

As stated earlier, the experiment stopped its \BBz -decay search in March 2021 and was converted into a search for decays of isomeric \nuc{180}{Ta} and dark matter. 
The results from the first year of running showed an improvement of 1–2 orders over exiting limits, making it the most sensitive searches for a single beta-decay and electron capture decay ever achieved.
Over all channels, the decay was excluded for half-lives below $0.29\times10^{18}$~yr~\cite{arnq24a}.
Data-taking is continuing into~2024 and further results are expected.

\subsection{Paper Organization}
\label{se:paper_org}

This paper presents a description of the \DEM\ and its operations. It covers the electroforming, materials processing, isotopic enrichment, detector fabrication, low-background techniques, electronics, data acquisition, databases, and data processing of the \DEM\ program.
It is organized as follows:
Sec.~\ref{sec:electroforming_machining} describes the electroforming process for making the ultra-pure copper at the heart of the \DEM 's low background levels, the cleaning and fabrication processes for components, and the underground infrastructure at SURF. 
Sec.~\ref{sec:geprocessing_and_fabrication} covers the germanium enrichment, material processing that minimized enriched material losses, and detector fabrication at vendors. 
Sec.~\ref{se:det_tranport} describes the acceptance testing and characterization of HPGe detectors before they were installed in the \DEM.
Sec.~\ref{se:det_arrays} covers the design of the detector modules and later improvements. 
Sec.~\ref{sec:cryovac} describes the cryogenic and vacuum systems, their performance, and the unique thermosyphon design.
Sec.~\ref{sec:calibration} describes the radionuclide-based calibration system.
Sec.~\ref{sec:shield} describes the passive shield design and muon veto. 
The electronics and data acquisition systems are covered in Sec.~\ref{sec:DAQ_electronics}, and the slow controls and monitoring systems are in Sec.~\ref{sec:slow_controls}. 
Sec.~\ref{se:databases} describes the many databases used to store calibration constants, environmental conditions, part histories, and other information. 
Finally, Sec.~\ref{se:data_prod} provides a high-level overview of the data-taking phases of the \DEM\ and how the data was staged and processed. 

\section{Electroforming, Machining, and Fabrication}
\label{sec:electroforming_machining}

The \MJD\ collaboration took great care selecting low background materials and employing surface cleaning and handling protocols to remove radioactive isotopes and avoid contamination. 
In this section we describe the key processes in the selection and production of construction materials with an emphasis on the underground electroforming and machining of copper.

\subsection{Electroforming}
\label{sec:electroforming}
\begin{figure}[ht]
\begin{center}
\includegraphics[width=0.95\textwidth]{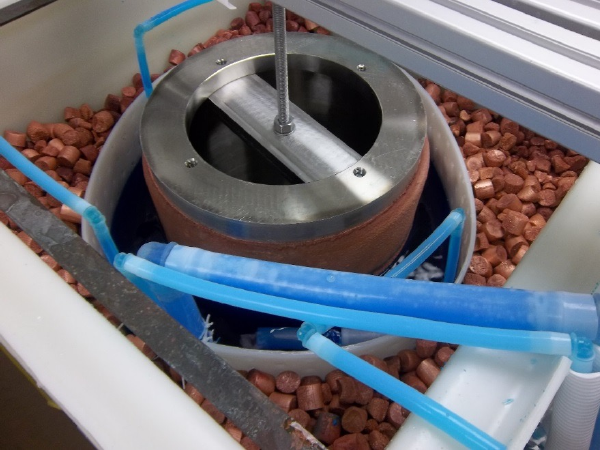}
\caption{An electroforming bath without electrolyte. The cylindrical stainless steel mandrel is already  covered in a layer of electrolytic copper. Also visible are the OFHC copper anode nuggets.}
\label{fig:EformingBath_Empty}
\end{center}
\end{figure}

Ultra-clean Electroformed Copper (EFCu) was the key material utilized in the \DEM\ for detector mounting parts, cryostat vessels, and inner shielding of the experiment. 
Since the copper contributes the most mass and was closest to the detectors, it had the most stringent radiopurity requirements. 
The collaboration built on the experience of the Pacific Northwest National Laboratory (PNNL) electroforming process, starting with the IGEX experiment ~\cite{aals99,aals07}, to make ultra-low background copper, which requires the highest-grade materials and cleaning techniques, a clean environment to reduce radioactive contaminants, and an underground location to prevent cosmogenic \nuc{60}{Co} accumulation~\cite{hopp08}. 
Fig.~\ref{fig:EformingBath_Empty} shows an example of an electroforming bath.  

In 2007, LANL, PNNL and UW set up and operated an early underground electroforming bath at the LANL-operated laboratory underground at the Waste Isolation Pilot Plant near Carlsbad, NM. 
This included the electroforming apparatus, clean-room infrastructure underground, and all the required safety infrastructure of a highly-regulated DOE facility. 
The underground environment also required operation under a mining safety envelope.
The primary goal of this project was to demonstrate safe operations of an underground electroforming facility, which was accomplished by producing a 660~gram copper part after ten days of operation. 

\begin{figure}[ht]
\begin{center}
\includegraphics[width=0.95\textwidth]{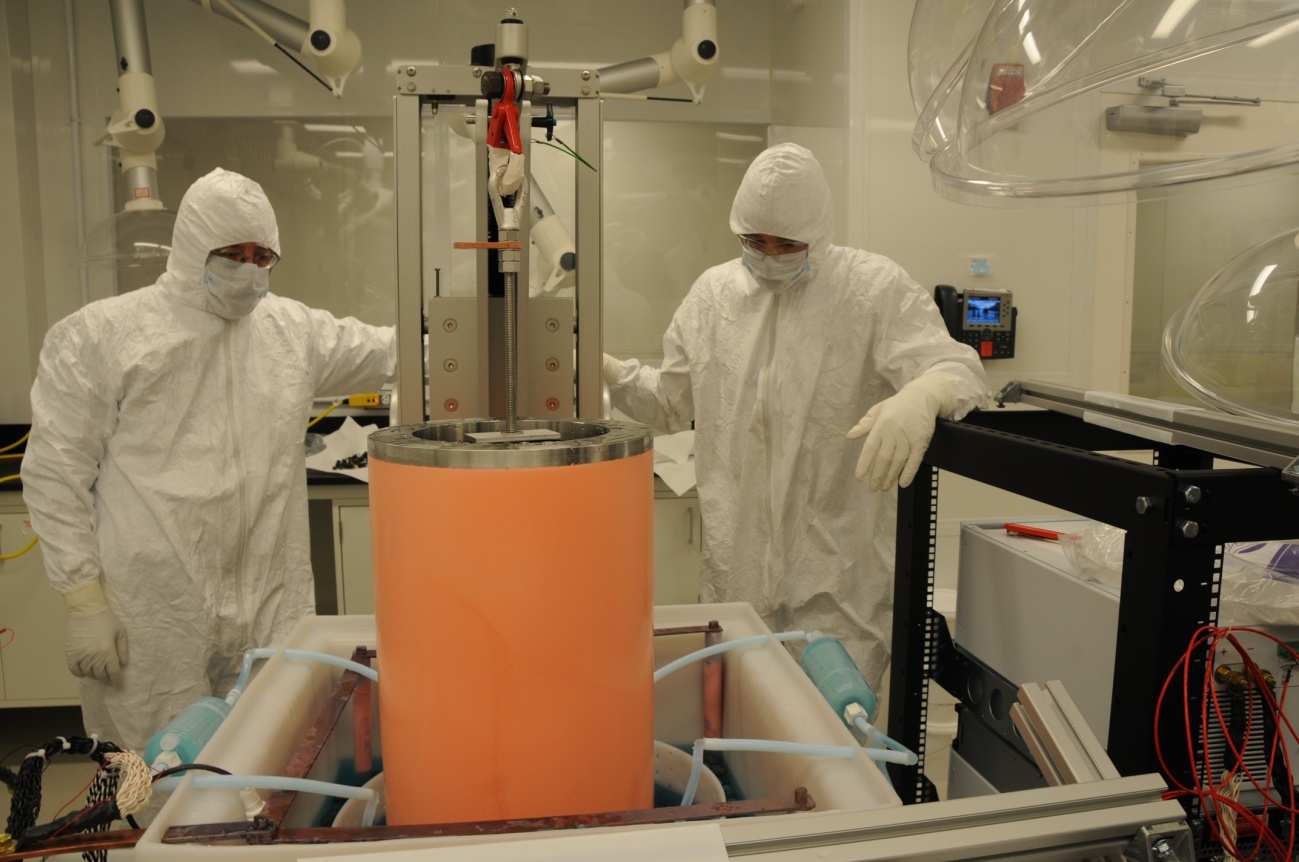}
\caption{A mandrel with copper undergoing inspection at the PNNL Shallow Underground Laboratory.}
\label{fig:PNNL_UG_Lab}
\end{center}
\end{figure}

In 2009 PNNL produced an electroforming bath capable of producing copper parts at a scale useful for  the \DEM .
It was based on larger scale PNNL baths operated above ground and modified using informed behaviors or operational requirements of the electroforming bath operated at WIPP. 
Electroforming for \MJ\ began in the middle of 2010~at PNNL with the construction and operation of six large scale baths in Shallow Underground Laboratory (SUL) (38~m.w.e.) facility (Fig.~\ref{fig:PNNL_UG_Lab}). 
These systems used 180~L High Density Polyethylene (HDPE) tanks to plate onto cylindrical mandrels made from 316~stainless steel with diameters up to 33~cm.
Plating durations of approximately 14~months were required to achieve a final thickness of 14~mm, corresponding to an average growth of 1~mm/month. 
The operation of the six baths at PNNL was briefly halted in 2011 when a radioactive plume from the  Japanese Fukushima Daiichi reactors, damaged by the Great Tohuku earthquake of March 11, was detected.
No contamination has been attributed to this plume to date in the \MJD .

\begin{figure}[ht]
\begin{center}
\includegraphics[width=0.95\textwidth]{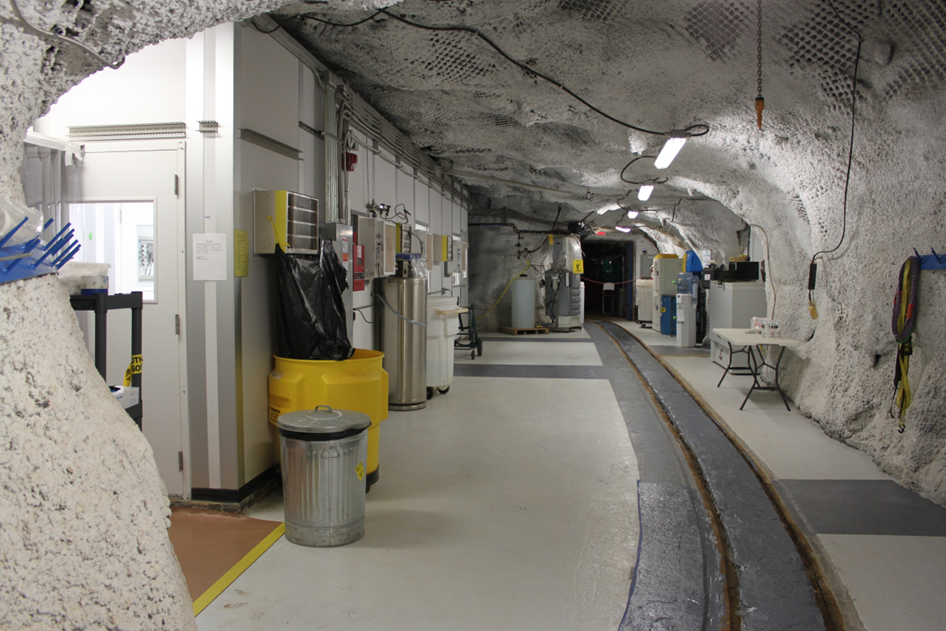}
\caption{The exterior of the temporary cleanroom (left), located at the 4850-foot~level of SURF, at the end of its construction.}
\label{fig:TCR_Exterior}
\end{center}
\end{figure}

\begin{figure}[ht]
\begin{center}
\includegraphics[width=0.95\textwidth]{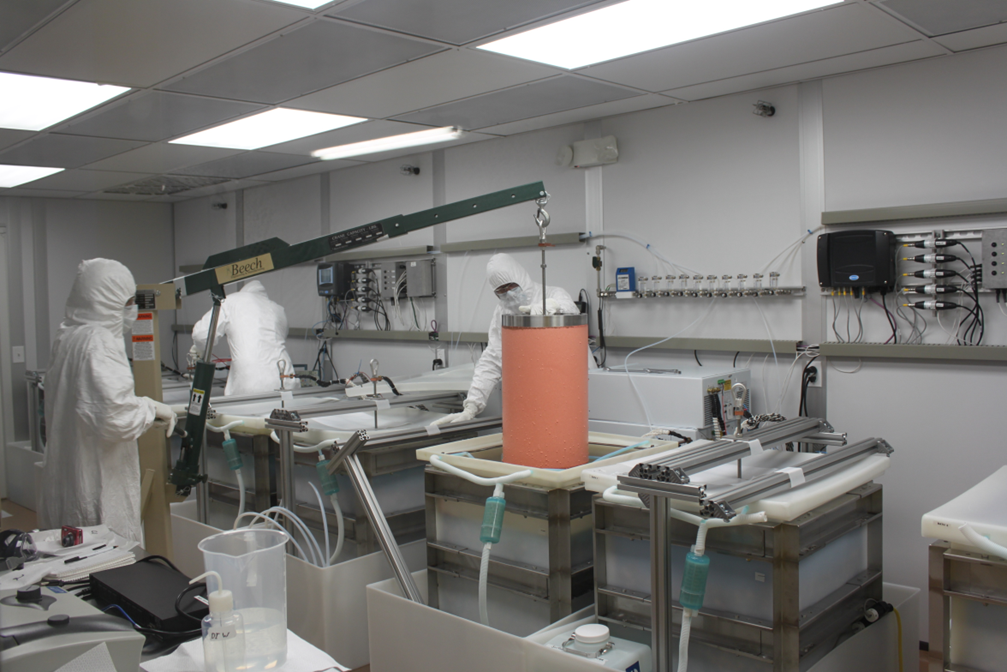}
\caption{A copper coated mandrel being inspected over its electroforming bath in the temporary cleanroom.}
\label{fig:TCR_Loading}
\end{center}
\end{figure}

In summer of 2011, electroforming also began at the 4850-foot~level of SURF in a temporary cleanroom (TCR) facility with ten baths of similar design. 
Construction and operation of the TCR facility at SURF was necessary because the \MJD\ main laboratory space in the Davis Campus was not ready for beneficial occupancy until May~2012 (Sec.~\ref{se:ug_infrastructure}).  
To stay on schedule, the collaboration had to start the lengthy electroforming process elsewhere and constructed a temporary hard wall cleanroom in a mine drift next to the Ross station, which became the TCR.
This prefabricated cleanroom building was $16'\times28'\times10'$ tall with an interior partition to create a small office space and the electroforming lab.
Extensive cleaning preparations of the inside and the outside surrounding mine drift had to be performed to establish a class 1000 space for the baths to operate.  
When the High Efficiency Particulate Air (HEPA) filters were activated, typical cleanroom protocol for a class~1000 room were adopted for all internal operations. 
Due to space issues, the reverse-osmosis (RO) tanks that supplied  the deionized (DI) water generator inside the cleanroom were stationed outside of the lab. 
See Figs.~\ref{fig:TCR_Exterior} and \ref{fig:TCR_Loading} for pictures of the facility.

While preparation of the cleanroom occurred underground, construction of the electroforming tanks was completed offsite at South Dakota School of Mines and Technology (SDSMT). 
Following the same cleaning and operational protocols used for the tanks at the PNNL SUL, the HDPE tanks were modified and all parts triple leached in 6M nitric acid and then rinsed with DI water. 
Once clean, the equipment was triple wrapped in plastic in a cleanroom and transported underground allowing for a multi-step process of bringing in all pre-cleaned items. 
Other equipment, like electroforming power supplies (Dynatronix 990-0430-110), slow control and monitoring systems, and uninterruptible power supplies were shipped directly from vendors and collaborators.

The TCR was maintained as a class 1000~cleanroom using 0.3~micron HEPA filters and positive pressure. 
Additional external prefilters were added to extend the life of the HEPA systems and changed as needed. 
Underground conditions at SURF initially required that the prefilters be changed monthly and then upgraded to weekly when heavy construction began in the nearby Ross shaft. 
The radon concentration was maintained at a typical (for an underground location) 12-15~pC/L with continuous airflow of surface air in the outside drift and within the cleanroom. 
Boil-off nitrogen served as cover gas to the baths to provide positive pressure against particulate and radon which could fluctuate greatly within the SURF environment. 
Temperature, humidity, radon, particle counts, and differential pressure were all monitored at the TCR, along with leak sensors within the secondary containment of the baths.  
All processes and and sensors were monitored remotely using the Object-oriented Real-time Control and Acquisition (ORCA) software package described in Sec.~\ref{se:DAQ_Software}.  

Once positioned inside the cleanroom, the baths were reinforced externally with stainless steel frames and an anode frame was positioned internally with a bus bar extending out the back of the tank, allowing for attachment to the power supply.
Next, 99.999\% OFHC (oxygen-free high thermal conductivity) copper slugs from multiple vendors were double rinsed in 6M nitric acid and DI water before being placed between the inner wall of the tank and a porous partition that maintained a void for the mandrel (see Fig.~\ref{fig:EformingBath_Empty}). 
Once full, the baths were allowed to generate copper sulfate in a dilute solution of sulfuric acid before plating could begin.
The largest diameter mandrel was 33~cm in diameter and 54~cm in length. 
Mandrels of similar diameter and length were used for electroforming the copper cryostat cans and IR shield. 
Finally, special mandrels were used for making the cross-arms and thermosyphon tubes. 
 
Prior to use each mandrel was given a skim cut on a lathe to provide a pristine surface for plating, remove contamination, and to ensure a successful release during the bake and quench. 
Hence, the mandrel welding was done with care to avoid the introduction of inclusions that would lead to surface pitting after skim cuts. 
Non-thoriated welding rods were also used to avoid contamination of \thttt .   

After cleaning and passivation of the mandrel surfaces using detergent and nitric acid, nucleation of the copper on the mandrel was established within the bath using an alternating voltage to control growth based on operational parameters from Ref.~\cite{hopp08}.
The baths were checked twice weekly to address natural electrolyte evaporation and power interruptions at SURF, along with adjustments in chemistry to allow constant plating. 
Copper nuggets were replenished every 4~months or once a new mandrel was inserted into a bath.

The PNNL and TCR systems ran from September~2010 to April~2016 and produced over 2700~kg of EFCu on over 60~different sized mandrels. 
These baths generated the inner copper shield plates, 2 cryostats used as modules 1 and 2 for the \MJD , and all the copper parts supporting the HPGe detectors.  

In addition to performing radiopurity assays, material properties of the EFCu were also evaluated.
Plated copper from each electroformed mandrel were sampled in regions of top, middle, and bottom along with both planes of orientation (in the direction of growth, perpendicular to the growth front).
Samples were tested to evaluate the mechanical response of the copper using tensile strength, optical metallography/Scanning Electron Microscopy (SEM), and Vickers hardness tests.

Optical metallography and SEM showed that the copper was of polycrystalline structure and had some small voids in early samples that did not have significant effect on hardness or resulting density. 
Tensile strength tests were plotted to the Ultimate Tensile Strength (UTS).
Although specimens with voids generated more noise in the data that reduced UTS in some cases, the voids were only observed in the first round of electroforming and interfaces did not impact the yield strength.  Subsequent batches did not have observed void features.
The overall results showed consistent mechanical response from the copper grown at both the TCR and PNNL facilities and an overall strength of (14~ksi~average), exceeding the required 10~ksi yield strength specified in the \MJD\ engineering design~\cite{over12}.

More details on the electroplating process, material properties, and assay of the EFCu are provided in Ref.~\cite{abgr16}.

\subsection{Machining and Fabrication}
\label{se:machining}

The underground-produced EFCu for the \DEM\ had to be kept underground to avoid cosmic-ray activation.
To meet this requirement, the collaboration had to outfit an underground machine shop in the Davis Campus in a clean-room environment.  
Before outfitting, test-machining was performed at PNNL, the University of Washington, and the University of North Carolina at Chapel Hill to develop machining procedures and test tools. 
The collaboration initially operated some of the machine tools in a collaboration-built soft-walled cleanroom located at a warehouse space in Rapid City, SD. 
This space was referred to as the Above-Ground Machine Shop (AGMS) and allowed for the training and procedural development for all mandrel handling by a contractor machinist that would eventually transition to underground when the Davis Campus space became available.  

\begin{figure}[t]
\begin{center}
\includegraphics[width=0.5\textwidth]{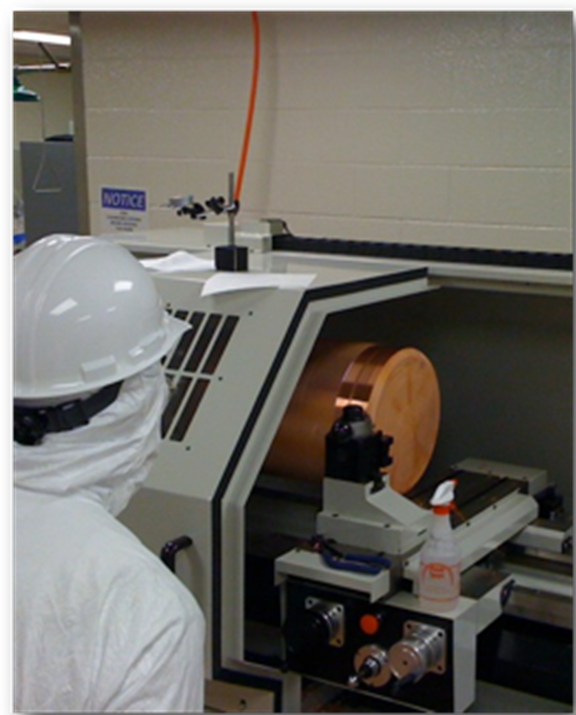}
\caption{EFCu on mandrel being machined on the large TRAK lathe in the underground machine shop.}
\label{fig:Cu_MandrelOnLathe}
\end{center}
\end{figure}

\begin{figure}[t]
\begin{center}
\includegraphics[width=0.8\textwidth]{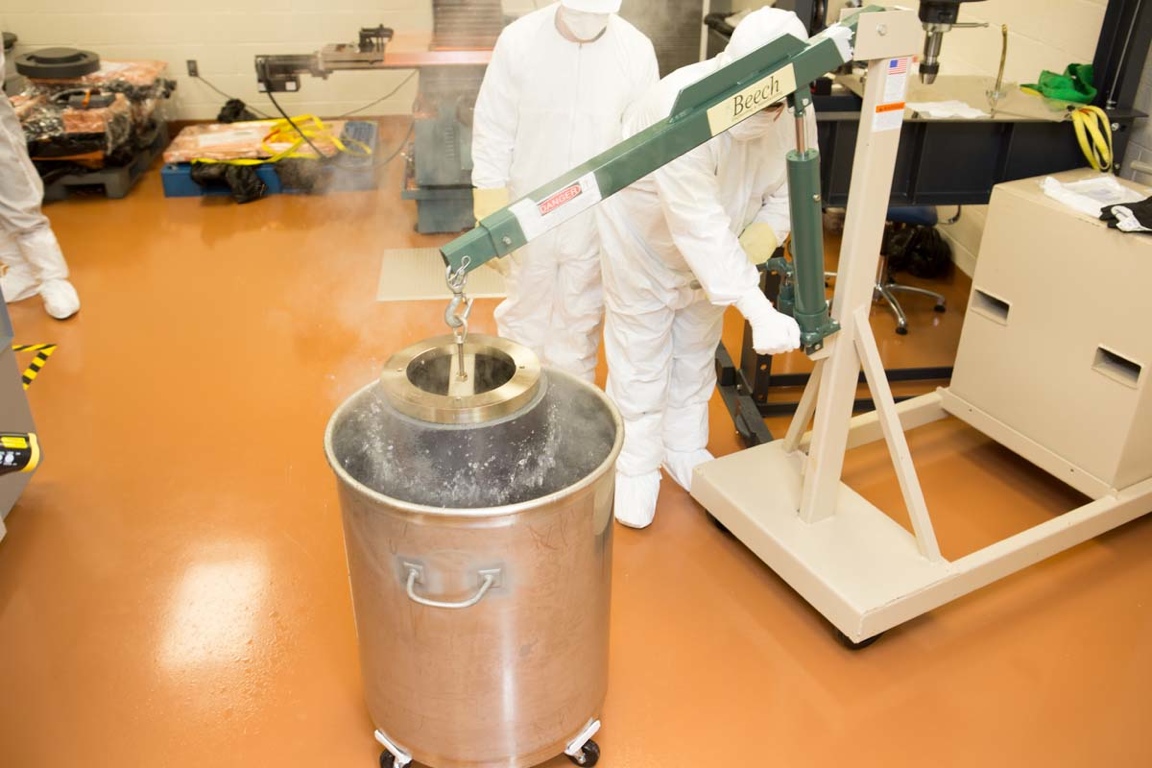}
\caption{Copper being removed from mandrel via thermal shock.}
\label{fig:dunking}
\end{center}
\end{figure}

The main purpose of the underground machine shop was to prepare and machine all of the EFCu produced in a low cosmic-ray environment. 
Once the electroplated copper was thick enough, it was removed from its bath, packaged, and transported from the production sites at PNNL or the TCR. 
EFCu from the TCR was transported underground via rail car to the machine shop located at the David Campus. 
The mandrel was mounted on a large lathe and the rough outer layer of copper was removed (Fig.~\ref{fig:Cu_MandrelOnLathe}). 
The copper and mandrel were then heated in an industrial oven, followed by an immersion in DI water (Fig.~\ref{fig:dunking}). 
The differing thermal contraction properties of copper and stainless steel allowed for separation of the copper from the mandrel. 
After the copper material was removed from the mandrel and the base removed, it was typically cut into two equal halves lengthwise and then flattened in a hydraulic press to make sheet stock material. 
Specialized mandrels were used to produce the copper cryostats, IR shields, hoops, and cross-arms. 

\begin{figure}[ht]
\begin{center}
\includegraphics[width=0.8\textwidth]{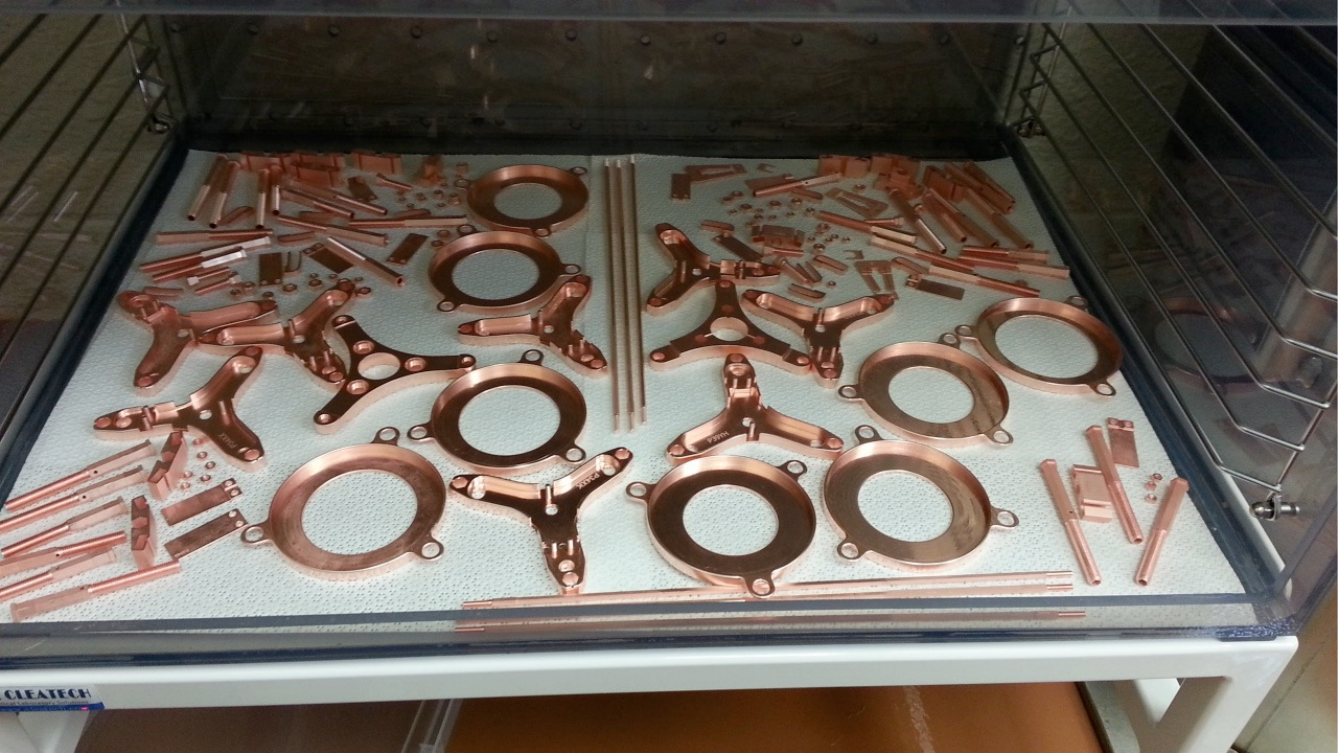}
\caption{A selection of small copper parts fabricated in the underground machine shop.}
\label{fig:small_cu_parts}
\end{center}
\end{figure}

Each final fabricated component was  cleaned, etched, passivated and stored under nitrogen purge gas until installation~\cite{hopp07,abgr16}. 
Each part was also assigned a unique database identifier and tracked using a parts tracking database that tracked its entire fabrication history (Sec.~\ref{se:parts_tracking}).
The identifier was laser engraved onto each part, except for parts that were too small.
Fig.~\ref{fig:small_cu_parts} shows a selection of fabricated copper parts.

Table~\ref{tab:underground_tools} provides a list of machine tools in the underground shop.
The underground machine shop was staffed by the specially trained machinists who followed cleanroom protocols and unique procedures to maintain cleanliness.
Only selected components and materials were allowed to be machined in this shop to reduce the chance of contamination. 
These included copper, clean plastics, and mandrels. 
The \MJD\ required approximately 8000~hours of underground machining time to fabricate the required parts. 

The cryostats required a small number of copper welds.
Electron-beam welding (EBW) is a demonstrated method for performing radio-pure welds.
Unfortunately the collaboration could not deploy such a welder underground because of its cost and size. 
Instead, the EBW was performed on a small subset of components at various locations across the US.
Once these components were brought to the surface, they were immediately driven to the welder, welded, and immediately driven back to SURF and returned underground.

\begin{table*}[ht]
\begin{tabular}{||l|l|l||}
\hline \hline
Description & Vendor and Model & Main Function\\
\hline
Small CNC Mill & HAAS OM-2A Office Mill & Precision machining of small components\\
Small CNC Lathe & HAAS OL-1 Office Lathe & Precision machining of small components \\
Medium CNC knee Mill & TRAK DPM SX3P & Machining of large sheet stock and shield \\
&&plates.\\
Wire EDM & Sodick VZ300L & Precision fabrication of small components \\
Large Lathe & TRAK TRL 2460SX & Surface cuts of copper on mandrels and \\
& & mandrels. Cryostat fabrication. \\
Laser Engraver & Epilog Fibermark & Engraving part numbers \\
Bandsaw & ACRA KB36 & Cutting and sizing \\
Drill press & ACRA MD-32MMF & Prepping of stock \\
75 Tonne Hydraulic press & Dake Model 75H & Flattening EFCu stock \\
Oven & Despatch RAD2-35-2E & Heating copper on mandrel prior to quench.\\
\hline \hline
\end{tabular}
\caption{List of major machine tools and other equipment in the underground machine shop.
\label{tab:underground_tools}}
\end{table*}

\subsection{Underground Infrastructure}
\label{se:ug_infrastructure}

\begin{figure}[ht]
\begin{center}
\includegraphics[width=0.99\textwidth]{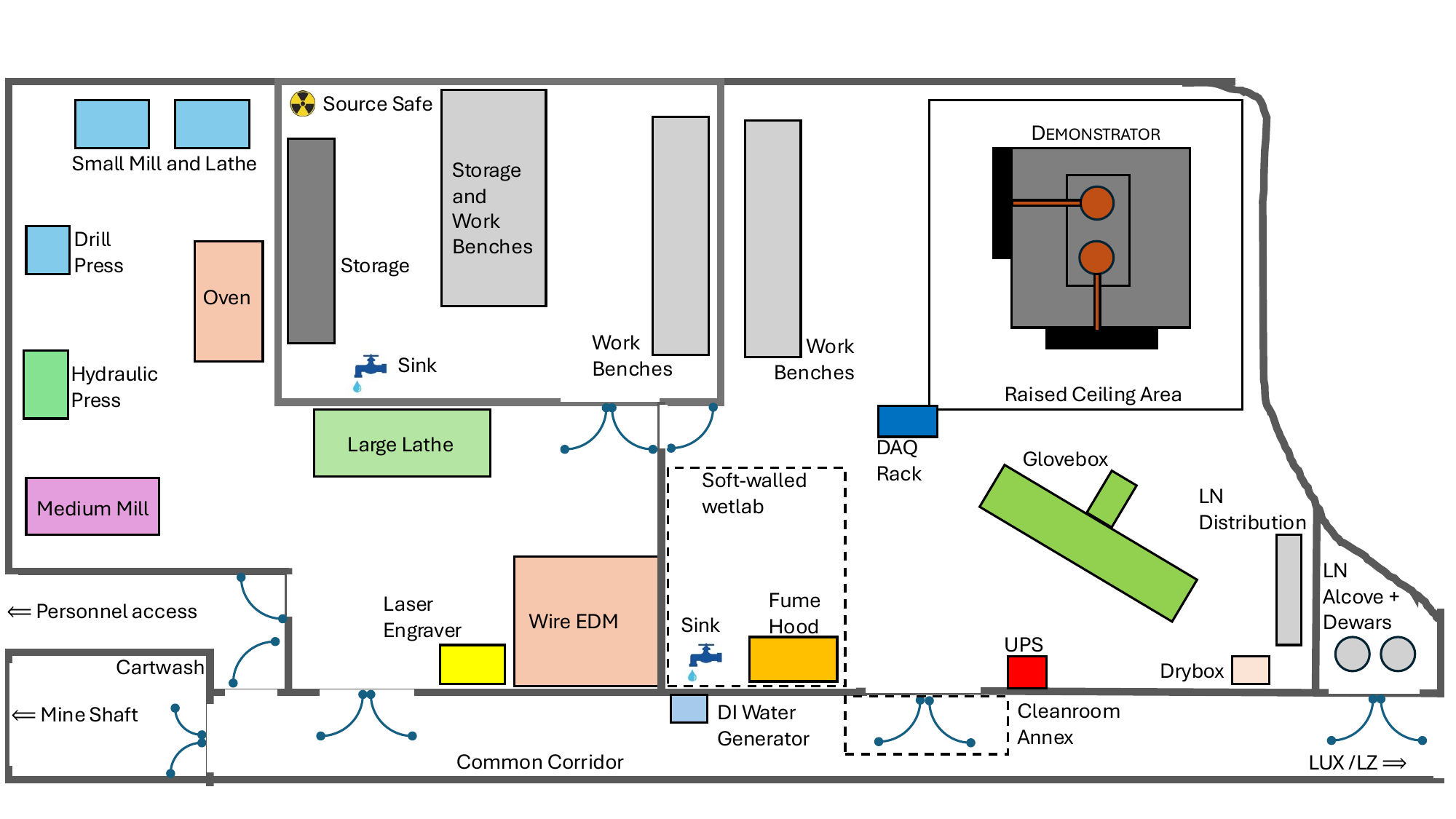}
\caption{The general layout of the 80' $\times$ 40' \MJD\ Davis campus lab. 
Shown are the major machine tools and locations of significant equipment.
On the bottom right is the LN alcove.}
\label{fig:davis_layout}
\end{center}
\end{figure}

The \DEM\ laboratory was located in the Davis campus on the 4850-foot level of SURF ($\sim 4260$~m.w.e.) in a specially excavated cavity that was outfitted with an 80' $\times$ 40' (24.38~m $\times$ 12.19~m) laboratory area, shown in Fig.~\ref{fig:davis_layout}.
The \MJD\ laboratory space consisted of three cleanroom areas: the machine shop (Sec.~\ref{se:machining}), a general lab area, and the main detector laboratory, as shown in Fig.~\ref{fig:davis_layout}. 
Outside the cleanroom was a small alcove that housed the Liquid Nitrogen (LN) dewars for the \MJD . 
Access to the Davis campus is via vertical mine shaft, specifically the Yates shaft, which placed significant constraints on the size of equipment transported underground. 

\begin{figure}[ht]
\begin{center}
\includegraphics[width=0.99\textwidth]{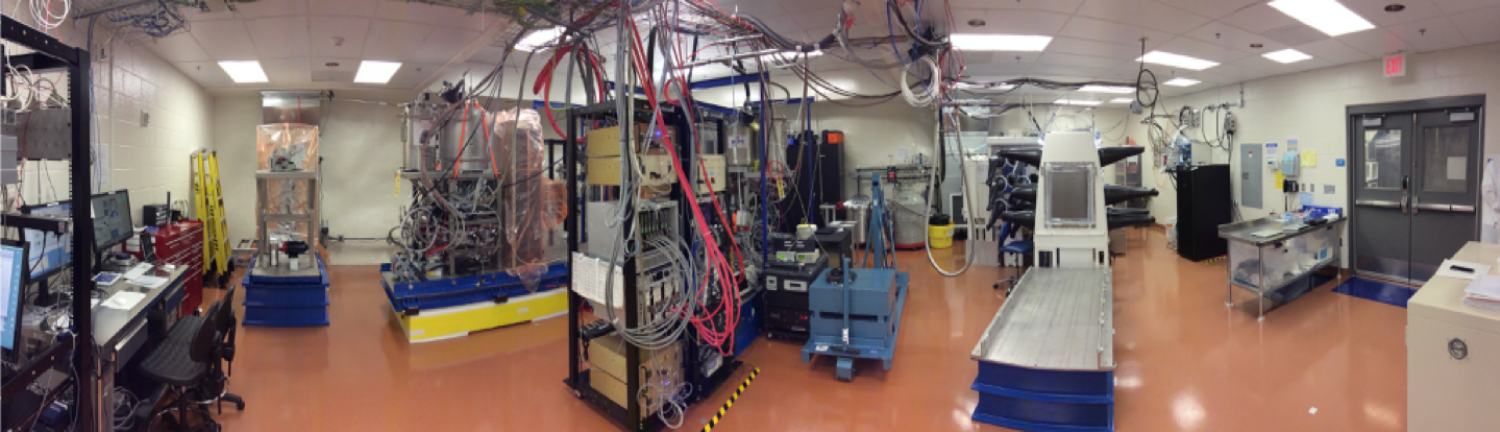}
\caption{A panorama of the detector lab. 
To the left is a shield monolith and module that have been removed and are resting on the Hovair (section~\ref{passive_shield}). 
The experiment itself is behind the DAQ rack at the center, and the portion of raised ceiling is visible above it. On the right is the glovebox and entry doors to the lab. The wet lab is off-camera to the right. }
\label{fig:det_lab_panorama}
\end{center}
\end{figure}

The detector laboratory housed the experiment, a small softwalled wetlab with fume hood, the glovebox for assembling parts that go inside the cryostats, a drybox flushed with LN purge gas for detector storage, and data acquisition systems.
A portion of the detector laboratory had a raised ceiling to accommodate the full height of the \DEM\ shield assembly.  
A panoramic view of the detector lab is shown in Fig.~\ref{fig:det_lab_panorama}. 
LN from two dewars in the alcove was distributed using vacuum-jacketed lines mounted to the ceiling. 

The  \MJD\ laboratory was designed as a class~1000 cleanroom, though during operations it was typically better than class~100.
For cleanliness, a positive differential pressure was maintained between the laboratory and the rest of the Davis campus. 
Utilities included three phase power for machine tools, compressed air, venting for fume hoods and the laser engraver, non-potable water, and 1000BASE-T ethernet. 
The collaboration installed a dedicated Millipore Super-Q water purification system to provide water for the EDM machine and parts cleaning.
An uninterruptible power supply (UPS) was also installed to mitigate unplanned power outages and prevent damage to the HPGe detectors and electronics from uncontrolled loss of high-voltage or vacuum. 
The UPS was a Smart-UPS\texttrademark\ VT\texttrademark\ 10-40 KVA operation of 200/208/220V APC unit by Schneider Electric.

The Davis campus also incorporates the cavity for the original Ray Davis solar neutrino experiment~\cite{Cleveland_1998}. 
During \MJD\ construction and operations that space was occupied by the LUX~\cite{LUX} and subsequently LZ~\cite{LZ} experiments. 
Various levels of cleanliness were maintained throughout the Davis Campus, depending on the needs of the collaborations. 
Prior to entry into the Davis campus, researchers removed dirty coveralls and boots, and donned clean hard toed shoes, bootcovers, and hairnets.
Large equipment was cleaned in a cartwash prior to being moved into the Davis Campus.
Entry into the main \MJD\ laboratory required full cleanroom garb, which was donned in a small cleanroom annex in front of the main doors. 
Most incoming items were double bagged, with the outer layer of bagging removed before being brought into the lab space.

A purge system built by the collaboration provided pure, low radon, and dry N$_2$ purge gas from boiloff LN. 
This gas purged the detector assembly gloveboxes, a part storage box, electronics boxes mounted on the modules, the space within the polyethylene shielding, and the vacuum system back-fill purge. 
A separate system purged the inner volume of the \DEM\ shield (section~\ref{sec:purge}).
Both LN boil-off purge system operated autonomously and drew LN from dewars located in the alcove. 

The TCR (section~\ref{sec:electroforming}) was located at the same level about 1~km away near the Ross shaft station. 
Materials and personnel were transported between the two sites on foot or via railcar. 


\section{Detector Enrichment, Fabrication, and Acceptance}
\label{se:detectors}

This section describes the enrichment and chemical processing of the of the germanium source material. 
It also describes  detector fabrication, handling, and characterization before installation into the \MJD .

\subsection{Germanium Processing and Detector Fabrication}
\label{sec:geprocessing_and_fabrication}
The \DEM\ used high-purity germanium, isotopically enriched in \gess , as both detection medium and source.  Prior to the \DEM, the most sensitive experiments used detectors fabricated from germanium enriched to 86\% in \gess\ ~\cite{aals07, kla04a}. For comparison, the natural abundance of \gess\ is 7.8\%.

Enriching germanium is expensive and a major cost driver for a future tonne-scale experiment. 
As part of its R\&D effort, the collaboration developed techniques to minimize the loss of enriched germanium when it is converted from oxide into HPGe detectors. 
Additionally, though HPGe detectors are intrinsically very radiopure, cosmic-rays at the earth's surface can interact with the germanium and produce radioactive isotopes, with \gese\ (271~d) and \cosixty (5.27~y) being of particular concern as backgrounds to a \BBz-decay search and \nuc{3}{H} (12.3~y) a concern to the low-energy BSM physics program. 
This required special handling and transportation of detectors and enriched germanium oxide in shielded containers. 
The enrichment, chemical processing and handling of the germanium is described in detail in Ref.~\cite{abgr18}, and we only provide a brief overview here.

The isotopic enrichment was performed at the large centrifuge facility,
Electrochemical Plant (ECP), in Zelenogorsk, Russia. 
A total of 42.5~kg of enriched \gess\ in the form of 60.4~kg of \geoxide\ was purchased from ECP.
The collaboration built a facility in Oak Ridge, Tennessee that was managed by Electrochemical Systems, Inc. (ESI) to carry out the reduction of the \geoxide\ to germanium metal.
This facility also zone refined the metal to $10^{13}$ electrically active impurities/cm$^3$ or better as required by detector manufacturers, and it reprocessed scrap germanium left over from the detector fabrication process.
The enriched isotope was transported as powdered oxide from Russia via truck and boat to Oak Ridge in a shielded container to reduce cosmic-ray activation.
The high cosmic-ray flux at commercial airline cruising altitude made air transport not viable. 

The enriched detectors used in the \MJD\ were a point-contact design manufactured by AMETEK-ORTEC Inc, at their facility in Oak Ridge. 
They zone-refined the metal to $10^{11}$ impurities/cm$^3$ ($47\,\Omega$.cm resistivity) and grew germanium crystal boules in a Czochralski crystal puller, which further purified the germanium metal.
The boules are machined and etched before being converted into detectors. 
The collaboration and ESI was able to recover  material normally rejected after zone refining and crystal growing.
The  fraction of enriched germanium mass converted to detector-grade germanium was 98.3\%. 
Thirty-five point-contact detectors having a total mass of 29.7~kg were fabricated for the \MJD, which represented an overall yield of detector mass to that of purchased material of 69.8\%, with
2.64\,kg of  germanium remaining that could be converted into future HPGe detectors. This is the largest yield to date for a germanium experiment~\cite{abgr18}.
The largest loss was from machining and etching the germanium at various stages of the detector manufacturing process. 
The collaboration took care to store germanium underground in a nearby cave (80~m.w.e.) when it was not being processed or converted into detectors. 
The average overall estimated sea-level equivalent exposure for all detectors, excluding detector manufacturing, was 12.5~days. 

An isotopic mass abundance of $88.1\pm0.7$\%  \gess\  was initially determined by inductively coupled plasma mass spectrometry (ICP-MS). 
This value was used in the first two \BBz-decay results~\cite{aals18,alvi19b}
Further refinement of the analysis using a maximum-likelihood technique that included other isotopes (see Appendix~\ref{se:abundances}) modified the enrichment fraction slightly to $87.4\pm0.5$\%, which was used in the final \BBz-decay result~\cite{arnq22}.

The collaboration also deployed 23~natural (unenriched) germanium modified Broad Energy Germanium (BEGe) detectors from CANBERRA/Mirion Industries with a total mass of 14.4~kg. 
These were modified to not have the thin front window that permits sensitivity to low-energy external gamma-rays. 
No measures were taken to decrease or track the exposure of these detectors or their stock material during fabrication, transportation, or storage. 
Finally, during the upgrade described in section ~\ref{se:CC_upgrades}, the collaboration installed four enriched \ICPC\ detectors (6.7 kg) enriched to 88$\pm$1\%. 
These were also manufactured by AMETEK-ORTEC Inc.

\subsection{Detector Acceptance, Transport and Storage}
\label{se:det_tranport}
Once the enriched detectors were fabricated, AMETEK-ORTEC mounted them inside their own vendor cryostats with their pre-amplifier and associated electronics. 
The collaboration performed a preliminary acceptance test at the AMETEK-ORTEC's facility  during which the detector energy resolution and multi-site/single-site (MS/SS) discrimination performance were determined with \cosixty\ and \thttt\ calibration sources. 
MS/SS discrimination was achieved in the \DEM\ using pulse-shape discrimination (PSD) techniques applied to digitized waveforms.
MS/SS discrimination was required to separate predominantly multi-site gamma-ray backgrounds from single-site \BBz -decays.
Its power was estimated using single-escape and double-escape events from the \thttt\ 2615~keV source~\cite{PhysRevC.99.065501}.  

Detectors with unsatisfactory performance were returned to AMETEK-ORTEC to be reworked.
Accepted detectors were kept in their vendor cryostats and driven in batches to SURF. The drive took two days and the overnight location was chosen to be at low altitude to minimize cosmic-ray exposure. 
When detectors arrived at SURF, they were immediately taken underground to the \MJD\ laboratory.
Prior to characterization at SURF, the detectors were removed from their vendor cryostats and a visual inspection was performed to check for damage during the transport, followed by geometrical and mass measurements. 
The detectors were then returned to their cryostats and cooled.
The depletion voltage and leakage current were measured, followed by radioactive source measurements at typically 500V above the depletion voltage. 
Data from these source tests were recorded with both a shaping amplifier and Multi-Channel Analyzer (MCA), and digitizers originally developed for the GRETINA~\cite{lee04, zimm12} experiment (see section ~\ref{se:DAQ_Software}). 
The energy measurements reported here were computed by applying a trapezoidal filter to the digitized waveforms.

\begin{figure}[ht]
\centering
\includegraphics[width=0.9\textwidth]{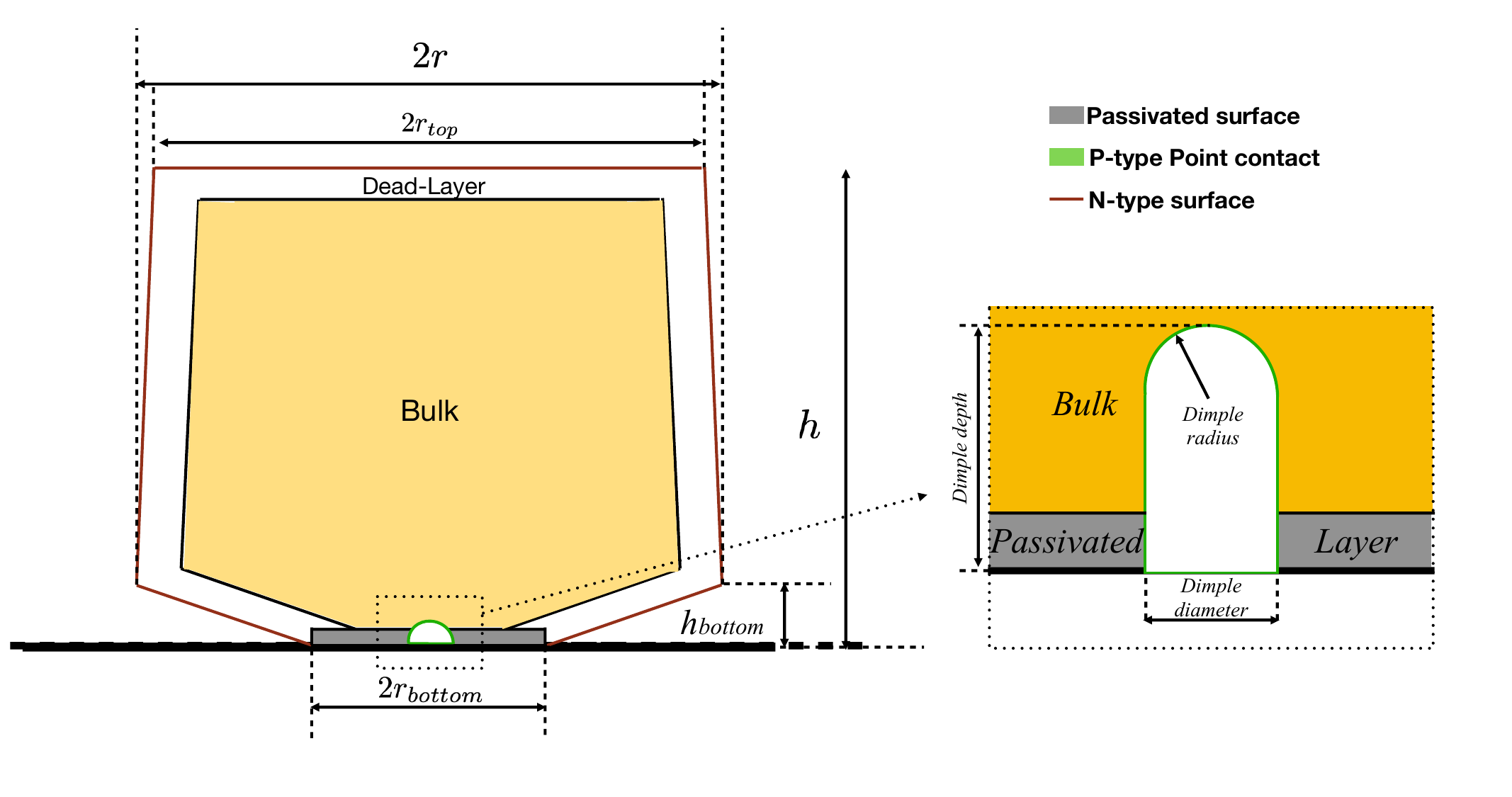}
\caption{The geometry of PPC detector that shows the dead layer outside the yellow shaded bulk or sensitive volume (left) and a zoom of the point-contact region (right). Some relative dimensions are exaggerated for clarity.}
\label{fig:PPC_Geo}
\end{figure}

The geometrical measurements provided all dimensions of the PPC detectors, as shown in Fig.~\ref{fig:PPC_Geo}. 
The dimension were provided by the vendor and verified using a Starrett Galileo EZ200 optical metrology instrument. 
The PPC detectors had a slight conical profile with an angled base approaching the passivated surface. 
The height and radius of the PPC detectors were unique to each detector, and 
the \DEM\ PPC detectors had heights that ranged from 31.0 -- 54.1\,mm and radii ranging from 30.1 -- 36.0\,mm. 
In the center of the bottom face was a small cylindrical dimple with a depth up to 2.2\,mm and a diameter up to 4.6\,mm, which forms a p$^+$ point contact where the germanium had been doped with boron.
At the corner of the face, a cut of approximately 45 degrees was made at a distance of between 4.5 -- 9\,mm from the cylindrical corner. 
The outer top, sides, and angled corner of the detector surface were doped with lithium to form the n$^+$ contact. 
The remaining surface between the n$^+$ contact and the p$^+$ was passivated.
The passivated surface was also an important source of degraded alpha events that had to removed using analysis cuts~\cite{Johnson2012,Arnquist_2022,Abt:2016trw}. 
At bias voltages above the depletion voltage, the full detector volume will be active, with the exception of small incomplete charge collection regions near the passivated surface and across the n$^+$ contact layer~\cite{agua13}. 
These regions are referred to as a dead layer.

Because of the incomplete charge collection, the $\sim 1$mm thick dead layer had to be excluded when the detector's active mass was calculated. 
The dead layer was comprised of a fully dead layer where no charge was collected and a transition layer where only a fraction of the charge was collected~\cite{agua13}. 
Energy deposits in the transition layer yielded energy-degraded events that accumulated at low energy, where they impact many BSM physics searches~\cite{GIOVANETTI201577, PhysRevD.88.012002}. 
In addition, the charge-trapping effect inside the active volume does not produce any reduction of the active mass but causes slight energy degradation~\cite{arnq22,arnq22f}. 
This degradation worsens the energy resolution and may shift a 2039~keV energy deposition to a lower energy outside the ROI. 
The radioactive source measurements quantified these energy degradation effects for each detector.

\begin{figure}[ht]
\centering
\includegraphics[width=0.48\textwidth]{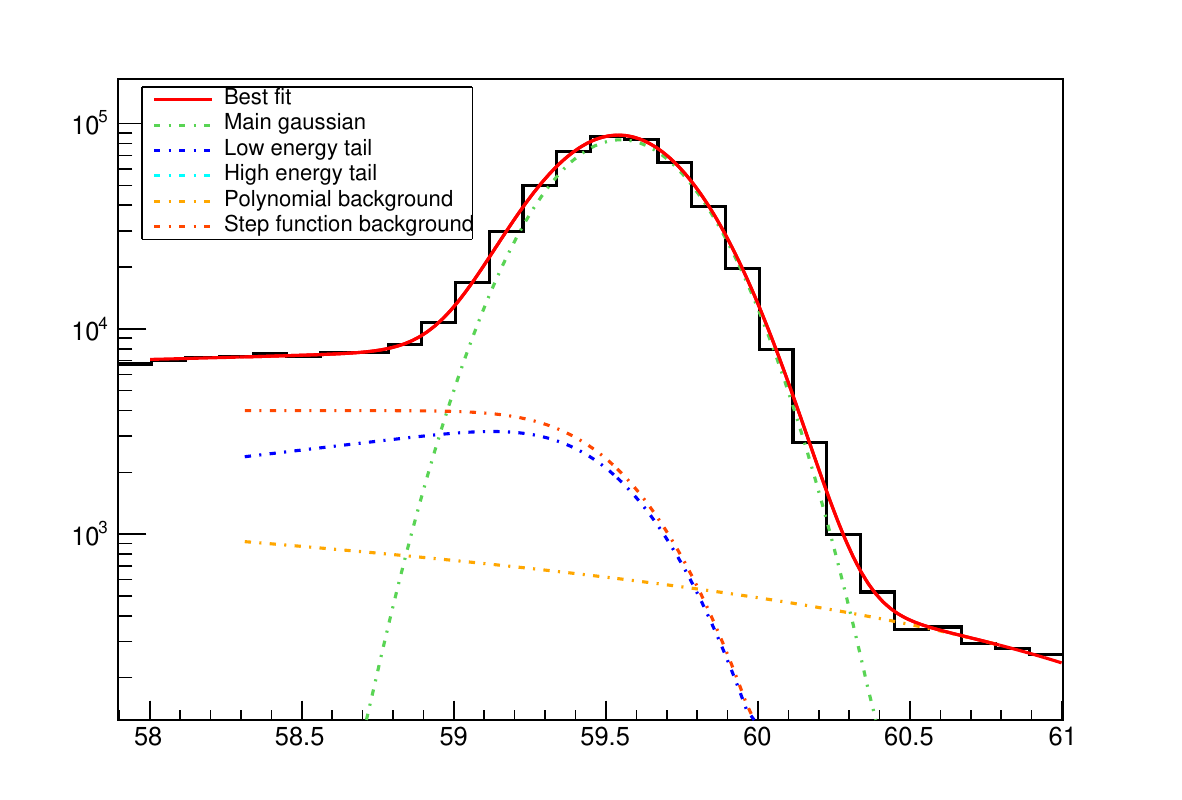}
\hfill
\includegraphics[width=0.48\textwidth]{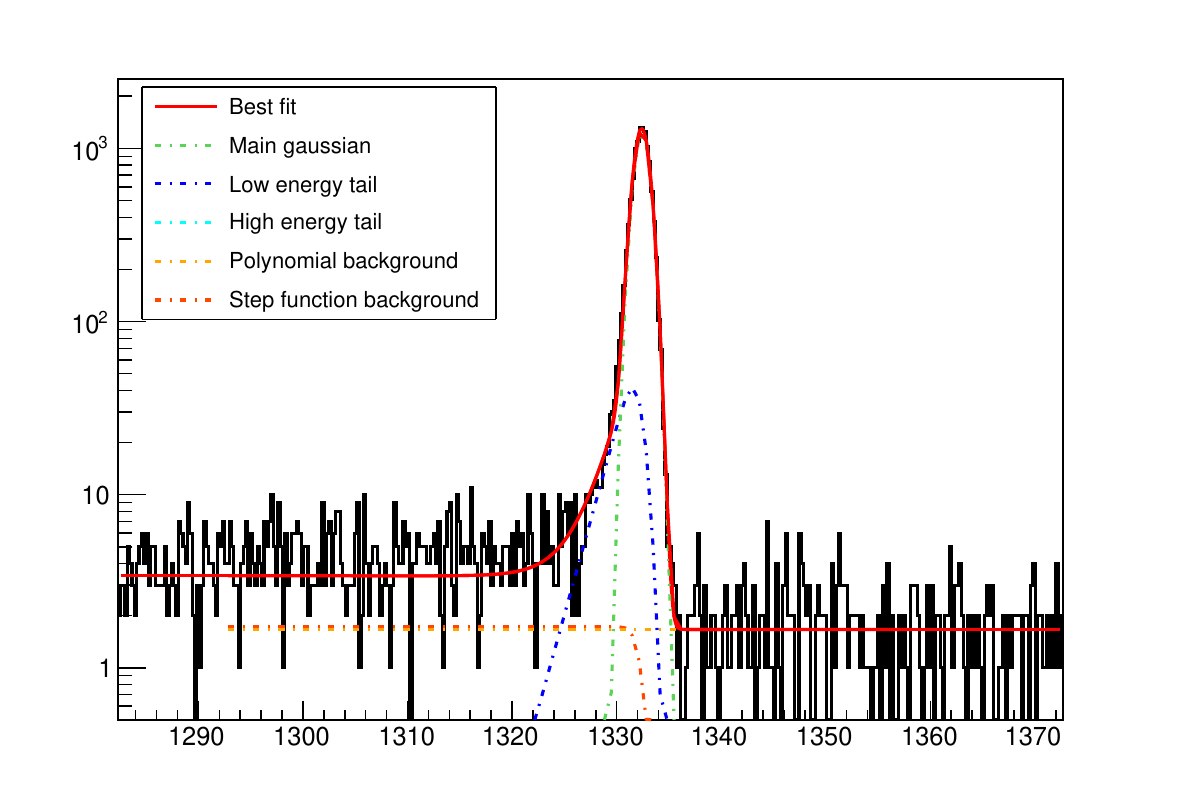}
\\
\includegraphics[width=0.48\textwidth]{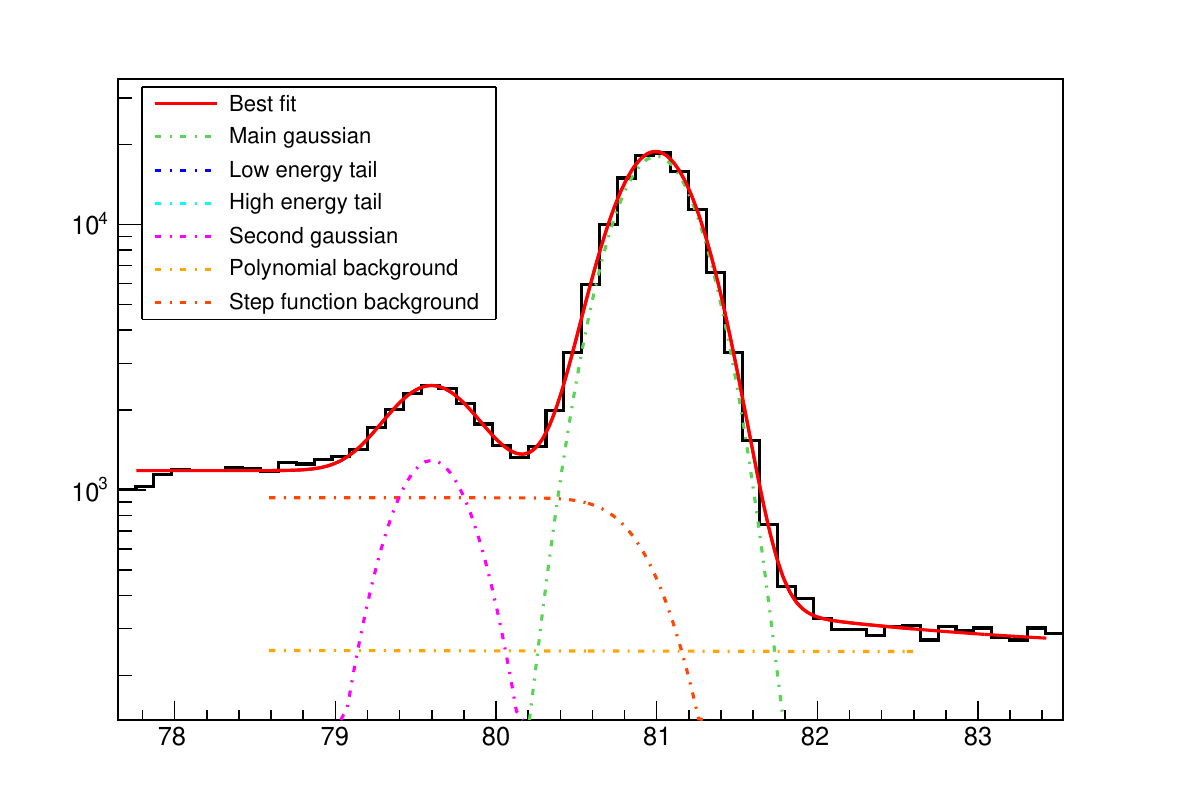}
\hfill
\includegraphics[width=0.48\textwidth]{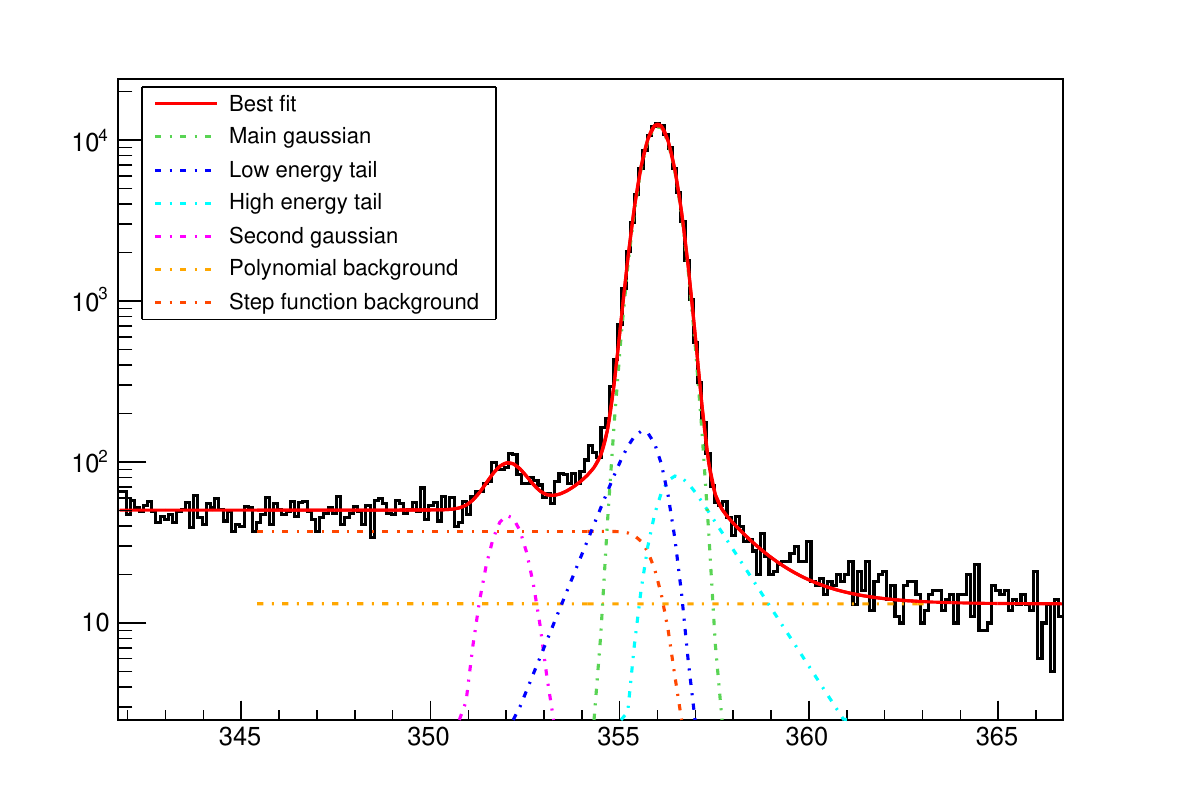}
\caption{An example peak-shape fit to the 56.5 keV $^{241}$Am peak (top left), 1.33 MeV $^{60}$Co peak (top right), 81.0~keV $^{133}$Ba peak (bottom left) and 356 keV $^{133}$Ba peak (bottom right) for a typical detector. The low-energy tail quantified the energy degradation effects in the detector. Note the logarithmic scale on the vertical axis.}
\label{fig:Fits}
\end{figure}

The radioactive source measurements consisted of flood measurements with $^{60}$Co, $^{133}$Ba, $^{241}$Am, and $^{232}$Th sources, and translational scans across the detector top and sides with a collimated $^{133}$Ba source.
These measurements provided energy resolution, dead-layer thickness, and relative efficiency.
The flood source measurements of the $^{60}$Co, $^{133}$Ba, and $^{241}$Am sources were used to determine the dead-layer thickness and energy resolution. 
Spectral fits of the main peaks, which were the 1332~keV peak from $^{60}$Co, the 59.5~keV peak from $^{241}$Am, and the 81~keV and 356~keV peaks from $^{133}$Ba, quantified the energy degradation effects. 
The spectral peaks were fitted to the standard \MJD\ peak shape, which was composed of a Gaussian peak, an exponential low energy tail, and an exponential high energy tail. 
A step background function and a linear background were also included, and an example is shown in Fig.~\ref{fig:Fits}.
In the case of the 1332~keV and 59.5~keV peaks, the high energy tail was suppressed, as typically found in \MJD\ data~ \cite{alvi19b,aals18}. 

For the last five detectors received at SURF, which were the ICPCs, no $^{133}$Ba flood measurements were made; instead a collimated source was scanned across the top surface. 
The results of these five detectors were consistent with the manufacturer specifications, but since their measurements followed different methods, they  were  not included in the summary plots presented here. 

\begin{figure}[ht]
\centering
\includegraphics[width=0.6\textwidth]{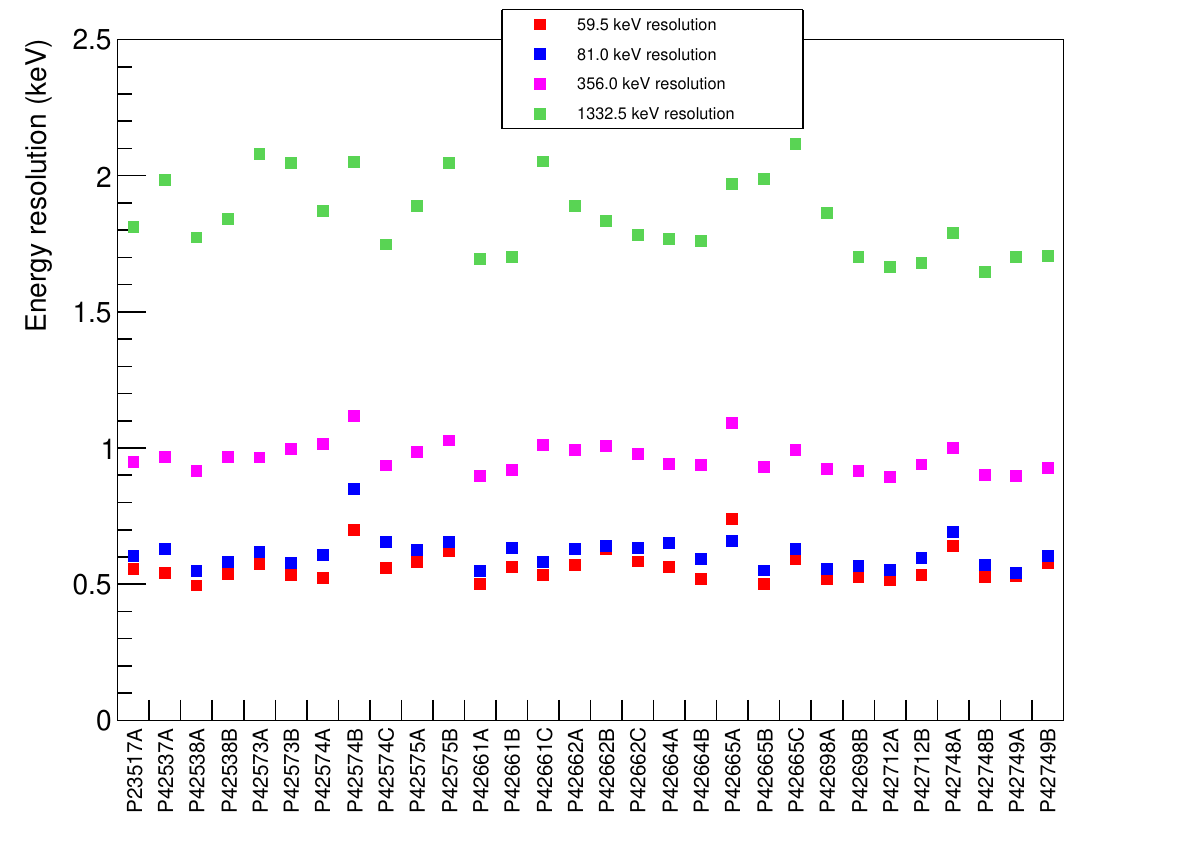}
\includegraphics[width=0.6\textwidth]{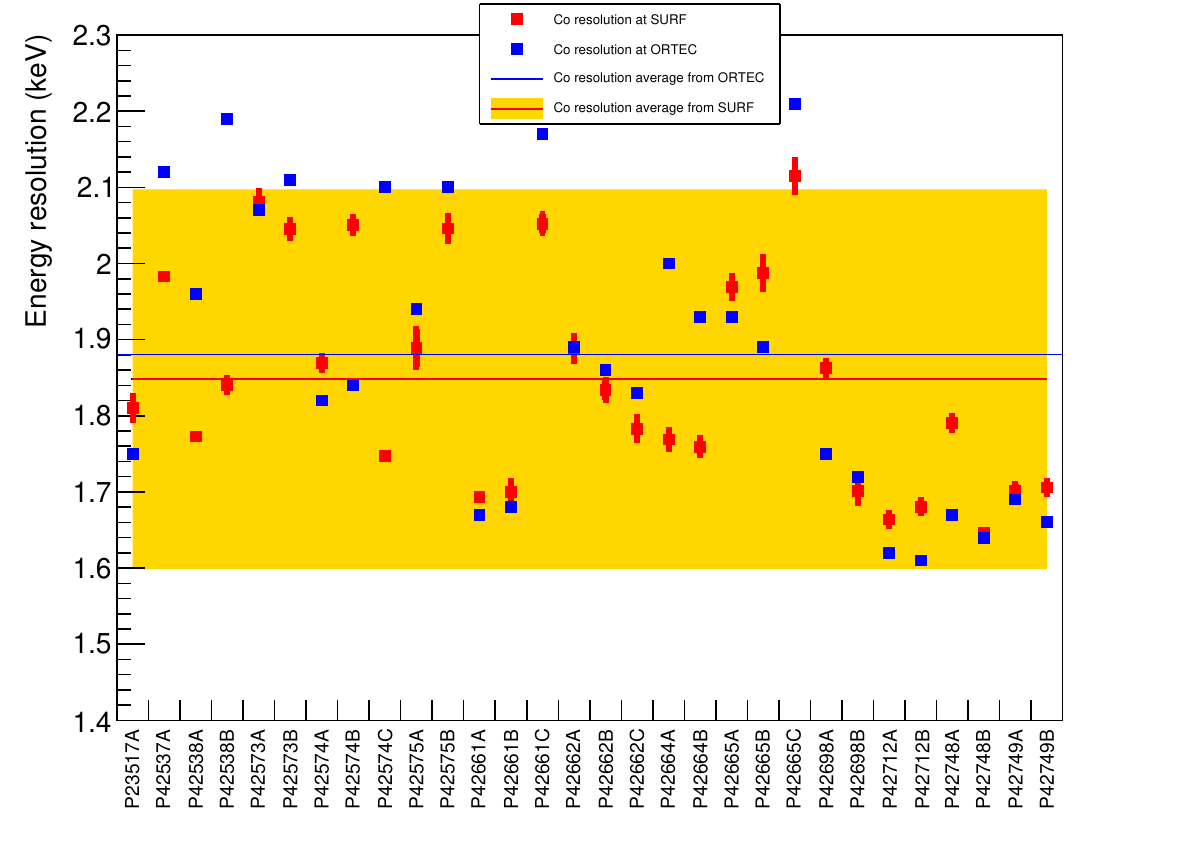}
\caption{Top: Summary of the energy resolution (in FWHM) measurements from the fits of the 59.5~keV peak (red), 81~keV peak (blue), 356~keV peak (purple), and 1332~keV peak (green).
Bottom: The 1.33~MeV FWHM gamma resolution for all detectors according to manufacturer measurements (blue) and \MJD\ acceptance measurements at SURF (red). The yellow band is the 90\% CL of the measurements done at SURF. The horizontal axis  shows detector serial numbers.}
\label{fig:Co}
\end{figure}

Measuring the detector energy resolution with a $^{60}$Co gamma source allowed a direct comparison with manufacturer specifications, shown in Figure~\ref{fig:Co}. 
The average energy resolution of 1.88~keV FWHM resolution at SURF was consistent with the 1.85~keV resolution provided by the manufacturer.

\begin{figure}[ht]
\centering
\includegraphics[width=0.48\textwidth]{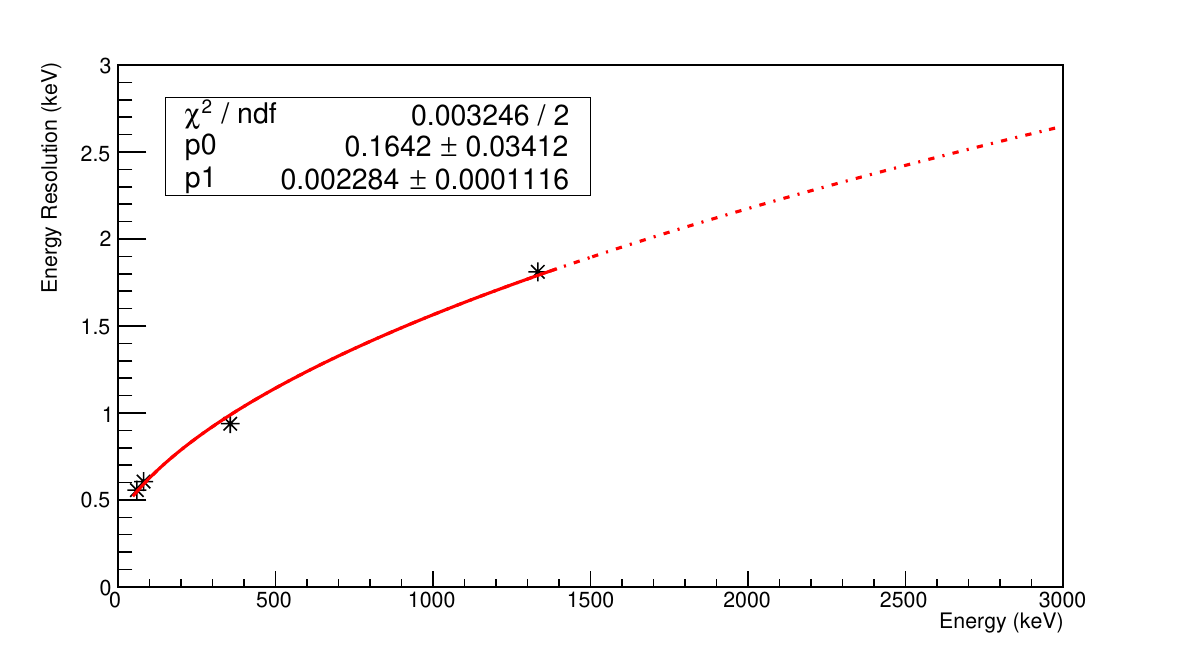}
\hfill
\includegraphics[width=0.48\textwidth]{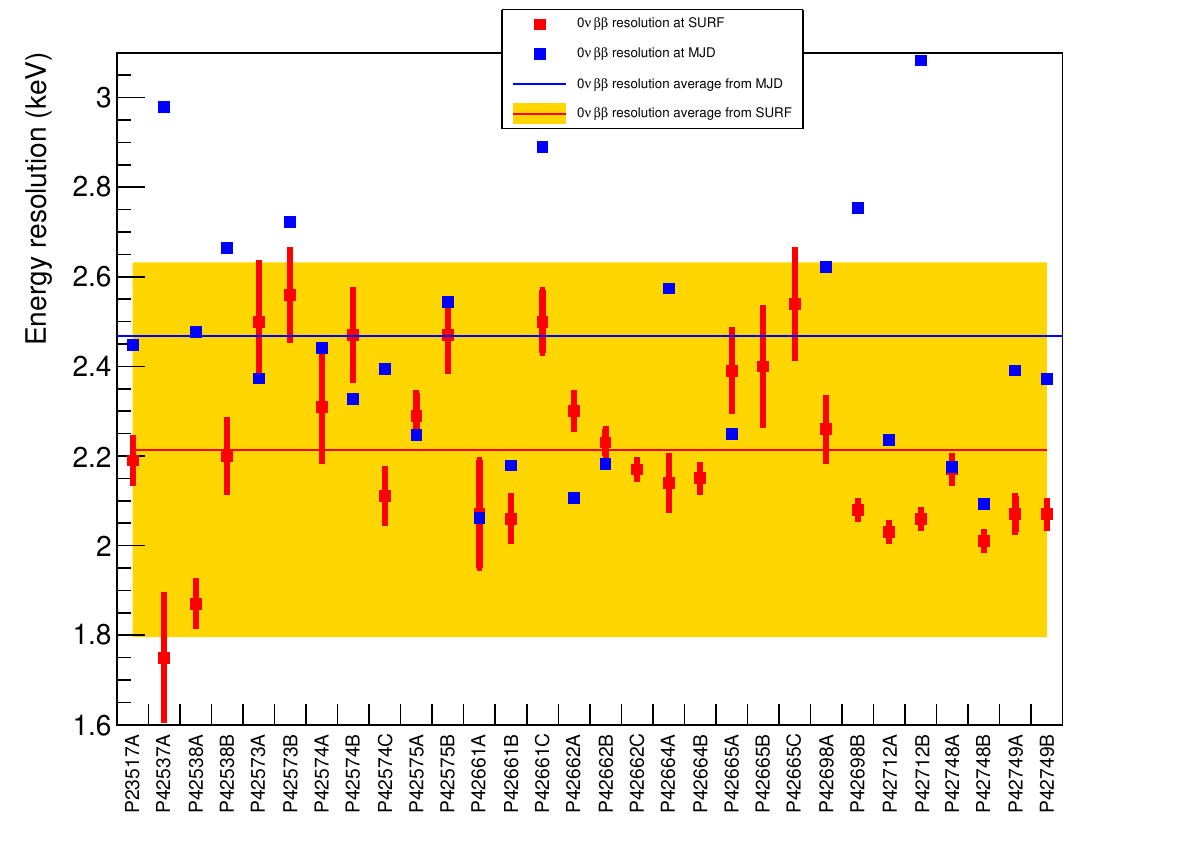}
\caption{Energy resolution curve for a detector (serial number P23517A) showing measurements and fit (left). On the right is shown the extrapolated resolution in the region of interest from the acceptance measurements done at SURF (red) and the in-situ measurements from \DEM\ calibration data (blue). }
\label{fig:ROIRes}
\end{figure}

The energy resolution data points shown in Fig.~\ref{fig:Co} provide an extrapolation to the energy resolution in the \BBz -decay region of interest at 2039~keV. 
The resolution as a function of energy was typically estimated using~Eqn.~\ref{eq:Reso_Compl}, where $p_{0}$ accounts for the electronic noise, $p_{1}$ for the Fano factor and the linear energy response of HPGe detectors, and $p_{2}$ for the charge trapping. 
\begin{equation}
\label{eq:Reso_Compl}
\sigma\left(E\right) = \sqrt{p^{2}_{0}+p^{2}_{1}E+p^{2}_{2}E^{2}} 
\end{equation}
In the acceptance measurements, only four source peaks were available.
As the SURF acceptance tests were intended to only show that the resolution was within the desired specifications, the approximation of $p_{2} = 0$ was used, generating functions like the one in the left side of the Fig. \ref{fig:ROIRes}.
The results of these extrapolations show an average of 2.2~keV FWHM resolution at 2039 keV at 77~K. 
The resolutions measured in-situ in the \DEM\ were on average 10\% higher~\cite{alvi19b}, most likely because the temperature in the array was slightly higher (78-80~K vs. 77~K) than in the vendor cryostats, see section~\ref{se:cryovac_performance}.

\begin{figure}[ht]
\centering
\includegraphics[width=0.48\textwidth]{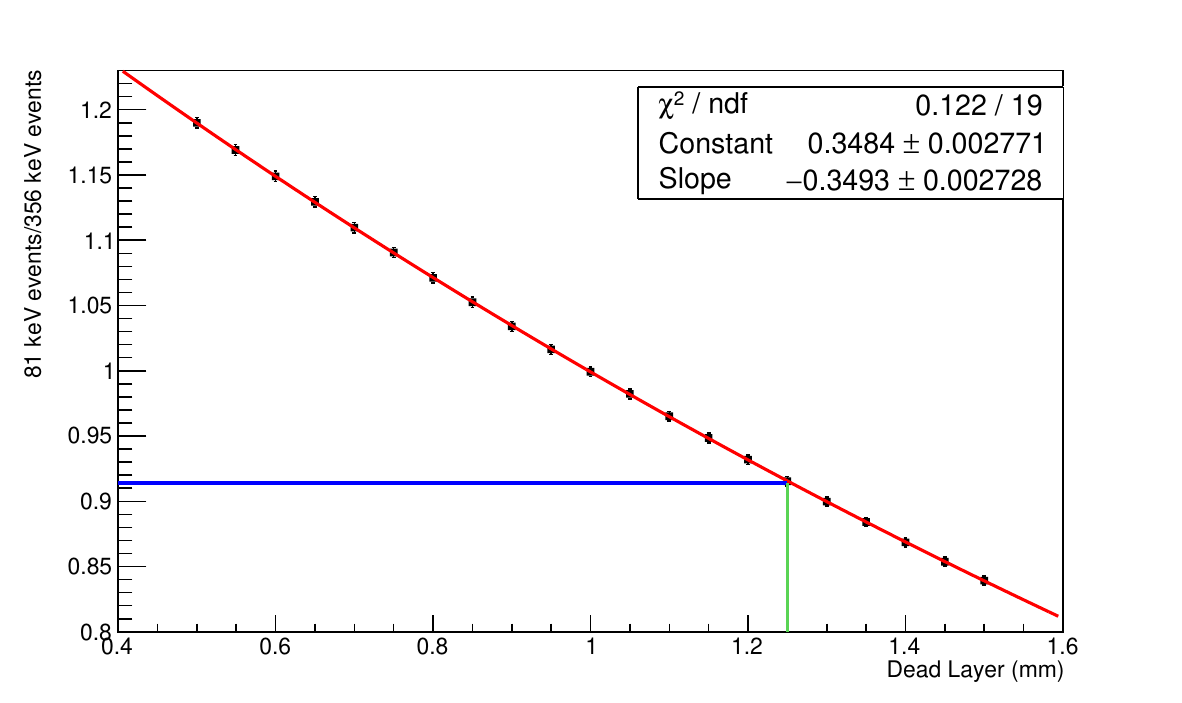}
\hfill
\includegraphics[width=0.48\textwidth]{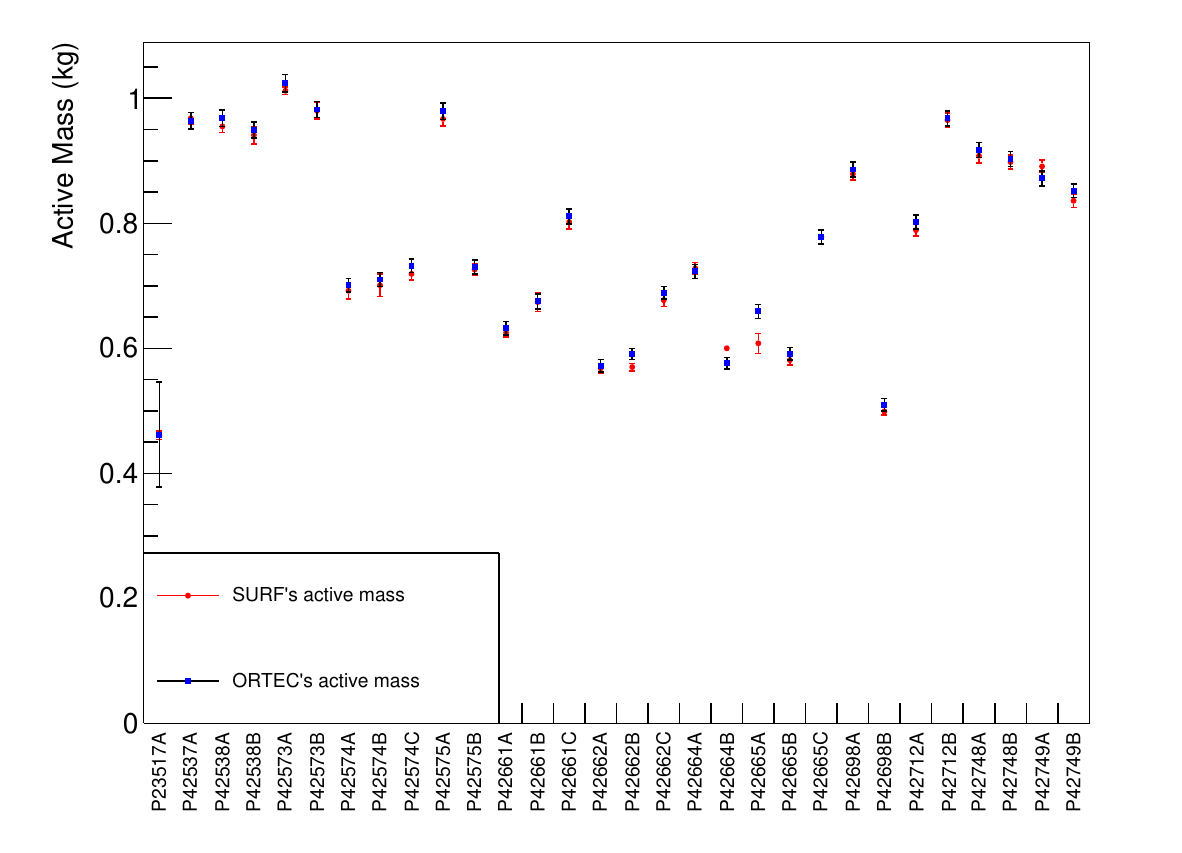}
\caption{Left: Fit of an exponential function (red line) to the 81 keV/356 keV ratio from simulations with different step function dead-layer thicknesses (black points). The measured ratio for a typical detector (horizontal blue line) and its intercept with the fit provide its estimated dead-layer thickness (vertical green line). 
Right: Active mass comparison for PPC detectors in the \MJD . The active masses were computed using two different techniques (see text).}
\label{fig:DL}
\end{figure}

The acceptance measurements also allowed a measurement of the dead layer profile by using the energy-dependent penetration depths of gamma-rays. 
For the 81~keV and 356~keV gamma-rays from the \nuc{133}{Ba} source, the mean-free paths in germanium are 1.9~mm and 17~mm, respectively, and for  a typical dead-layer of $\sim$1~mm, the 81~keV gamma-rays were attenuated significantly, while the 356 keV gamma-ray was relatively unaffected. 
Therefore, the dead-layer thickness could be determined from the ratio of the peak strengths between these two peaks and compared to that predicted by simulations with varying dead-layer thicknesses. 
$^{133}$Ba source simulations were produced using the MaGe simulation package.
The simulation results  were fitted to an exponential function to determine the dead-layer thickness from the 81~keV/356~keV ratio, as shown in Fig.~\ref{fig:DL}. 
The uncertainty of the simulations was estimated to be 10\% \cite{Schubert2012}.
The figure shows the measured peak ratio (blue line) and the determination of the dead layer (green line).

The active mass of each detector was calculated using the measured dimensions, total mass, and the dead-layer thickness using two different methods. 
The first uses dead-layer values estimated by the manufacturer. 
The geometrical calculations were made by building a conical frustum and removing the dimple, dead-layer and passivated layer in \textit{MATHEMATICA}.
High precision was not required for the passivated layer as its contribution was negligible compared with the 10\% dead-layer thickness uncertainty. 
A second method of determining the active mass uses the dead-layer measurements described above. 
Both methods were used in published results~\cite{arnq22}, and Fig.~\ref{fig:DL} shows the results for each PPC detector.
By summing the active mass of all detectors, the active mass of the \MJD\ could be calculated.
The sum of the errors considered the active masses of the detector to be correlated since they all share the same method. 
The active fraction of PPC detectors for most of its operation was $92.0^{+1.3}_{-1.7}$\%, though these numbers varied slightly as analyses were improved.
This active mass corresponds to data taken before the 2020 hardware upgrade (Sec.~\ref{se:CC_upgrades}) where six of the PPC detectors were replaced by four ICPC detectors. 
Characterization measurements at SURF were not performed for these ICPC detectors. 
Instead, the dead layer analysis from AMETEK-ORTEC for each detector was used, yielding an active mass of $90.9^{+1.2}_{-1.6}\%$.
After the ICPC installation, up to 27.2~kg of enriched detectors were operational~\cite{arnq22}.

The rest of the detectors in the \DEM\ were natural BEGe detectors, which were not used for the \BBz -decay search directly but were helpful to identify background sources and provide vetoes. 
They were shipped by CANBERRA directly to Los Alamos National Laboratory, where they underwent similar acceptance testing. 
They were removed from their vendor cryostats and shipped to SURF inside sealed stainless steel vacuum vessels. 
Since they were procured before the laboratory was ready, they were stored underground near the TCR (see~section \ref{sec:electroforming}) inside an industrial freezer  to slow the diffusion of their lithium dopant. 
Once they were required for assembly, the detectors were moved via underground rail to the Davis campus.
As expected, these detectors show significantly more cosmogenic activation~\cite{abgr16b}.

As described in section \ref{se:databases}, each step of a detector's transport and storage from time of manufacture to installation into the module was tracked using a database, enabling the calculation of each detector's integrated exposure to cosmic-rays.

\section{Detector Arrays}
\label{se:det_arrays}

This section describes the design, assembly and layout of the \DEM\ HPGe detector arrays and cryostat modules. 
We also discuss upgrades made during the operation of the \MJD .
We note that all EFCu parts described in this section  were made from copper grown underground in the TCR (Sec.~\ref{sec:electroforming_machining}). 
The copper grown in the shallower location at PNNL was used to make shield plates, discussed in Sec.~\ref{passive_shield}, which had less stringent requirements on \cosixty\ activation.

\subsection{Detector Mounts and Electronics}
\label{se:mounts_LMFE}

\begin{figure}[ht]
\centering
\includegraphics[width=0.6\textwidth]{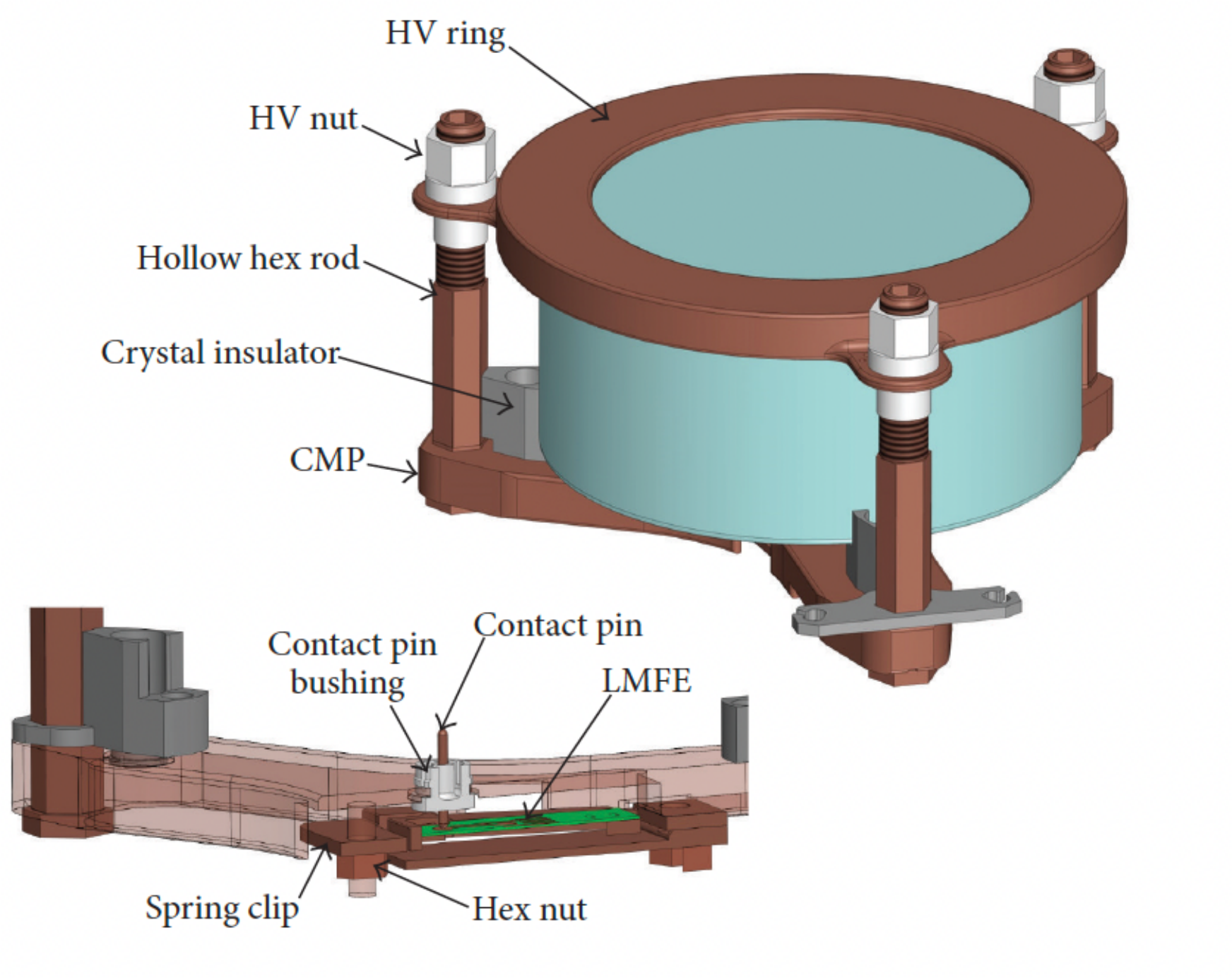}
\includegraphics[width=0.35\textwidth]{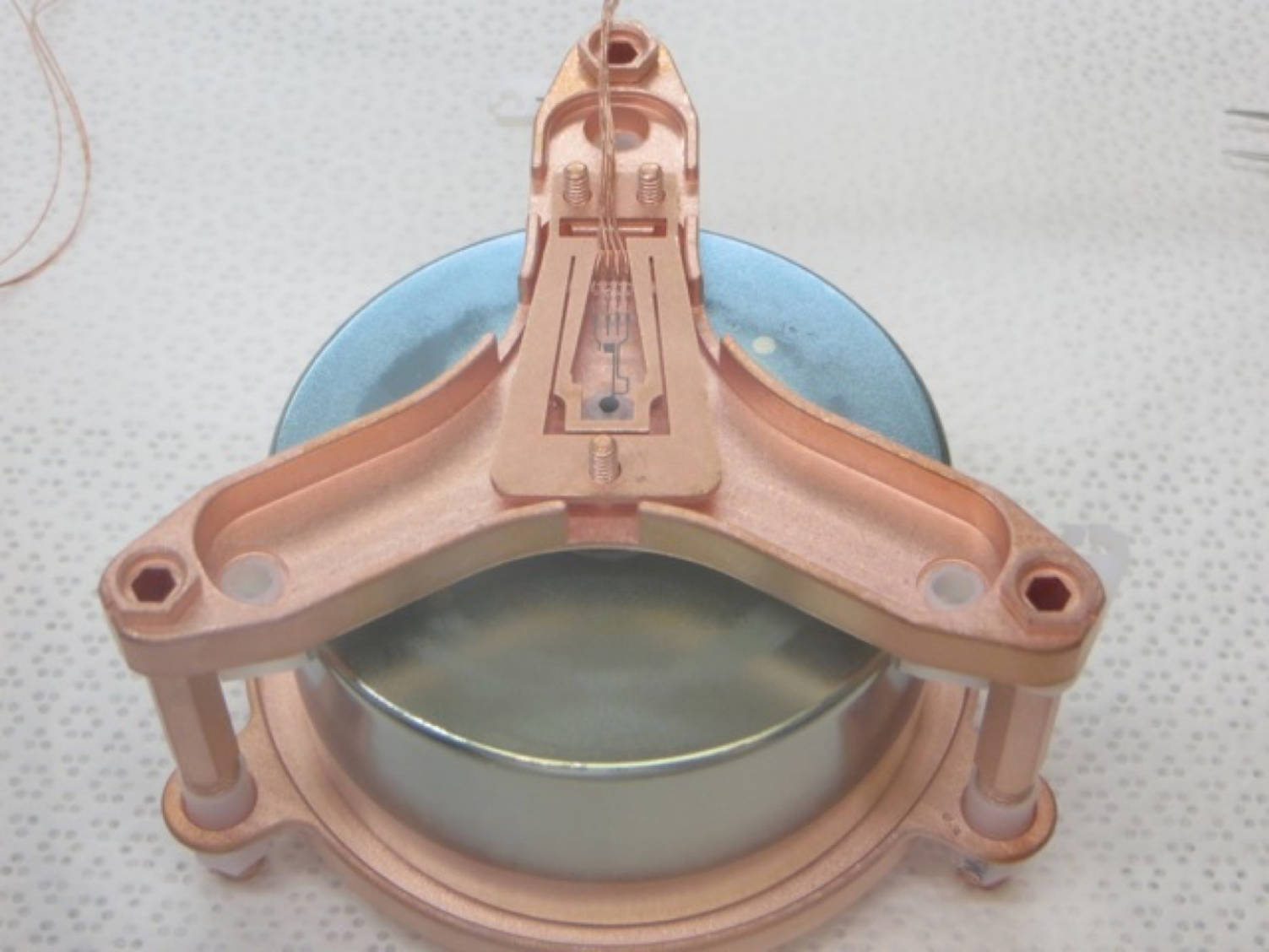}
\caption{(Left) Schematic of Detector Unit indicating major components, taken from ~\cite{abgr14}.
CMP refers to Crystal Mounting Plate and LMFE to Low Mass Front-End board.
(Right) A detector unit with a mounted BEGe detector. 
Note that it is upside down compared to the schematic. 
At the center of the ``Mercedes-logo" shaped CMP is the LMFE in its EFCu spring clip.}
\label{fig:DU}
\end{figure}

The \DEM\ HPGe detectors required mechanical support, bias voltages, and electronic readouts. 
Each detector was mounted in an individual Detector Unit (DU), as shown in Fig.~\ref{fig:DU}. 
DUs were made primarily from EFCu components fabricated in the underground machine shop. 
Their design minimized the mass of material near the detectors to reduce backgrounds, while providing the required mechanical support. 
They could accommodate a wide range of detector sizes from  different vendors, with diameters from 50~–~77~mm and heights up to 65~mm.
The threads of EFCu fasteners were coated with parylene as a lubricant prior to DU assembly. 

\begin{figure}[ht]
\centering
\includegraphics[width=0.4\textwidth]{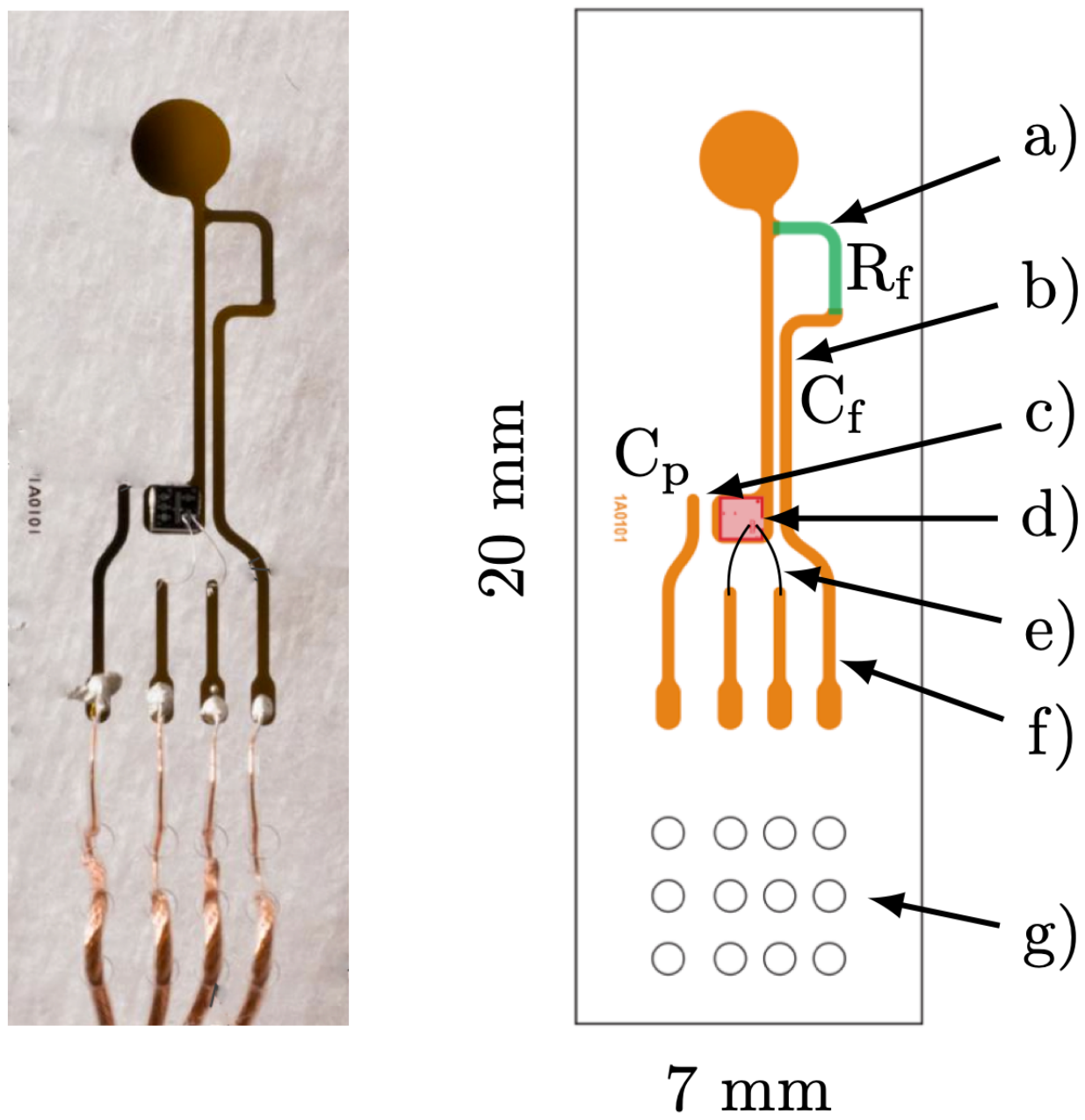}
\hspace{.3in}
\includegraphics[width=0.34\textwidth]{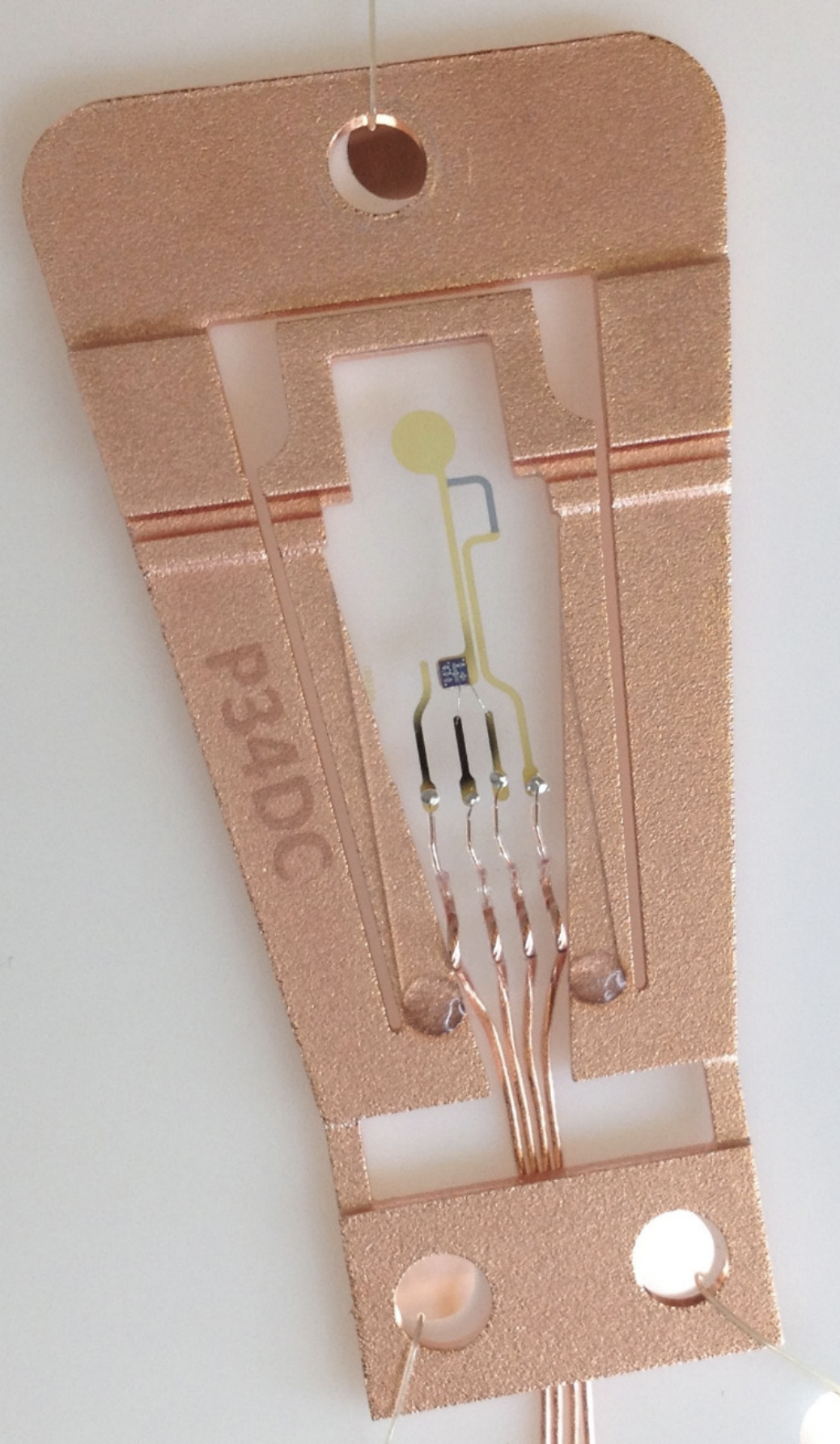}
\caption{ 
A photo and labeled diagram of an LMFE is shown on the left.
The circular pad at the top connects to the
contact pin, and the four traces at the bottom are from left to right: pulser, drain, source, and feedback. The labels
are a) feedback resistor, b) feedback capacitor, c) pulser capacitor, d) MX-11 JFET bare die, e) wire bonds, f) Ti/Au traces. and g) strain relief
holes in fused silica substrate.
The capacitance was provided by the capacitance of the gaps between traces.
On the right is a picture of an LMFE mounted in its EFCu spring clip. 
Figure and image taken from~\cite{Majorana:2021mtz}.}
\label{fig:LMFE}
\end{figure}

The \DEM\ utilized a resistive-feedback charge-sensitive preamplifier to amplify signals from the HPGe detectors.
The first stage had to be as close as possible to the detector to maximize signal-to-noise, which also required it to be low background. 
To this end, the collaboration developed a custom Low Mass Front-End (LMFE) board, mounted on the DU using an EFCu spring clip, as shown in Figs.~\ref{fig:DU} and~\ref{fig:LMFE}. 
The LMFE consisted of a fused silica substrate with clean gold and titanium traces. 
A three-terminal n-channel JFET bare die manufactured by MOXTEK with low built-in input capacitance ($C{gs}\sim 2.7\,\mathrm{pF}$) was mounted on the substrate with silver epoxy.
The LMFE also contained a (10~G$\Omega$ at 90~K) amorphous germanium feedback resistor.
To avoid loading the board with extra components, the feedback and charge injection capacitors
were provided by the capacitance between the on-board circuit traces.
The detector's p-type point-contact was connected to a circular pad on the LMFE using an EFCu pin coated with low-background tin, which was pushed against the detector using the springboard action of the spring clip. See Refs.~\cite{Barton:2011vja, Abgrall:2015hna} for more details about the LMFE design.
For details on the design and implementation of the signal readout electronics system, see~\cite{Majorana:2021mtz}.

Inside the modules, the \DEM\ used low mass and low radioactivity cables manufactured by Axon' Cable SAS.  
Four Axon' picocoax\textsuperscript{\texttrademark} signal wires were attached with silver epoxy directly to the LMFE board for the source, drain, pulser and feedback. 
These were true 0.4~mm diameter coaxial cables with a OFHC Cu central conductor and stranded OFHC Cu shield. 
The n-type exterior of the detector was held at positive bias voltage via an EFCu High Voltage (HV) ring around the upper corner of the detector (see Fig.~\ref{fig:DU}), on which a thin aluminum layer was deposited during the detector manufacturing process for improving the electrical contact with the EFCu.  
PTFE insulators provided electrical insulation between the HV ring and its mounting points on the conducting DU supports. 
The HV ring was biased via a 1.2~mm diameter Axon' coaxial HV cable with a maximum rating of 5~kV.
Its conductors were also made from OFHC copper. 

The full specifications of the Axon' cables are tabulated in Appendix~\ref{se:axon_specs}.

\subsection{Detector Strings and Modules}
\label{se:string_Modules}

\begin{figure}[ht]
\centering
\includegraphics[width=0.5\textwidth]{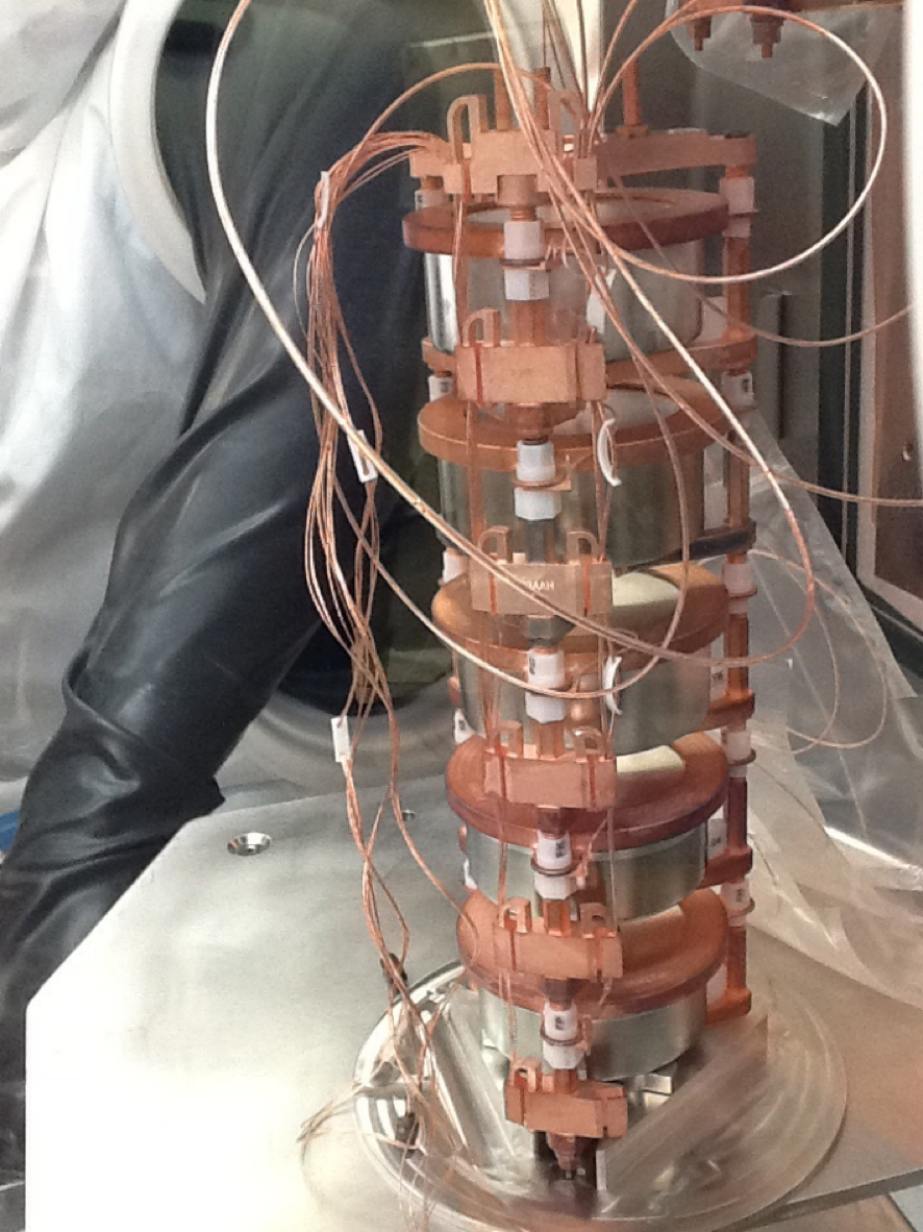}
\caption{A detector string consisting of five detector units after stacking in the glovebox.   
}
\label{fig:string_picture}
\end{figure}

\begin{figure}[ht]
\centering
\includegraphics[width=0.7\textwidth]{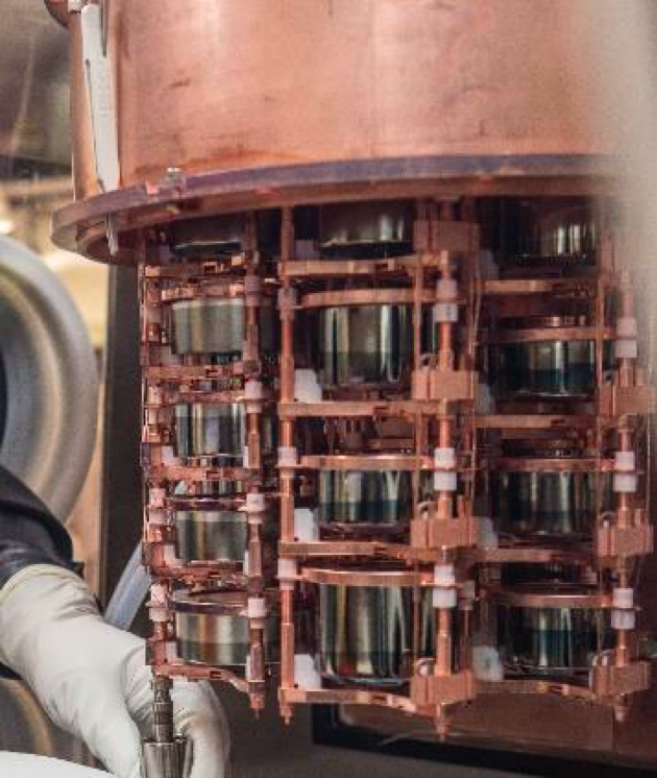}
\caption{Module~1 with loaded strings inside glovebox and prior to closure of the Module~1 cryostat and IR shield.}
\label{fig:loaded_Module}
\end{figure}

\begin{figure}[ht]
    \centering
    \includegraphics[width=0.9\textwidth]{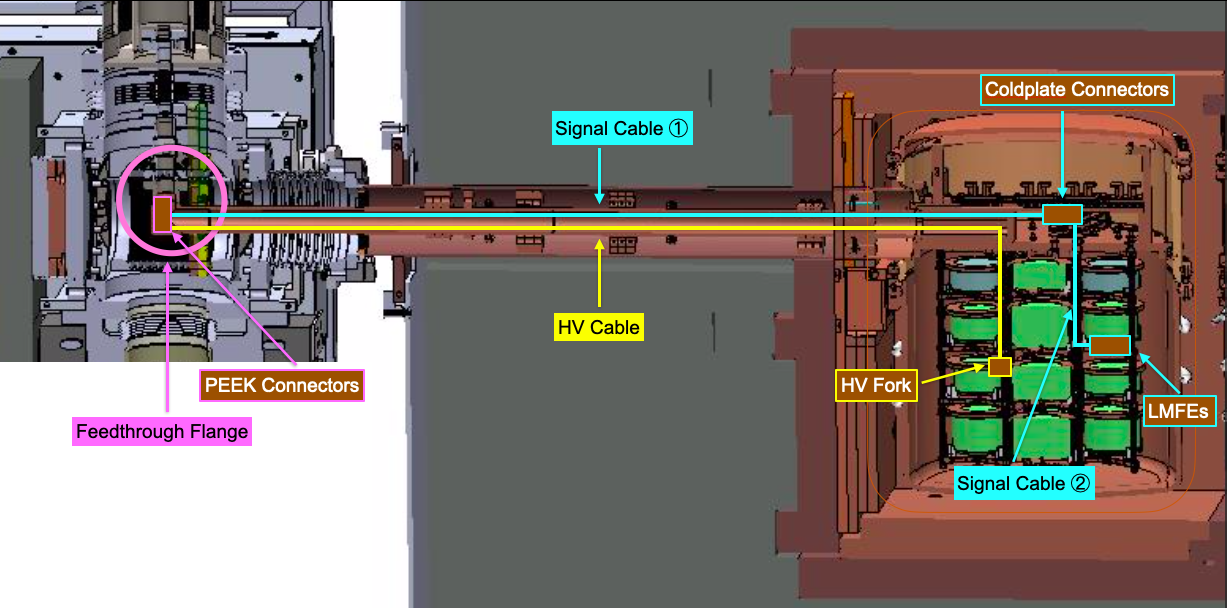}
    \caption{Cross sectional view of a \DEM\ module, highlighting the locations of cables and connectors. The baffle plate configuration is the original layout before the upgrade.}
    \label{fig:ccd}
\end{figure}

Three to five detector units were stacked into detector strings and clamped together with EFCu tie rods, as shown in Fig.~\ref{fig:string_picture}. 
HV and signal cables were guided and held in place with special EFCu spring clips mounted on the strings. 
Strings were mounted inside EFCu cryostats, as shown in Fig.~\ref{fig:loaded_Module}, and a cryostat loaded with strings was referred to as a ``module." 
Each module had seven strings, with one string at the center of a hexagonal arrangement of the other six strings. 
The strings were mounted to a thick EFCu disc, called the ``coldplate,"  at the top of the cryostat.
Each module was connected to electronic, vacuum, and cryogenic services via a copper crossarm tube that runs through the copper and lead shields, as shown in Fig.~\ref{fig:ccd}. 
A thermosyphon (see Sec.~\ref{sec:thermosyphon}) also ran through the crossarm and connected to the top of the coldplate to provide cooling.  

Inside the crossarm were EFCu ``baffle" plates that provided additional shielding from radiation propagating down the crossarm  while also allowing vacuum pumping and cable routing. 
Signal picocoax\textsuperscript{\texttrademark} cables from the detectors terminate at an EFCu terminal block on the coldpate, where they were connected to other coaxial cables that run the the length of the crossarm and terminate on a stainless steel feedthrough flange, outside the shield.
HV cables run directly from the detectors to the feedthrough flange via the crossarm.
Between the room-temperature cryostat wall and detectors was a thin-walled EFCu IR shield to reduce IR shine onto the detectors, which would otherwise increase detector leakage current and noise.

\subsection{Operational Configurations}
\label{se:operational_configurations}

\begin{figure}[ht]
\centering
\includegraphics[width=0.45\textwidth]{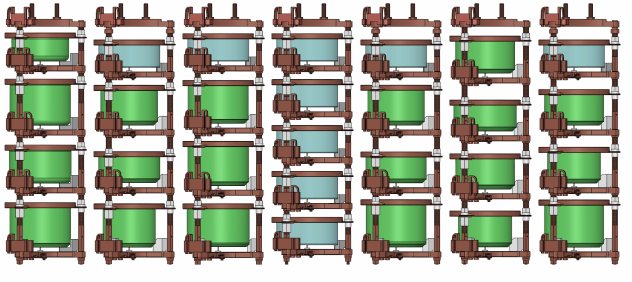}
\hspace{0.1\textwidth}
\includegraphics[width=0.45\textwidth]{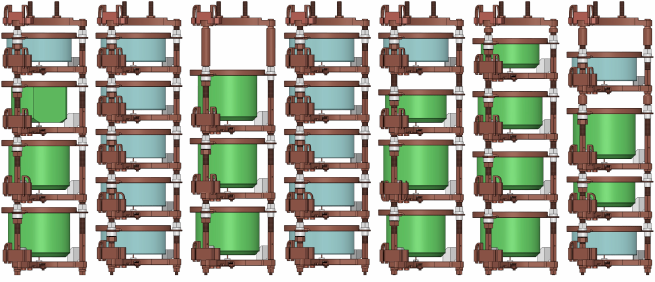}
\caption{Shown at the top is Module 1 and on the bottom is Module~2 configuration. Natural (BEGe) detectors are show in light blue. Enriched (PPC) detectors are show in green. The string at the left is located at the center of the module.  
}
\label{fig:Module_detectors}
\end{figure}

The \DEM\ was operated in several configurations from 2015 to 2021. 
Module~1, first deployed in June 2015, had 16.8~kg of enriched detectors (20 units) and 5.6~kg of natural detectors (9 units), whereas Module~2 was deployed in August 2016 and had 12.9 kg of enriched detectors (15 units) and 8.8 kg of natural detectors (14 units).  
Fig.~\ref{fig:Module_detectors} shows schematically the distribution of the detectors between the two modules and the design flexibility required to accommodate strings with different detector sizes and numbers.

The total enriched and natural masses were 29.7~kg and 14.4~kg, respectively but only up to 22.1~kg of enriched detectors and 10.0~kg of natural detectors were operational.
This was because, during commissioning, several issues with the reliability of electrical connections were identified.  
Eight detectors had issues either with custom-made Vespel\textsuperscript{\texttrademark} connectors that link signal cables together or with a damaged LMFE.
Nine detectors remained unbiased due to problems with their HV cables. 
One detector was found to be defective after installation. 

The collaboration decided to operate in this configuration while performing parallel R\&D on correcting these issues, and in Nov. 2019, Module~2 was removed from the shield and upgraded with improved cables, connectors and shielding, as described in Sec.~\ref{se:CC_upgrades} below. 
At the same time, 5.5~kg of PPC detectors were removed: five detectors were used for early testing in \LM\ and a non-operational detector returned to the vendor for rework.
After the upgrade, up to 27.2~kg of enriched detectors and 13.2~kg of natural detectors were operational~\cite{arnq22}.

\begin{figure}[ht]
\centering
\includegraphics[width=0.45\textwidth]{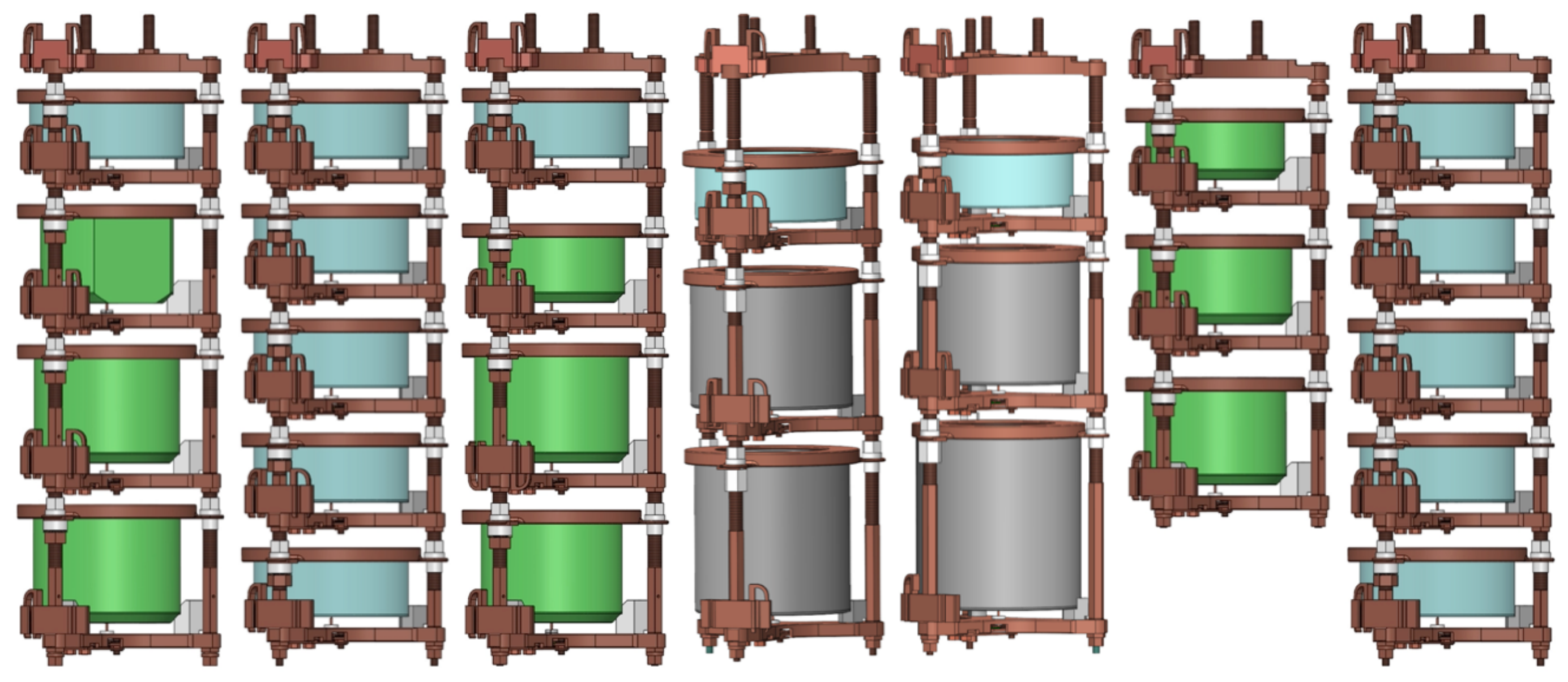}
\caption{The Module~2 detector configuration after the upgrade. The newly installed ICPC detectors are shown in grey.   
}
\label{fig:postupgrade_detectors}
\end{figure}

The collaboration also installed four enriched \ICPC\ detectors (6.7 kg). 
The ICPC design is the baseline for \LM , and the \DEM\  provided an early opportunity to study these novel detectors in a well-characterized, low-background vacuum environment.
Fig.~\ref{fig:postupgrade_detectors} shows the post-upgrade detector configuration.
 
In March~2021 operation with enriched detectors was stopped, and the modules removed from the shield.
The detectors were removed, and the enriched detectors were packaged and shipped to LNGS for installation in \LM .
The 23 remaining natural BEGe detectors were consolidated in a single module that was re-deployed into the shield in April 2021 for a period to study backgrounds without the enriched detectors.
After another reconfiguration in March 2022, the experiment was converted into a search for decays of isomeric \nuc{180m}{Ta}, including possible stimulated decays from dark matter~\cite{lehnert17, lehnert20, arnq24a}.

\subsection{Cryostat, Cabling, and Connector Upgrades}
\label{se:CC_upgrades}

The collaboration decided to upgrade Module~2 because it contained 11~of the 17~inoperable detectors
After the upgrade Module~2 had 27 detectors and all were operable, ie. had good HV and signal connections.
Its total available enriched mass became 14.1 kg (13 detectors) with 8.8 kg (14 detectors) natural germanium. 
The upgrades are discussed in this section. 
A more detailed account is provided in Ref.~\cite{haufe_thesis}.

\subsubsection{HV Cables Upgrades}
HV cables were run as one unbroken length from the feedthrough flange to detector, and each detector required exactly one HV cable (Fig.~\ref{fig:ccd}).
As mentioned earlier, the Axon' HV cables are rated up to 5~kV DC and demonstrated sufficiently low microdischarge rates during testing~\cite{abgr16c}.
However, during initial operation of Module~1 and Module~2, several detectors suffered from occasional high voltage breakdowns.
This led to several interruptions in data taking, as breakdowns would automatically trigger a ramp-down of bias voltages to all detectors in the same module. 
A temporary solution was found by disconnecting the outer conducting shield of the high voltage cables of problematic detectors from ground. 
Investigation indicated that the cable discharges were likely from kinks or deformed sections of cable caused where it was pressed against sharp edges of the thermosyphon baffle plates during installation.

\begin{figure}[ht]
\centering
\includegraphics[width=0.95\textwidth]{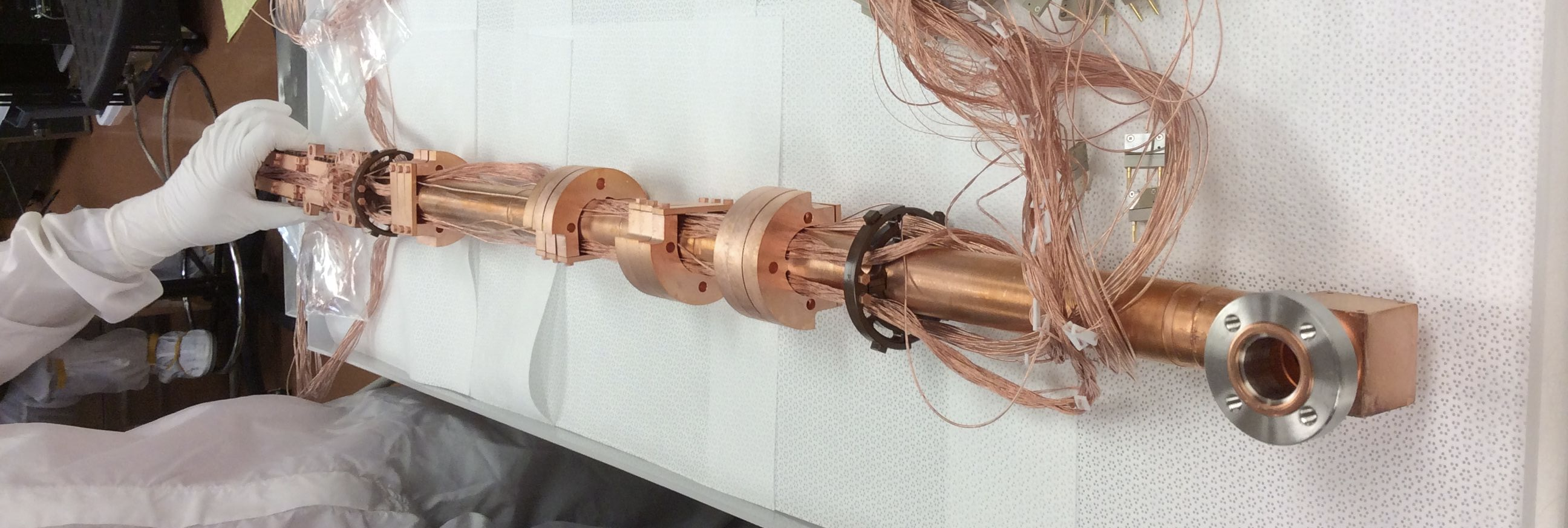}
\includegraphics[width=0.95\textwidth]{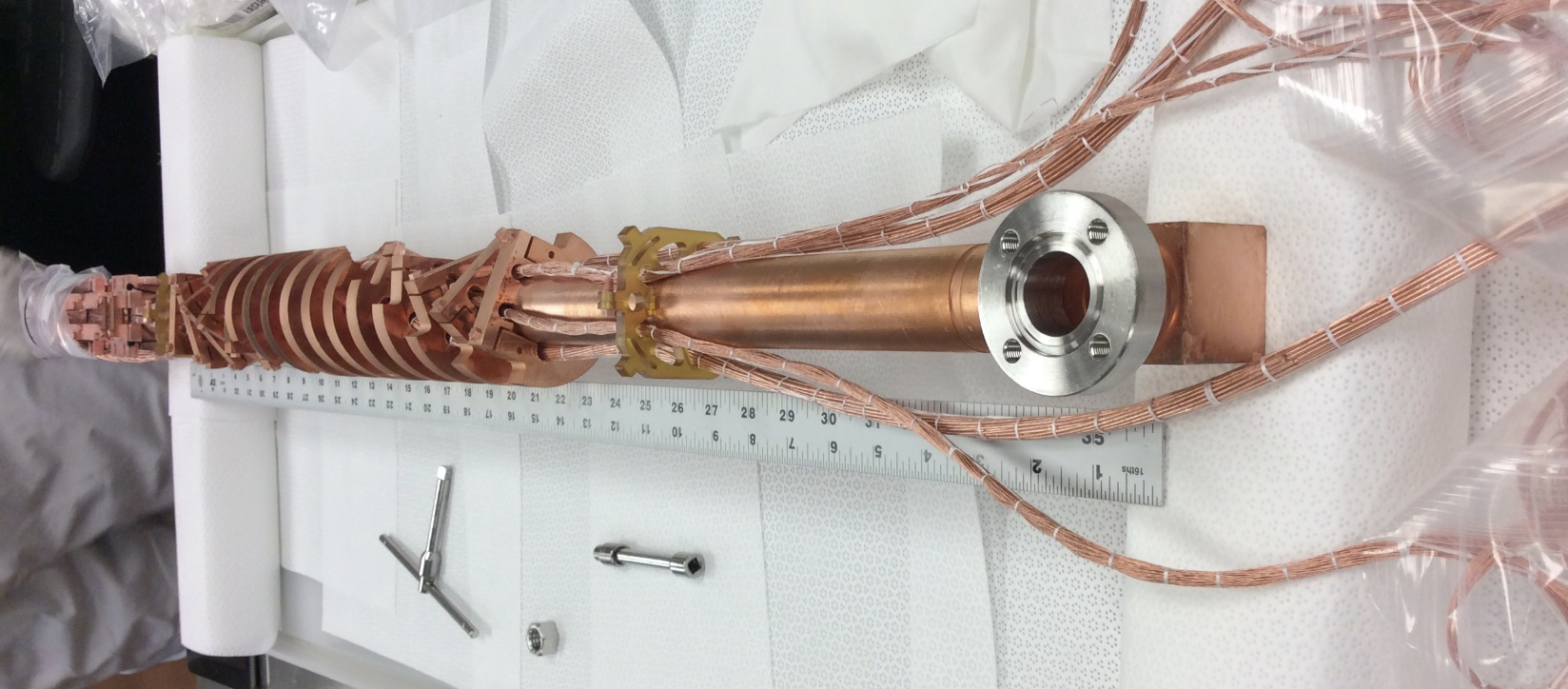}
\caption{(Top) The original baffle plate arrangement on the thermosyphon with 12~baffle plates, split into 4~groups of 3~at orthogonal angles.  Cables were not bundled.
(Bottom) The new baffle plate arrangement after the upgrade with 16~baffle plates, each plate at a staggered angle from the previous plate.
Cables were bundled with plain dental floss per NASA specifications.}
\label{fig:baffles}
\end{figure}

During the upgrade, new cables were handled as little as possible during installation and wound at a uniform radius when stored. 
The cables were bundled together following NASA specifications prior to quality testing~\cite{milspec}.
The baffle plates were also replaced with a new design that featured rounded edges and a spiral pattern to optimize radiation shielding. 
See Fig.~\ref{fig:baffles} for a comparison of the two baffle plate designs.
After the upgrade no detector suffered from HV breakdowns. 

\subsubsection{Signal Connector Upgrades}
\DEM\ signal cables consisted of two cables that were connected together above the coldplate (Fig. \ref{fig:ccd}).  
One cable terminated at the feedthrough flange and above the coldplate.  
The other cable traveled  from above the coldplate to the LMFE of its detector.  

A custom Vespel\textsuperscript{\texttrademark}   connector, fabricated in-house, initially connected cables above the coldplate. 
It was found that this design was not robust enough to withstand repeated temperature cycling because the connector lacked conventional CuBe spring components that had unacceptably high backgrounds.
This led to intermittent or permanent connectivity issues, rendering some detectors inoperable.

\begin{figure}[ht]
\centering
\includegraphics[width=3in]{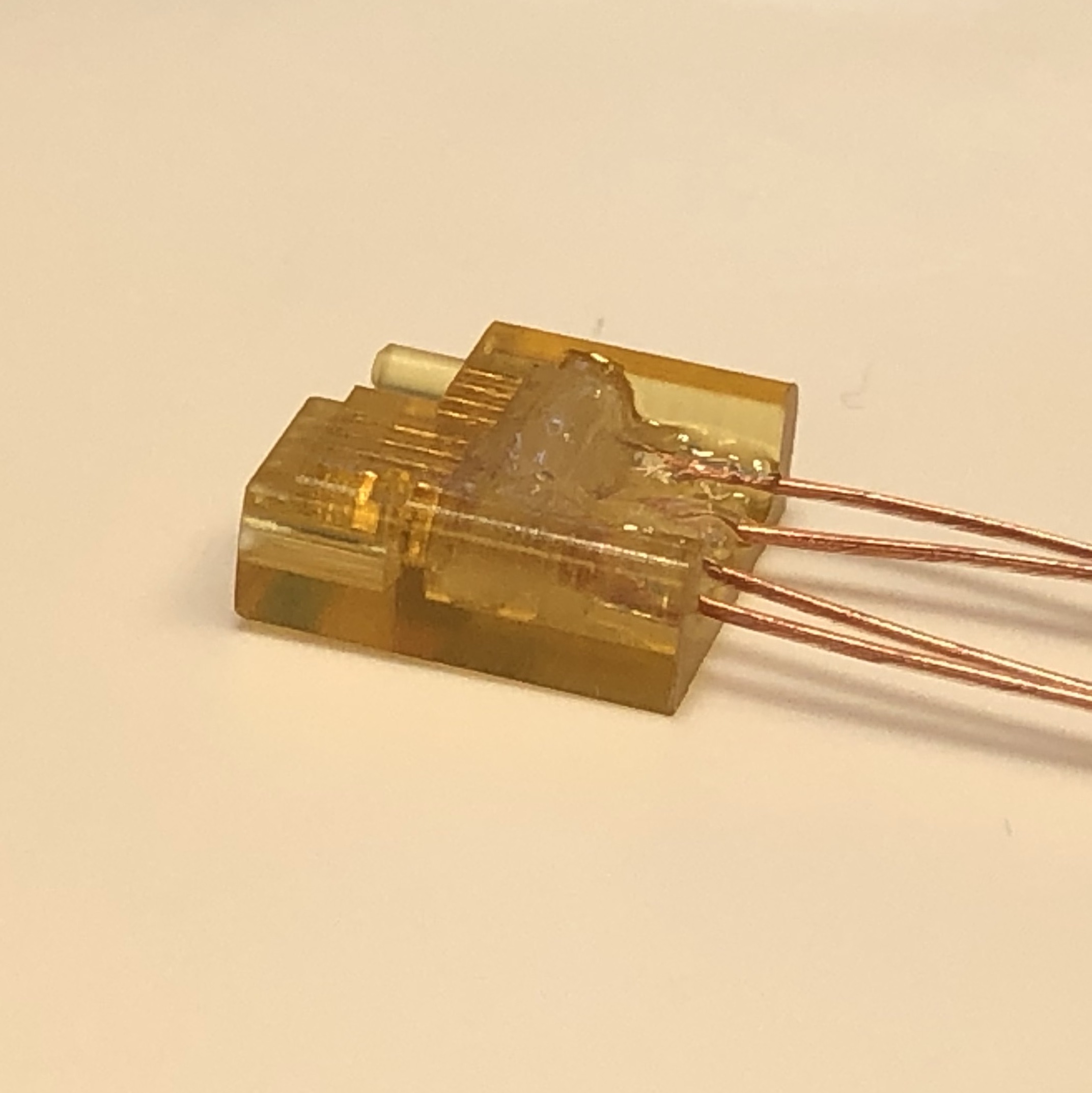}
\caption{One half of an Axon' Nano twist-pin connector.  Axon' delivered these connectors pre-terminated to signal cables.}
\label{fig:ntp}
\end{figure}

The solution was a low-mass connector designed by Axon'.  
These ``nano twist-pin'' connectors (Fig.~\ref{fig:ntp}) featured gold alloy contacts encapsulated by an Ultem\textsuperscript{\texttrademark} shell.  
Such connectors surpassed the radiopurity of the previous Vespel\textsuperscript{\texttrademark} connectors and passed thermal performance tests at Axon's production site in France, as well as at the cable fabrication site at the University of North Carolina at Chapel Hill (UNC).
Axon' pre-terminated these connectors to signal cables for the collaboration.

\subsubsection{Feedthrough Flange Connectors}

The signal cable termination at the feedthrough flange was a custom-built 50 pin D-sub made from PEEK material to meet radiopurity requirements.  
The connectors had incorrect pin-spacing due to a machining error, which made the D-subs difficult to plug and unplug into the feedthrough flange.  
The upgrade used the original PEEK design with correctly machined pin-spacings to solve previous installation difficulties. 

The HV cables were terminated at the feedthrough flange via custom low-mass PEEK connectors.  
The original connectors consisted of two sockets  terminated to the cable ground shield and central conductor with silver epoxy inside a PEEK two-body shell.  
Unfortunately, the sockets used were too small and also not completely covered by the PEEK housing, allowing for potential bending or catching the sockets on other cables during installation. 
All these issues resulted in connectivity problems.

The upgrade used a revised design that extended the PEEK two-body shell to encapsulate both sockets and secure them in properly-sized bore holes. 
This protected the sockets from bending and improved their alignment and holding force.
The sockets were also replaced with a product from Glenair that provided a stronger and more consistent clamping force. 

\subsubsection{HV Fork Connectors}
The HV cables were terminated at the detector by way of an EFCu ``HV Fork".  
This custom connector was clamped to a copper plate, called the ``HV Ring", via an ``HV Nut". 
The HV Ring makes contact with the n-type surface of the detector. 
See Fig. \ref{fig:DU}. 

\begin{figure}
\includegraphics[width=0.75\textwidth]{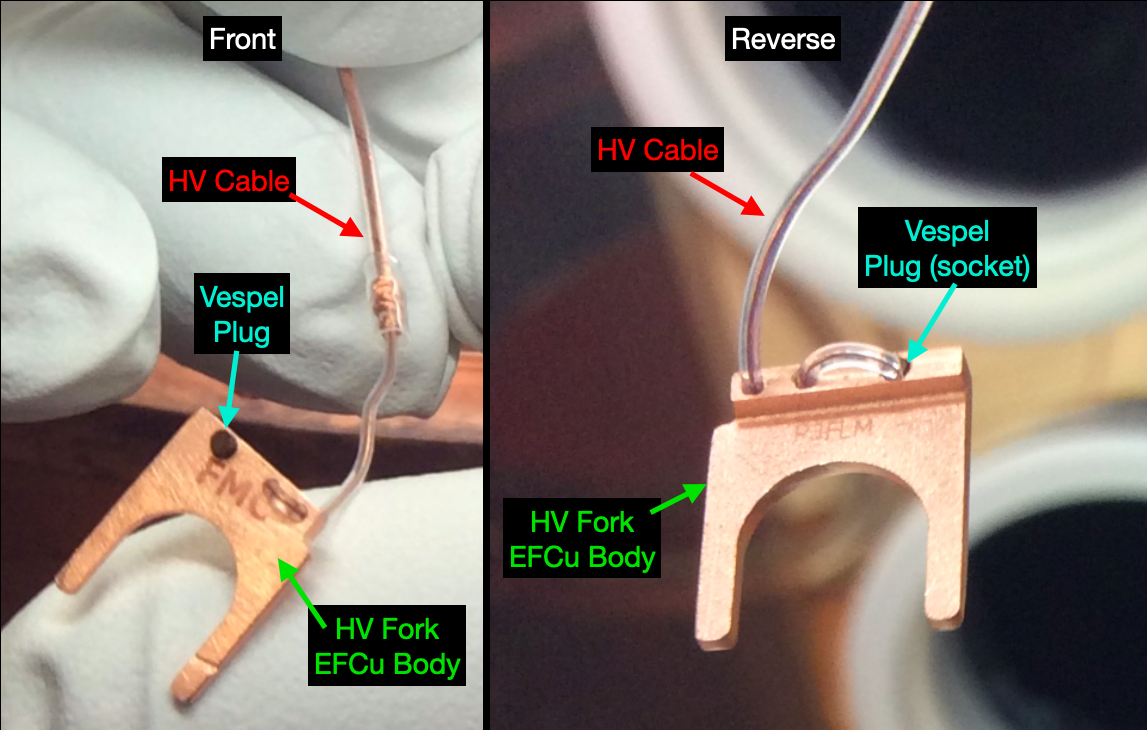}
\caption{The old design of the HV fork, where the HV cable was threaded through the body of the fork and the central conductor was terminated to the fork by way of a small Vespel\textsuperscript{\texttrademark}   plug.}
\label{fig:oldh}
\end{figure}

\begin{figure}
\includegraphics[width=0.75\textwidth]{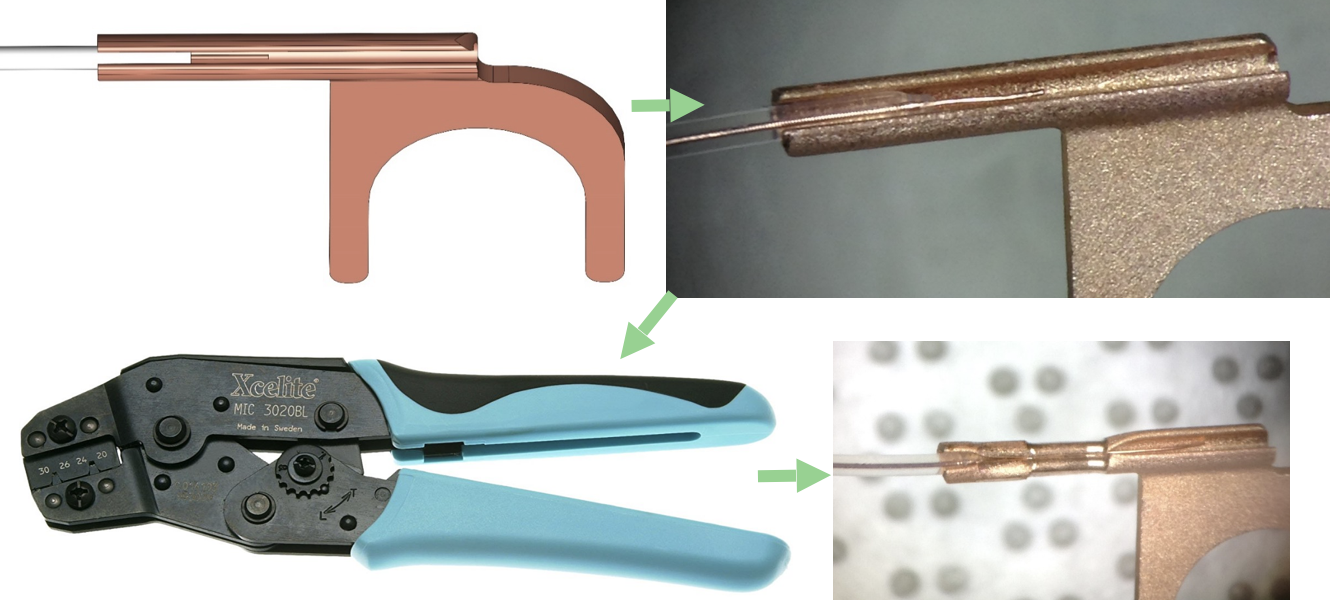}
\caption{The new design of the HV fork, where the central conductor of the HV cable is crimped to the fork through a copper receptacle. Crimping was done with a specially cleaned MIC3020BL Mini-Crimper manufactured by Xcelite.}
\label{fig:newh}
\end{figure}

The original design terminated the HV cable at the HV Fork with a custom plug made of Vespel\textsuperscript{\texttrademark}  that pinned the exposed central conductor to a bored hole through the HV Fork's surface (Fig. \ref{fig:oldh}).  
This Vespel\textsuperscript{\texttrademark}  plug did not always stay in place, leading to connectivity issues.
The upgrade utilized a new HV Fork design that featured a crimped connection, as shown in Fig. \ref{fig:newh}.  
The crimped connection of the central conductor directly to the HV Fork simplified assembly and resulted in robust connections.

\subsubsection{Fabrication of Signal Cable Bundles}
\label{sigcablefabrication}

\begin{figure}[ht]
\includegraphics[width=0.75\textwidth]{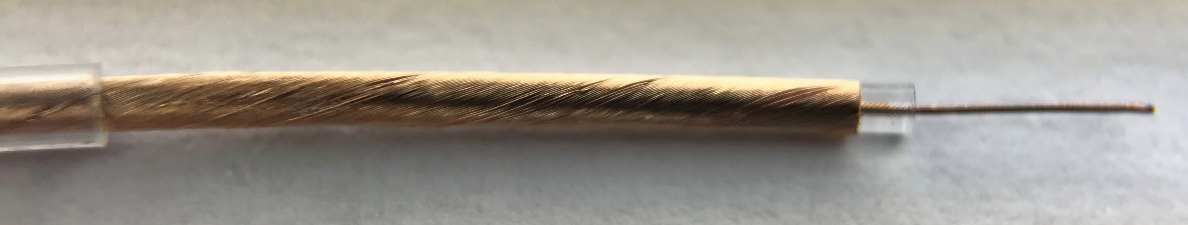}
\caption{Example of the high quality strip-back obtained using the CoaxStrip Machine}
\label{fig:stripback}
\end{figure}

The signal cable arrived from Axon' pre-terminated to the nano twist-pin connectors.
The remaining terminations at the feedthrough connector and LMFE were fabricated by the collaboration. 
At both terminations, the cable was stripped with a Schleuniger CoaxStrip 5300 RX machine using a programmable rotary cutter.
This machine vastly improved the quality of the stripped cables compared to the hand-stripped cables used in the initial installation (Fig. \ref{fig:stripback}).

At the feedthrough connector, the central conductor and ground of the signal cable were crimped to the Glenair sockets before installing them into the two-body PEEK 50 pin D-sub.  
At the LMFE, the signal cable was adhered to the terminals of the LMFE using silver epoxy as before.



The four wires per nano twist-pin connector were braided by hand, and braided groups were bundled with plain dental floss according to NASA specifications \cite{milspec}. 
Bundles were temperature cycled and tested for continuity before installation.
The HV cables were bundled and terminated to their respective connectors before testing in the same manner. 

\subsubsection{Results and Impact}

As stated earlier, Module~2 had 11 inoperable detectors prior to the upgrade.  With 29 total detectors, Module 2 had an operational efficiency of 62\%. 
The primary goal of the upgrade was to increase this operational efficiency to greater than 90\%.  
At the conclusion of the upgrade, 27 out of 27 detectors operated in Module 2 for an operational efficiency of 100\% (the upgrade reduced the total number of detectors in Module 2 from 29 to 27).

\section{Cryogenic and Vacuum Systems}
\label{sec:cryovac}

\begin{figure}[ht]
\begin{center}
\includegraphics[width=15cm]{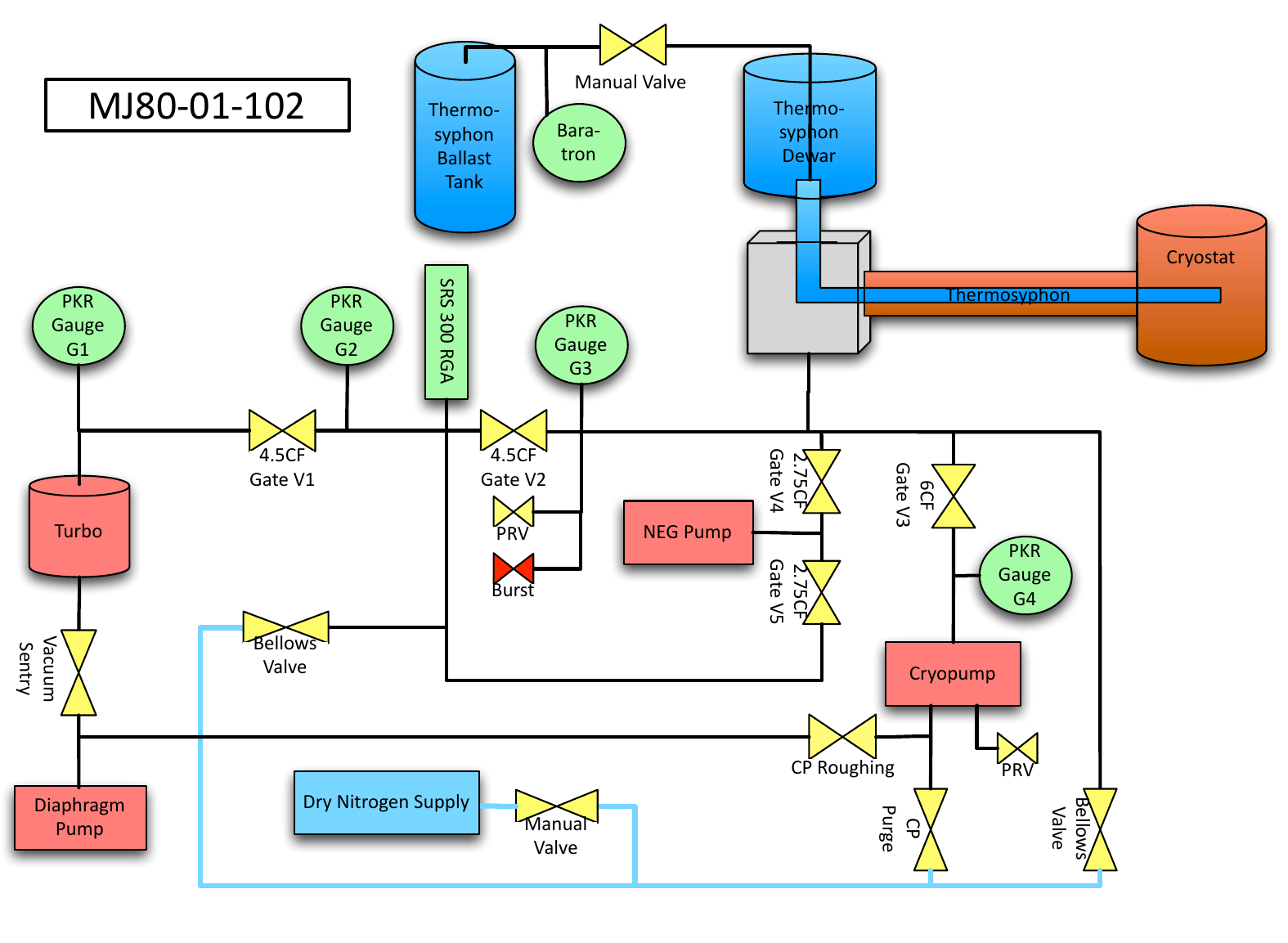}
\caption{Process and Instrumentation Diagram (P\&ID) of the \DEM\ cryogenic and vacuum systems. Each modular cryostat was cooled via its own horizontally-oriented LN driven thermosyphon.  Vacuum was generated through a turbopump and cryopump which were solenoid-controlled via pneumatically-actuated valves.}
\label{fig:cryovacPNID}
\end{center}
\end{figure}

Germanium detectors require cryogenic operating temperatures to limit thermal excitation of charge carriers in  the semiconductor material.
The \MJD\ cryogenic and vacuum systems were designed with the aims of stable, low-vibration, low-background operation.
An overview of the cryogenic and vacuum systems are shown in Fig.~\ref{fig:cryovacPNID}.
Cooling of the array was provided through the use of a horizontally-oriented thermosyphon, contained within the crossarm tube described in Sec.~\ref{se:string_Modules}.
The crossarm tube also provides a vacuum pumping path for the modules. 
The copper cryostat and all internal components were isolated electrically from the remainder of the cryogenic and vacuum systems through a set of glass vacuum electrical isolators, and isolated vibrationally through a set of thin-walled vacuum bellows.

\subsection{Cryogenic System Overview}
\label{sec:cryovac_overview}
Thermosyphon-based cooling was chosen for the \DEM\ due to the cooling power it could deliver, the limited vibration it would induce at the detector array, and the passive temperature stability of using evaporating cryogen for cooling.
The \DEM 's thick shield also required a long and thin cross-arm, which was incompatible with conduction cooling methods.  
To meet the stringent radiopurity requirements for materials located within the \DEM's shielding, the thermosyphon itself was constructed out of underground EFCu.
Here we will describe the design and operation of the thermosyphon, as well as the use of a mechanical cooler coupled to the thermosyphon ``cold finger" during \DEM\ commissioning.

\subsubsection{Thermosyphon Design and Operation}
\label{sec:thermosyphon}
The thermosyphon operates as a gravity-driven closed-loop system, where nitrogen gas was continuously condensed in a heat exchanger placed in contact with a liquid nitrogen (LN) bath.
The condensed LN would run into the nearly-horizontal EFCu thermosyphon tube, and flow in a shallow layer to the coldplate of the detector array.
Nitrogen evaporated at the coldplate then travels in the same tube, counter to the direction of the liquid flow, back up to the condenser.
The closed-loop recirculation system prevented the introduction of radon, present in laboratory air and concentrated in open-air LN dewars, into the interior of the detector cryostats.
It also simplified operations.

The steady-state volume of LN present in the thermosyphon during operation was determined by the initial charge of N$_2$ gas, measured by the pressure in the system at initial filling.
A 100-liter ballast tank was connected to the enclosed nitrogen volume to keep the thermosyphon pressure below 2 bar (required by vibration and electrical isolation hardware) while allowing for sufficient liquid mass for adequate refrigeration. See Ref.~\cite{AGUAYO201317} for more details.

The fabrication of the custom LN dewar with the built-in condenser volume proved to be a challenging task for vendors.
Delivery of the first dewar incurred substantial delays totalling nearly a year, and in order to move forward with initial commissioning, a temporary, mechanical pulse-tube cooler (PTC) was mounted to the copper thermosyphon tube, which was used as a cold finger. 
The PTC was replaced with the thermosyphon before production data-taking began.

\subsection{Vacuum System Overview}
The operation of cryogenic detectors requires a high-vacuum environment or immersion in an inert fluid to inhibit the condensation or freezing of gases on sensitive detector surfaces.  
Specifically, the passivated surfaces of high-purity germanium detectors can be damaged by excessive exposure to water, leading to increased leakage current or detector inoperability.  
Additionally, at detector operating temperatures, radon in the surrounding environment will be rapidly adsorbed onto the surfaces of detectors and surrounding components; the radon decays and subsequent decays of radioactive daughters constitute a potentially significant source of background for the \Demo, and so radon content in the residual gas must be limited.

The \MJD~module vacuum systems were designed with the goals of suitable high-vacuum performance, reliable long-term operation, control of major system operations from off-site, redundancy and automatic intervention in the case of equipment failure, and low radon emanation from / permeability through hardware components.
Stainless steel, metal-sealed vacuum fittings were used throughout the assembly to limit radon permeation through sealing materials, the only exception being the cryostat gaskets (see Sec.~\ref{se:LB_seals}). 
Vacuum was generated through a set of pumps: a 200-lpm oil-free diaphragm pump to bring a module to rough vacuum, and a 300-lps turbo-molecular pump  to establish high vacuum ($\sim10^{-7}$~mbar).  
A 1500-lps cryopump provided an additional 10x reduction in base pressure, and was used for steady-state operation once base pressure was established.
A zirconium-based sintered Non-Evaporable Getter (NEG) pump was installed in each module with the aim of removal of non-condensable gases during steady-state operation.
A 300-AMU-range residual gas analyzer was used to determine gas composition. 
A set of solenoid-controlled, pneumatically-actuated all-metal gate valves provide isolation of the detector array from each pump, enabling regeneration of the cryopump and NEG pump without exposure of the array to evaporated gases.  
The valves also provided protection against pump mechanical failures, with valve closure triggered by out-of-tolerance pressures or reporting of a fault status.

\subsection{Low-background Vacuum Seals}
\label{se:LB_seals}
The EFCu cryostat enclosing each of the modules' detector arrays required a pair of vacuum seals.
As the vacuum vessel sealing surfaces were entirely copper, any o-ring or gasket material used to create a seal must be softer than the metal so as not to permanently deform the sealing surfaces.
This restricted the material choices to very soft materials such as lead or indium, elastomers such as FKM (Viton\textsuperscript{\texttrademark} ) or Nitrile (Buna-N) rubbers, or synthetic materials.
The proximity of these seals to the detectors required them to conform to strict radiopurity requirements, disqualifying commercially-available metallic or rubber seals, and necessitating an investigation into alternative materials

\subsubsection{Thin-film Gaskets}
Initial designs of the \DEM 's vacuum vessel seals were based on a thin-film gasket design developed and successfully tested on half-scale prototypes prior to module fabrication.
Mating tapered surfaces on the vacuum vessel components would seal around a gasket made from parylene-C, a polymer used to produce thin coatings on threaded components used elsewhere throughout the \DEM .
Keeping the parylene thickness low would minimize the gaskets' contribution to the radioactive background.
Unfortunately, initial testing showed that gaskets made from parylene material were not sufficiently elastic to conform to tapered surfaces on the full-scale seals without introducing wrinkles or tears which would yield vacuum leaks.
Viton\textsuperscript{\texttrademark} o-rings were substituted for the parylene gaskets during commissioning while alternatives were sought.

Commercially-sourced 0.13-mm PTFE sheets provided an acceptable substitute for the parylene.  
The PTFE material is more deformable than parylene and could be leached of metallic impurities using nitric acid solutions.
Testing with Module~1 showed the base pressure achievable with this seal ($5\times10^{-7}$~mbar typical) was not as low as with Viton\textsuperscript{\texttrademark}  o-rings ($7\times10^{-8}$~mbar typical) but was deemed acceptable for detector operation.

\subsubsection{PTFE O-rings}
For the deployment of Module 2, a sealing solution supporting lower vacuum pressures was found in PTFE o-rings.
The PTFE o-rings offered similar radiopurity and chemical resistance  as the PTFE sheet material with better sealing capabilities.
This came at the expense of increased mass; each o-ring had a mass of 20 grams, increasing the background contribution from the seals by a factor of 10, though still a minor contributor to the overall background budget.

PTFE creep was considered as a potential hazard of PTFE-based seals, especially under the 9.3-kN force exerted by 34.3-cm diameter flanges under 1 atm of pressure difference.
Neither the PTFE sheet installed in Module~1 nor the PTFE o-ring installed in Module~2 have shown any signs of vacuum performance degradation over the lifetime of either module. 

\subsection{Cryogenic and Vacuum Systems Slow Controls}
\label{se:cryovac_SCM}
Each module's cryogenic and vacuum systems were controlled and monitored by its own Mac Mini PC, running a dedicated ORCA-based (see Sec.~\ref{se:DAQ_Software}) vacuum system control application developed to interface with all cryogenic and vacuum systems and to provide a graphical user interface for controlling those systems.
The vacuum gauges and pumps on each system were equipped with their own controllers, supplied by the various manufacturers, which were interfaced through RS-232 or RS-485 communication standards via USB to serial adapters.
Sampling of the vacuum pressure within the cryostats directly was not possible because of the radioactivity a gauge within the shielding volume would introduce; 
instead, pressures were measured outside of the shielding and inferred for the cryostat internal volumes.
Vacuum valves were either 120~VAC solenoid-actuated or 120~VAC solenoid-controlled, pneumatically actuated, and positions were set through the use of a 16-channel RS-232-enabled relay controller.
The gate valves on the system were equipped with position-indicating switches, which allowed confirmation that the valve was in the desired position, and was not in between fully-opened or fully-closed conditions.
This served to identify mechanical malfunctions or loss of compressed air supply, the latter of which could arise from facility power outages.

The potential for damage to electronics, detectors or vacuum pumps in the case of a vacuum or cryogenic failure, or the accidental biasing of detectors while at room temperature or pressure motivated the development of a set of ORCA-controlled software interlocks.
Within the vacuum system's ORCA application there were a set of prohibited behaviors determined by current pressures and pump operating conditions.  
These were clearly illustrated on the user interface.
This prevented accidental opening of valves or deactivation of pumps which could lead to equipment damage.
Additionally, two way communication existed between the ORCA instances on the computers controlling the vacuum systems and the data acquisition systems, the latter of which controlled detector HV.  
As a result, the vacuum machines would prevent actions that would expose biased detectors to potentially damaging pressures, and the data acquisition system would prevent biasing detectors if the current temperatures and pressures were not suitable for operation.  
The data acquisition machine would also unbias detectors if vacuum or temperature conditions deteriorated, in case of an extended power outage, or if there was sustained loss of communication with a vacuum machine and conditions were unknown as a result.

\subsection{Performance and Stability}
\label{se:cryovac_performance}



\subsubsection{Cryogenic Performance}
A series of measurements were performed to evaluate the cryogenic performance of the \DEM~modules, both in thermosyphon and cryo-cooler configurations, with the aims of establishing array cool-down times, relative differences in detector temperatures during steady-state operation, and the temperature stability during steady-state operation.

\begin{figure}[ht]
\begin{center}
\includegraphics[width=15cm]{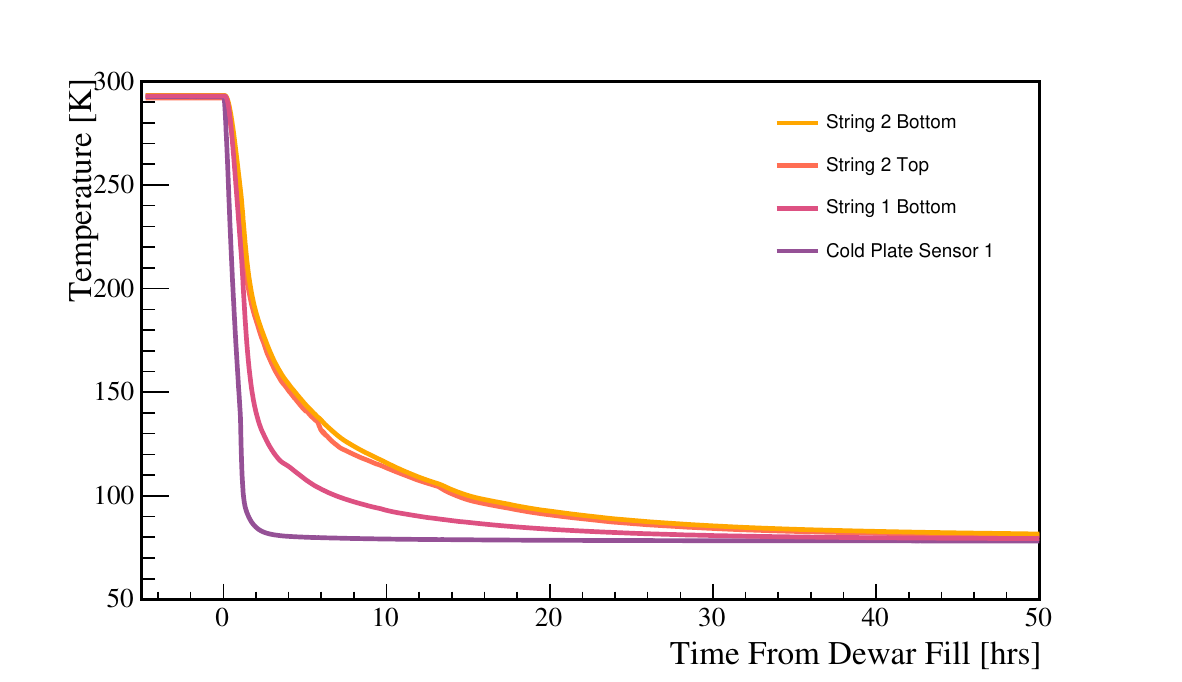}
\caption{Temperatures measured throughout Module 1 during a thermosyphon-driven cooldown cycle.  The thermosyphon and coldplate reach operating temperature rapidly ($<2$~hrs), while the high-impedance thermal coupling of the detectors' EFCu string hardware slows their cooling.  The detector array reaches steady-state operation after 48~hrs of cooling.} 
\label{fig:TS_cooldown}
\end{center}
\end{figure}

A typical array cooldown operation is depicted in Fig.~\ref{fig:TS_cooldown}.  During this May 2014 commissioning run the Module~1 thermosyphon, coldplate and a detector string were instrumented with Si-diode temperature sensors.  
Due to the high effective cooling power of the thermosyphon, cooling of the thermosyphon and coldplate occurs rapidly during the first 2 hours following filling of the thermosyphon dewar with liquid nitrogen.  
The cooling of the detectors lags behind the coldplate; their cooldown time was entirely dominated by the detector masses and their high-impedance thermal coupling to the EFCu string hardware through which they were cooled.  
The germanium detectors reach the final operating temperature after $\sim48$~hours, confirmed by observation of the stabilization of detector baseline voltages as the leakage current and feedback resistance vary during the cooling process.

\begin{figure}[ht]
\begin{center}
\includegraphics[width=15cm]{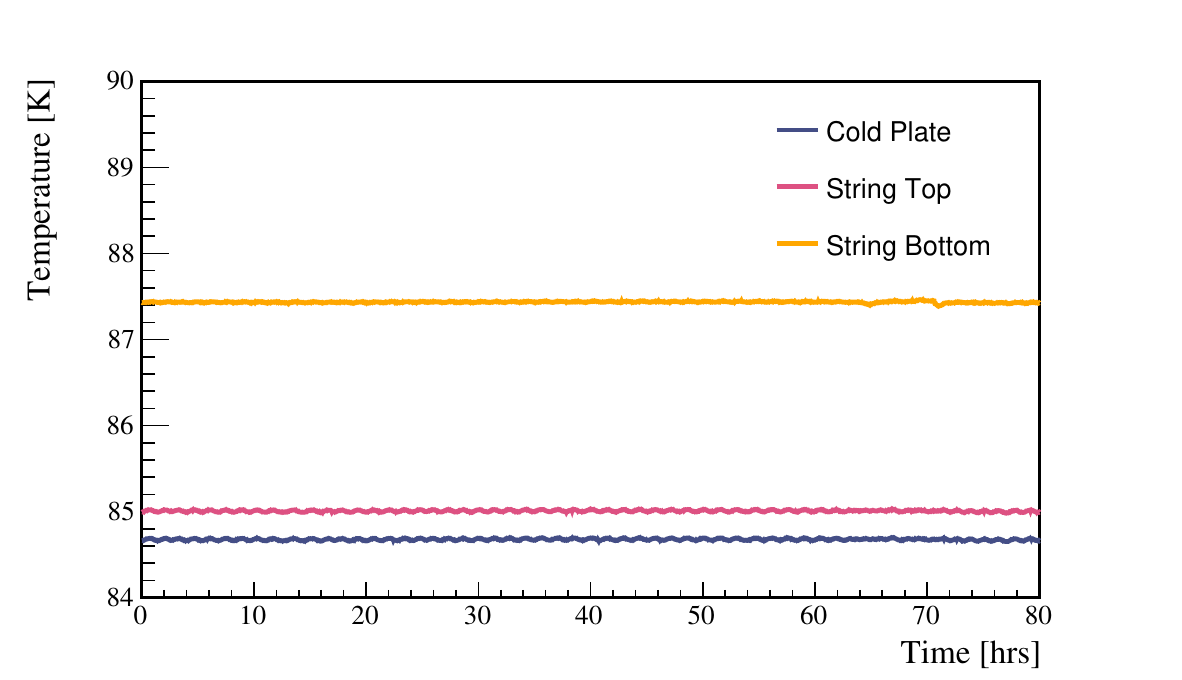}
\caption{Temperatures recorded in a prototype module equipped with a pulse-tube cooler for the coldplate, the top of a deployed detector string, and the bottom of a deployed detector string during 80 hours of operation in March 2014.  The PID temperature control showed excellent long-term stability, though generated small ($<0.1$~K) fluctuations with hour-long timescales.  The temperature differential along the string was measured to be 2.4~K.}
\label{fig:ptc_stability}
\end{center}
\end{figure}

There was a temperature gradient produced along each string resulting from heat generated in the Low-Mass Front Ends, with larger overall thermal impedance for lower detectors positioned farther from the coldplate.  
This gradient was 2.4~K along the length of the string in a prototype module in pulse-tube cooler mode, as seen in Fig.~\ref{fig:ptc_stability}.
The temperature stability of the proportional-intergal-derivative(PID) controlled module temperature in pulsed-tube cooler mode can also be seen in Fig.~\ref{fig:ptc_stability}.
Short-timescale oscillations with an amplitude of $<0.1$~K, can be seen on the temperature traces; these were associated with the PID feedback loop.
Long-term stability was excellent, with no discernible temperature drift over several-day timescales. 

\begin{figure}[ht]
\begin{center}
\includegraphics[width=15cm]{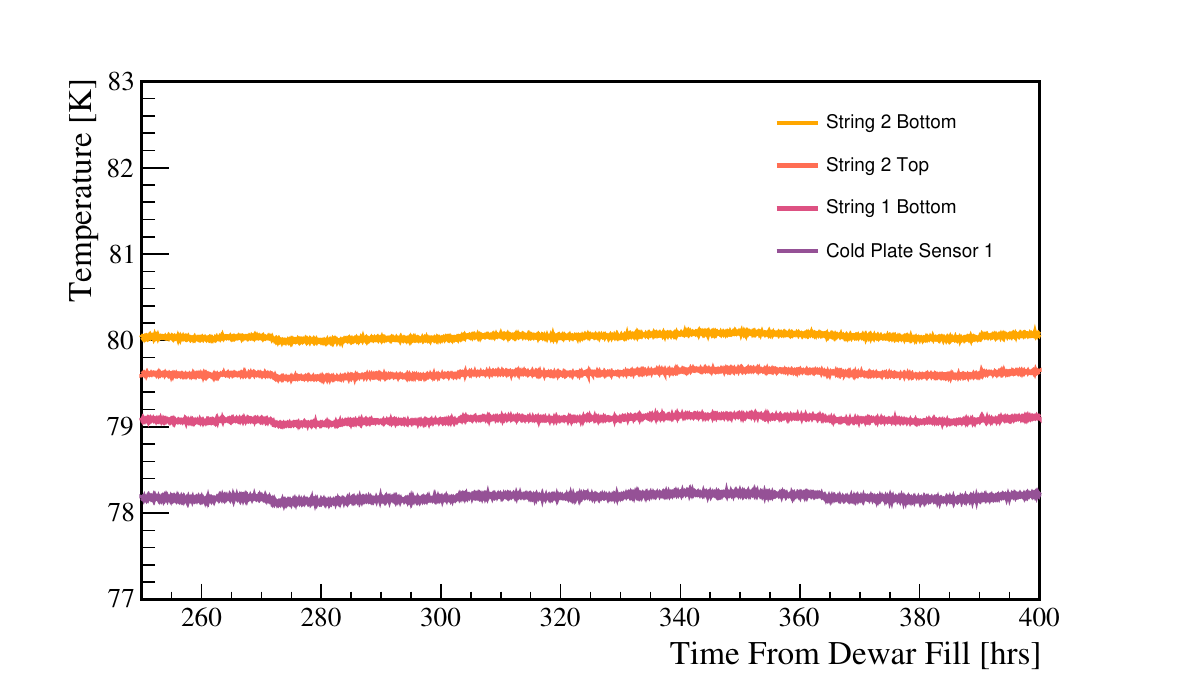}
\caption{Temperatures measured in Module~1 during a 150-hour period in May 2014 when it was operated with the thermosyphon. More temperature variation was seen when operating with the thermosyphon than with a pulse-tube cooler, due to the lack of active feedback.  Long-term temperature variation was measured to be $<0.25$~K, dominated by environmental conditions and the fill level in the thermosyphon dewar.
During this time 2 fills of the thermosyphon dewar were performed at hours 270 and 365, and were responsible for abrupt and small drops in temperature.}
\label{fig:TS_Stability}
\end{center}
\end{figure}

The thermosyphon operation had no active temperature feedback; cooling was performed by evaporation of LN in the thermosyphon volume, with the base temperature set by the boiling point of the liquid.
This provided better operational stability (e.g. there was no risk of losing PID lock), but left the module susceptible to changes in the environmental temperature.  
This can be observed in Fig.~\ref{fig:TS_Stability} depicting temperature variations over the course of 6~days of operation.

\subsubsection{Operational Stability}
The passive temperature stabilization provided by the thermosyphon allowed for excellent operational efficiency over the operating period of the \DEM.
Generally, the few disruptions in operation were the result of power interruptions at the Davis Campus which depleted our UPS system, requiring a shutdown of vacuum systems for a complete warmup and subsequent cooldown of the modules.
The vacuum systems operated with both a turbo-molecular pump and cryopump operated in parallel; this provided robustness in the case of failure of a single pump, or required maintenance. 
The custom dewars which contain a heat exchanger for the thermosyphon volume were challenging to fabricate; one of the delivered dewars required constant operation of a small ion pump connected to the insulation vacuum space to maintain reasonable LN consumption rates.

\section{Germanium Detector Calibration System}
\label{sec:calibration}

The design and installation of the \DEM\ germanium detector calibration system is described in Ref.~\cite{abgr17a}. 
In this section we provide a synopsis of the system and discuss its operational performance over the lifetime of the experiment. 

\begin{figure}
    \centering
    \includegraphics[width=0.8\textwidth]{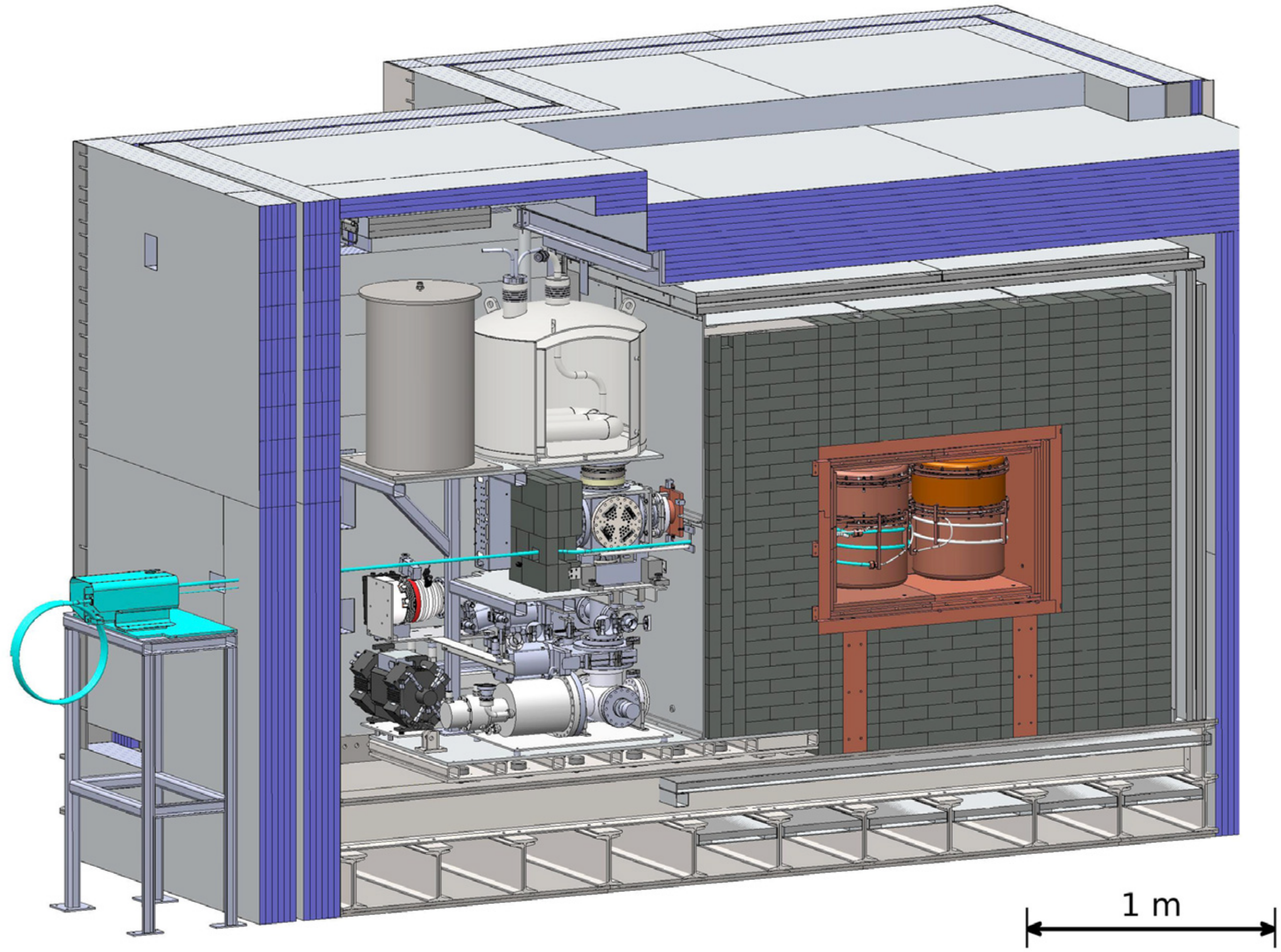}
    \caption{Drawing of the \DEM\ showing the calibration system and shields. The entire calibration system for a module is highlighted in light blue. The line source was stored in the mirror tracks on the left when not in use and deployed by motorized rollers via a sealed calibration track into a helical tube around each module.}
    \label{fig:calibration_layout}
\end{figure}

\begin{figure}
    \centering
    \includegraphics[width=0.8\textwidth]{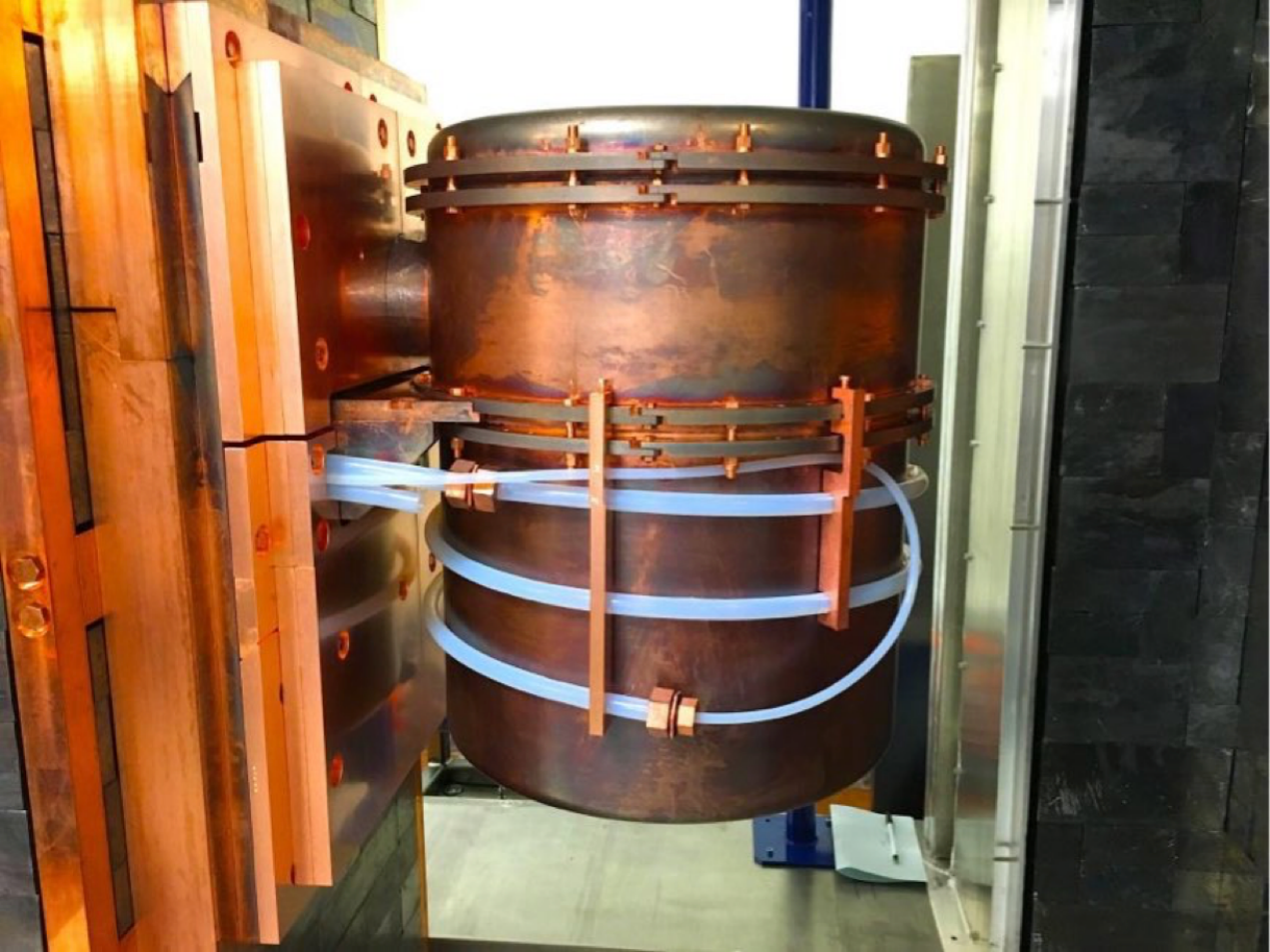}
    \caption{Picture of helical PTFE calibration track mounted around a module. The thinner tube was for the nitrogen purge.}
    \label{fig:calibration_track}
\end{figure}

The main requirements of the \DEM\ calibration system were to verify the proper functioning of the germanium detectors, determine energy calibration constants, provide data sets for training data cleaning and pulse shape cuts, and Monte Carlo bench-marking. 
These requirements were achieved using radioactive line sources deployed through the \DEM\ shielding into a helical track tube  or ``track" surrounding each module, as shown in Fig.~\ref{fig:calibration_layout} and ~\ref{fig:calibration_track}. 

Each line source consisted of a radioactively-doped epoxy injected into a 3~mm~diameter tube sealed at both ends, produced by Eckert \& Ziegler Analytics, Inc.
Four \thtte\ calibration sources were prepared in 1-m lengths, each with an integrated activity of $5.18\pm0.30$~kBq. These were used for regular calibrations. 
A single \cosixty\ source for was prepared with an integrated activity of $6.3\pm0.30$~kBq and used for special calibration studies.
A \cofs\ line source was deployed in both tracks in January 2019 to collect calibration data to fine-tune the analysis. 

The line source assembly was transported through a gate valve into a track from outside the shield using two drive rollers, one of which was driven by a motor. 
A track followed a curved path through the lead shield to prevent a direct shine path for external gammas. 
Inside the inner copper shield, the track was wrapped in a helical shape around the cryostat (Fig.~\ref{fig:calibration_track}), and the active region of the line source wrapped twice around the cryostat, allowing simultaneous calibration of all detectors in that module.
Each track was made of 1/2" diameter polytetrafluoroethylene (PTFE) tubing, and LN boiloff gas was fed into the track so that the whole track was flushed during calibrations to minimize radon intrusion.
Cleaning of the track was performed using acid leaching, as was done for other \DEM\ PTFE components. 

A positioning system using a combination of Hall effect sensors and weak magnets inside the source was implemented to determine the source position reliably and prevent accidental closure of the gate valve over the source assembly. 

When not deployed, sources were stored in ``mirror" tracks outside the shield so that they do not contribute to the background in the detector. 
Mirror tracks were wound in the same shape as the track inside the shield, thereby
``mirroring" the shape of the deployed source as it was stored. 
Because this part of the calibration system was outside the shield and accessible, the whole mirror track with source inside could be exchanged in only an hour if another source was required with negligible impact on live-time. 
It also simplified routine maintenance and part replacement, such as when drive rollers fail. 
All mechanical parts of the calibration system were commercially available, making this a cost-effective design. The motors and sensors were controlled and monitored by an Arduino micro-controller, which in turn was controlled via the ORCA data acquisition (DAQ) software (see Sec.~\ref{se:DAQ_Software}).

Mechanical and electrical components of the system were tested individually before assembly. 
Individual parts were cleaned following the cleanliness protocols of the \DEM\ (see Sec.~\ref{se:machining}). 
Parts close to the detectors were selected based on assay results.
The system was assembled in the clean room and test fitted when the assembly of individual modules was finished and initial detector tests in the glovebox were performed. 
The sensors, motor and valves were tested, and test deployments and retractions of the radioactive source were also performed.

After the assembly of the first module, one hundred deployments and retractions were done to demonstrate the reliability of the mechanical parts before operation. 
After installation, the calibration system was operated from the DAQ computer, where
ORCA routines monitor and control the deployment and retraction, and notify experts in case of failures.
 
During routine operations, the collaboration performed approximately hour-long weekly calibrations with a \thtte\ source, which has multiple gamma-lines useful for energy calibration, ranging from 238~keV to 2615~keV. 
The duration of the calibration was increased over time to account for activity loss due to the 1.92-year half-life of $^{228}$Th.
Gamma-rays below the 238~keV line were mostly shielded by the cryostat walls and other methods were developed for low energy calibration~\cite{Wiseman_2020}.
Special runs using \cosixty\ and \cofs\ sources were also performed, as well as occasional long $\sim$24~hour runs. 
During calibration, the rate was 30-40~Hz per detector, which was well within the limit of the DAQ at $\sim$100~Hz per detector. 
The offline procedure for calibration of the HPGe detectors, including the simultaneous fitting of multiple spectral peaks, estimation of energy scale uncertainties, and the automation of the calibration procedure is described in~\cite{Arnquist_2023}.

\begin{figure}[ht]
\centering
\includegraphics[width=0.9\textwidth,keepaspectratio=true]{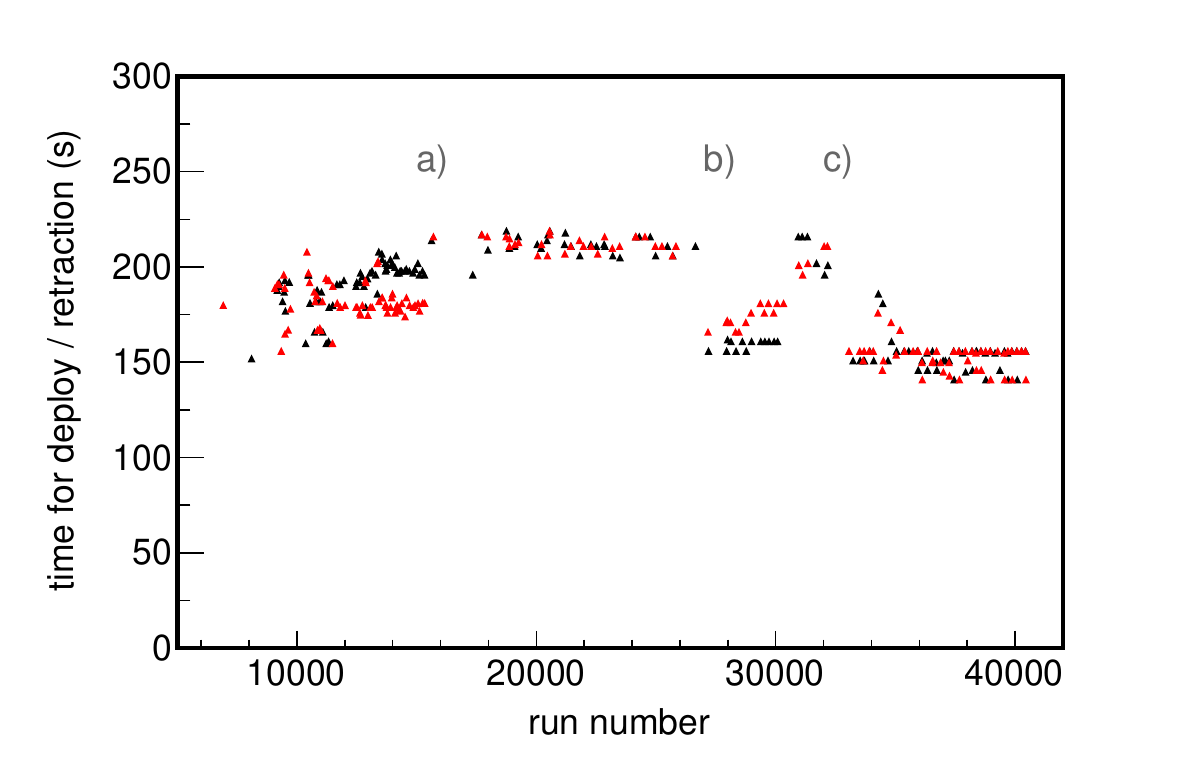}
\caption{Times needed for source deployment (red) and retraction (black) as measured during four years of operation. The three indices a,b,and c indicate times when the motor speed was tuned. For index a this happened after the second module was deployed and the deployment times of the individual calibration system were synchronized. At index b the calibration system was pulled back to its final position behind the poly shield. After that one of the drive rollers started to slip. This slowly lengthened the deployment period. After the roller was replaced  at index c, the speeds were close to the set points of b again.}
\label{fig:calibration_times}
\end{figure}

Source deployment and retraction times were about 3~minutes, as shown in 
Fig.~\ref{fig:calibration_times}, which was only a small fraction of a typical hour-long calibration run. 
If the motion of the sources took more than 300 seconds, a notification was sent out and the closure of the valve which seals the track was disabled. 
This feature prevented accidental destruction of the source when still inserted. 
Over four years of operation, each module's calibration source was deployed and retracted about 200~times without significant issues (Fig.~\ref{fig:calibration_times}), and the source position was reproducible to within two millimeters. 

\section{Shield Systems}
\label{sec:shield}

The \MJD\ detector array was contained in a low background shield, as shown in Fig.~\ref{fig:ShieldOverview}. 
The shielding attenuated neutrons and gamma-rays originating in the experiment hall (from rock, construction materials, and from the shielding materials themselves).
It also provided an active veto against cosmic-ray muons and a barrier against radon gas in the laboratory. 
The design of the shield is given in Ref.~\cite{abgr14} while the source materials are listed in Ref.~\cite{abgr16}. 
We describe here a summary of the final shield system in place during \MJD\ operation. 

An inner 5~cm layer of underground EFCu was used as the innermost shield for most of the operation of the \DEM . 
Surrounding this was an outer shield of commercial copper (5~cm) and lead (45~cm).
Next, a sealed aluminium enclosure defined a radon exclusion volume flushed with LN$_2$ boil-off purge gas. 
Two layers of active cosmic-ray anti-coincidence (veto) detectors enclosed the entire passive shield. Finally, 30 cm of high density polyethylene --- 5~cm of which was doped with boron as a neutron absorber --- reduced the neutron flux.  

\subsection{Passive Shield}\label{sec:passive}
\label{passive_shield}

\begin{figure}[ht]
\centering
\includegraphics[width=0.9\textwidth,keepaspectratio=true]{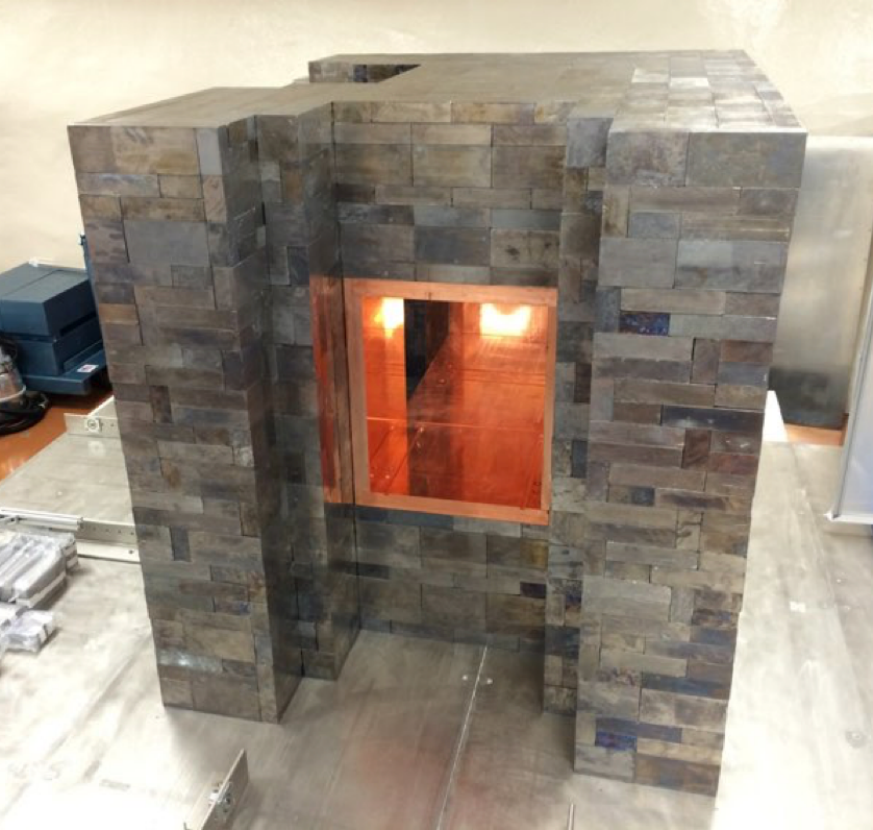}
\caption{The \DEM\ outer copper and lead shield showing the keyed openings for the monoliths. Only the outer copper shield was installed when this photograph was taken.}
\label{fig:pb_cu_shield}
\end{figure}

The lead shield was assembled by stacking over 5000~individual lead bricks, some machined to custom sizes, that had gone through a cleaning and processing stage offsite as preparation for use in the cleanroom laboratory.
Two sources of lead bricks (nominally 2~in$\times$ 4~in$\times$ 8~in) were
used for the shield. 
One was a new production run from virgin Doe Run Mine lead formed into bricks by Sullivan
Metals, Inc. and the other was from a discontinued, low-background counting facility at the University of Washington.
The outer copper shield was constructed from low-background commercially sourced C10100 oxygen-free high-conductivity copper provided by Southern Copper \& Supply Company, which sourced the plate material from KME in Europe.
The original copper cake material was supplied to KME by Aurubis and Mitsubishi Materials and selected based on assay results~\cite{abgr16}. 
The commercial copper plates were machined underground at SURF, chemically cleaned, and assembled, forming a box defining the center of the shield supported by copper legs. 
The lead shield was assembled around the commercial copper box. 
Bricks shapes and placements were chosen so that there were no cracks running from the outside to the inside of the lead shield (see Fig.~\ref{fig:pb_cu_shield}). 
Due to the long time required to electroform copper, the innermost electroformed copper shielding plates only became available after the start of operations and therefore the initial physics runs, which lasted from July 2015 to Dec. 2015, had only the commercial copper shield in place (see~Sec.~\ref{sec:data sets}).
The EFCu plates were chemically etched following a newly validated cleaning protocol to address potential surface contamination during part handling and cleaning~\cite{chri18}. 
This inner copper shield layer was installed into the existing shield by Jan.~2016 when operations were restarted. 
The passive shield materials selection, assay and cleaning is described in detail in Ref.~\cite{abgr16}.

The polyethylene shielding consisted of 2.54-cm-thick sheets tiled and stacked around the exterior of the active muon shield. 
The outermost layer of the polyethylene shield was enclosed in aluminum cladding to provide an ignition and fire barrier. 
Since the polyethylene shield fully enclosed the heat load of the detector cryogenics and electronics hardware, a chilled water loop running to two heat exchanger units cooled the air space surrounding the detector modules within the shield. 
There were  two smoke detectors and two temp/humidity sensors operating inside of the shield. 

\begin{figure}[th]
\centering
\includegraphics[width=0.9\textwidth,keepaspectratio=true]{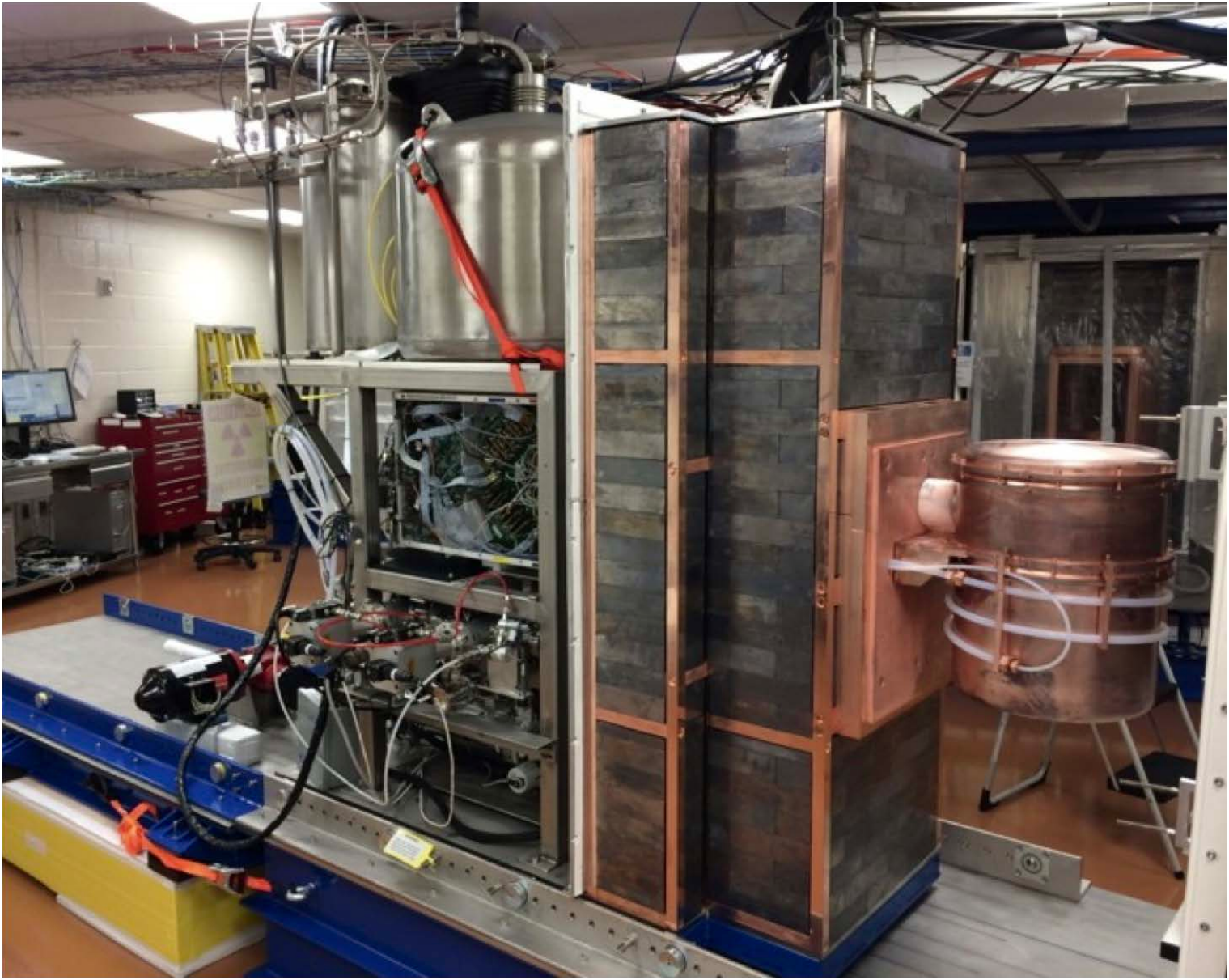}
\caption{A shield monolith in transit with cryogenic services, shielding, module and calibration track.}
\label{fig:monolith_in_transit}
\end{figure}

Each module and its associated cryogenic and vacuum systems was mounted on a section of the copper and lead shield, known as a monolith (Fig.~\ref{fig:monolith_in_transit}), to allow phased implementation of detector modules.
A blank monolith, which had no module, could be exchanged should one of the modules be absent from the shield (e.g., during installation or for servicing).
Monoliths were mated with the glovebox to allow detector installation and moved using a Hovair air bearing transporter. 
Each monolith was mated with a keyed opening in the shield (Fig.~\ref{fig:pb_cu_shield}).
The radon exclusion box had two openings that mated to `doors' integrated into the installed monoliths. 
Upon deploying the detector module into the shield, the monolith's radon exclusion door was attached to the radon exclusion box and sealed with a PTFE gasket around the door's perimeter and an inflatable rubber seal to fill the clearance gap underneath the monolith.

\subsection{Active muon shield}
\label{sec:veto}
\begin{figure}[ht]
\begin{center}
\includegraphics[width=0.8\textwidth]{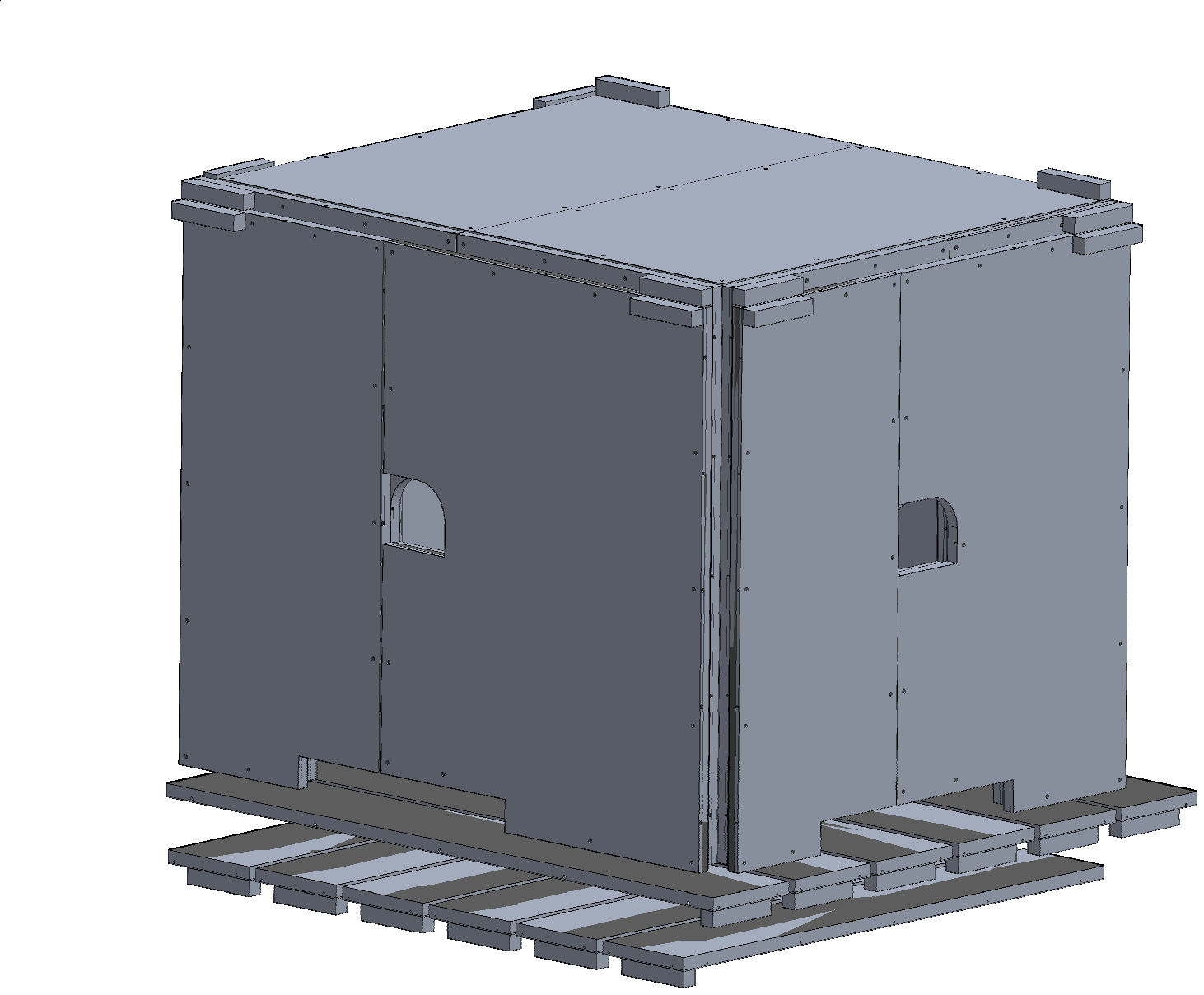}
\caption{The muon veto panels that surround the lead shield. Muon panels on the bottom were installed in channels in the overfloor that supports the passive shield.}
\label{fig:veto}
\end{center}
\end{figure}

The cosmic-ray anti-coincidence (veto) detector array consisted of 32~polyvinyl toluene scintillating (PVT)  panels surrounding the passive shield.  
Two layers of 2.54-cm-thick EJ-204B scintillating PVT sheets, encapsulated within aluminium cladding, covered each side of the shield with a total area of 37~m$^2$ and almost a 4$\pi$ solid angle coverage. 
The system operated in various configurations since June~2014 depending on the deployment status of the germanium detector modules. 
The final configuration of the muon veto panels is shown in Fig.~\ref{fig:veto}. 
The four panels on each of the four sides and the top were  overlapping to minimize gaps. 
The 12~panels that reside on the bottom in two orthogonal  layers were  narrow to fit within the channels of the steel overfloor that supported the weight of the passive shield. 

Light from each individual panel was read out by a single 1.27-cm photomultiplier tube (PMT) with wavelength shifting fibers embedded into grooves machined in the scintillator.
Each panel contained a pulsing light emitting diode (LED) to provide a consistent light pulse to all channels simultaneously to monitor the stability of the system. 
The veto system readout was triggered whenever any two panels have signal amplitudes above the hardware threshold, at which time raw signals from all 32 panels' PMTs were  read out by QDC cards and assigned a timestamp from a 100~MHz clock common with the germanium detector readout for offline analysis.
See~Sec.\ref{se:veto_DAQ} for details about the veto DAQ and readout. 
A full description of the muon veto system is given in Ref.~\cite{abgr17c}, including an initial measurement of the muon flux using data collected up through Nov.~2014.
Further work measured the in-situ cosmic-ray activation and the differential muon flux~\cite{PhysRevC.105.014617}. 

\subsection{Shield purge system}
\label{sec:purge}
A sealed aluminum radon exclusion box surrounded the lead shield. 
A dedicated shield purge system delivered LN boil-off N$_2$ gas directly to the inner shield cavity to displace laboratory air. 
The purge gas was allowed to vent out of the radon exclusion box to maintain a slight over-pressure of purge gas.
A 50-L, sealed LN cylinder produced the N$_2$ gas through an internal pressure builder set to maintain a fixed head pressure. 
The exhaust gas was routed through a heat exchanger to bring the N$_2$ gas up to room temperature, and its flow was  controlled by a metering valve. 
The purge flow rate was displayed and logged by the slow controls system (see Sec. \ref{se:DAQ}) for data quality checks. 
A radon purge model has been developed to characterize the behavior of the radon flow within the shield and to assist in determining any trace levels of radon backgrounds present in the \DEM\ data. This model was based on data from two tests using set durations of high, low, and no nitrogen purge to the shield.

\section{Readout Electronics and Data Acquisition System}
\label{sec:DAQ_electronics}

The ionization signals from the HPGe detectors were amplified using a charge-sensitive preamplifier. 
The first stage of the preamplifier was located near the detector and is described in section~\ref{se:mounts_LMFE}. 
The second stage was in an ``electronics box'' mounted outside the lead shield, and its output signals were recorded using VME-based digitizers developed for the GRETINA experiment~\cite{lee04, zimm12}.
Charge and timing information from the muon veto PMTs were acquired using commercial charge to digital (QDC) and time to digital (TDC) converters.
The ORCA software package was used for all data acquisition and control~\cite{howe04, orcaweb, orcagit}.
This section describes these subsystems. 

\subsection{Germanium Detector Readout Electronics and Data Acquisition System}
\label{se:DAQ}
The readout electronics are described in detail in Ref.~\cite{Majorana:2021mtz}, and we provide a summary here.

\begin{figure}[ht]
\centering
\includegraphics[width=0.75\textwidth,keepaspectratio=true]{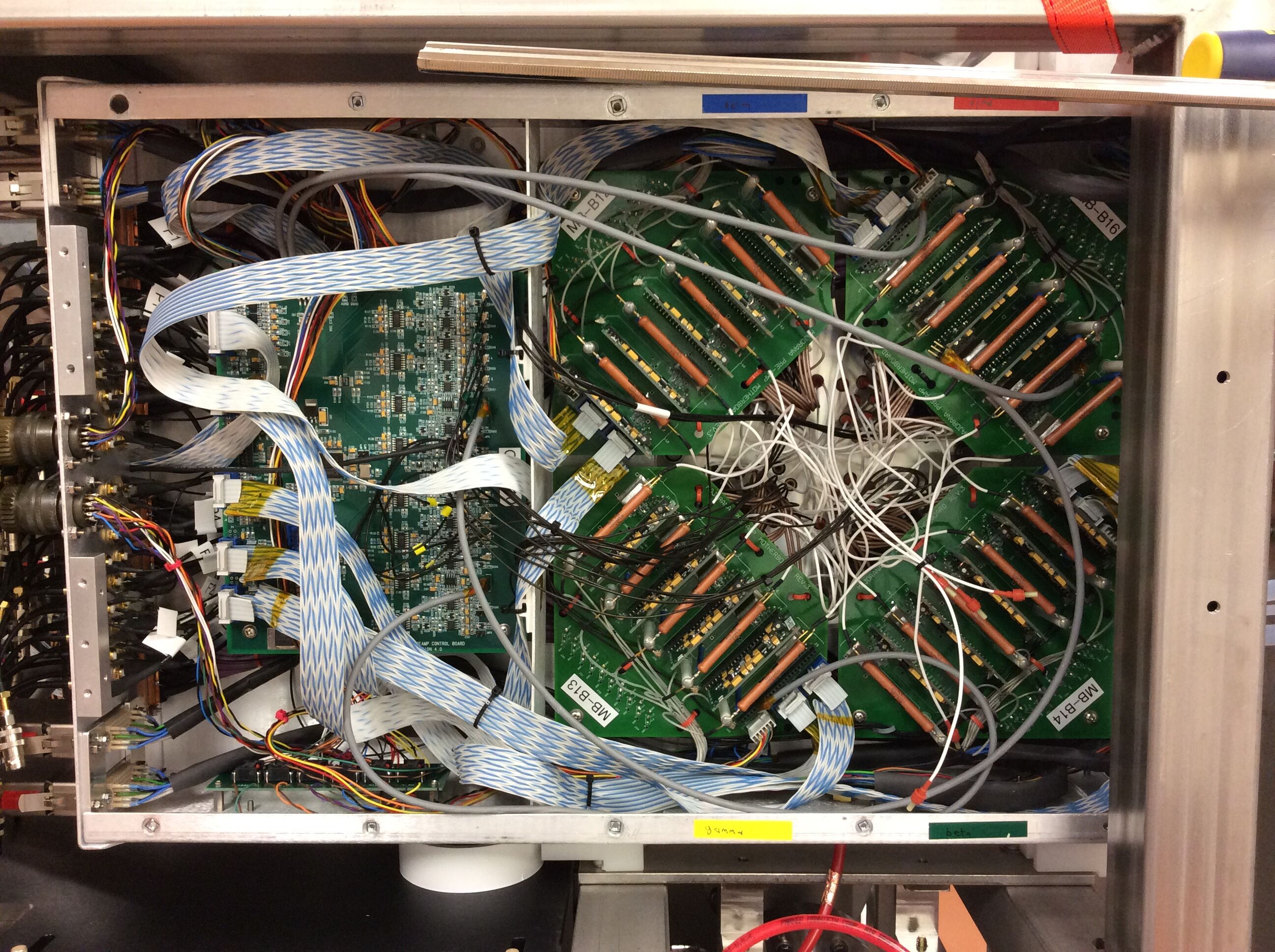}
\caption{Picture of the inside of an electronics box, showing a controller card on the left, and four motherboards, each with five mounted preamplifier cards, on the right. 
Connections to the vacuum feedthroughs were made in the central space between the motherboards. Figure taken from~\cite{Majorana:2021mtz}.}
\label{fig:electronics_box}
\end{figure}

The signal cables from the detectors were terminated on  feedthrough flanges that provided the electrical interfaces between the laboratory and vacuum spaces of the cryostats, as described in Sec.~\ref{se:CC_upgrades} and shown in Fig.~\ref{fig:ccd}.
An electronics box was mounted directly to the warm-side of each feedthrough flange and each module had two independent electronics boxes. Fig.~\ref{fig:electronics_box} shows an example of an electronics box.   
Each electronics box contained a custom controller card that in turn contained sixteen 12-bit ADCs for monitoring detector baseline voltages and sixteen 16-bit digital-to-analog converters for pulsing the detector FETs. 
These pulsers were used to monitor the detector livetime, gain stability, and trigger efficiency. 
Each electronics box also contained four ``motherboards" that housed the preamplifiers and routed HV and low-voltage power.
Each motherboard was connected to the front-end electronics cabling (Sec.~\ref{se:mounts_LMFE})  via one 50-pin DSUB connector on the flange. 
Up to five preamplifier cards were mounted on each motherboard, and each preamplifier card provided second stage signal amplification for one detector. 
The preamplifier cards provided both high and low gain differential signals to the digitizer that were used in different analyses.   

These signals were recorded in the \DEM\ by digitizers developed for the GRETINA experiment~\cite{lee04, zimm12}. 
Each digitizer card had ten digitizer channels and a field-programmable gate array (FPGA) digital signal processing (DSP) module.
Each channel was sampled at a frequency of 100~MHz with a 14-bit precision analog-to-digital converter (ADC).
The DSP provides on-board algorithms of which several were used by the \DEM . 
These included on-board triggering, trapezoidal shaping, energy estimation, and pre-summing of the pre- and/or post-rising edge for an effective variable sampling rate. 
The on-board trapezoidal filter-based trigger with thresholds programmable for each channel was critical to achieve a low trigger threshold of $\sim0.5$~keV, and each detector's high and low gain outputs were recorded by separate channels on the digitizer. 
This allowed a wide dynamic range of physics events to be recorded, from $\sim0.5$~keV to $\sim10$~MeV.

The GRETINA-card ADC chip (Analog Devices AD6645) exhibited small non-linearities in its response. 
To fully utilize the excellent energy resolution and pulse shape analysis capabilities of the HPGe detectors, these had to be accounted for. 
A simple measurement protocol was developed by the collaboration, and a slope-dependent hysteresis was measured and corrected for, as described in Ref.~\cite{abgr21}.  

The GRETINA digitizers were housed in a standard VME64x crate that was read out and controlled by a single board computer (SBC) running the Linux operating system. 
Each of the two detector modules was read out by a dedicated SBC and crate instrumented with digitizer cards. 
The SBCs for both modules were controlled and read out by a main DAQ computer described in Sec.~\ref{se:DAQ_Software}. 
During routine operations the trigger rate was tens of hertz and during calibration it would exceed 1~kHz.
Each time the detector configuration was changed, the detectors were unbiased, or some other change occurred which required re-initialization of the digitizers, the trigger thresholds were reset based on the filter's value at initialization.  
In order to maintain low thresholds just above the electronics noise, an automated script running on the main DAQ computer allowed the thresholds to be automatically set relative to the new reference value of the filter by performing a binary search on the threshold for each channel while monitoring the trigger rate.

A multi-board synchronization system specific to the GRETINA digitizers was utilized to maintain synchronous acquisition across the system.  Each GRETINA digitizer within a module was connected to a router trigger card, and the router cards for each module were driven by a global trigger card for the system. 
The global trigger card also provided clock and reset signals for the veto system which is described in the next subsection.  On initialization, the clock counters were reset for each digitizer in the system, and the initial time reference was taken from the NTP server running on the primary DAQ machine.  

The bias voltages of the detectors were provided by 8-channel ISEG EHS8260p\_105 modules hosted in a 344~WIENER MPOD crate.
Each channel voltage was set individually, based on detector manufacturer specifications or operational requirements. 
The MPOD modules were monitored and controlled by the main DAQ computer. 

\subsection{Muon Veto Data Acquisition System}
\label{se:veto_DAQ}
The 32~signals from the muon veto PMTs were distributed between two 16-channel CAEN792 QDC cards and two 16-channel logic discriminators. 
The QDC provided energy information and the discriminators provided hit patterns for the veto trigger. 
For most of the \DEM 's operation, the veto system triggered if two or more panels were hit.
Once triggered, all 32 channels were read out by the main DAQ computer. 
Each panel was constantly monitored with LEDs embedded in the scintillator. 
Reconstructed LED events were also used to measure the live time of the system. 
Clock and reset signals for the muon veto system were provided by trigger modules for the GRETINA digitizer synchronization cards.
The trigger condition allows $>$99\% of muon tracks to be accepted while keeping the accidental background rate from $\gamma$ rays at a reasonable level.

During offline analysis, clear muon-induced tracks in the HPGe detectors were used to demonstrate clock synchronization of the two systems to within 0.2~ms.
Most muon-induced events in the HPGe detector array occur promptly while over 99\% of delayed events from excited state decays deposit energy within 1~s of a muon interaction. 
Therefore, the muon veto cut applied to the final HPGe detector analysis for \BBz\ was within 0.2 ms before and 1 s after a candidate muon detection. 

\subsection{Data Acquisition Computer and Software}
\label{se:DAQ_Software}

\begin{figure}[ht]
\centering
\includegraphics[width=0.95\textwidth,keepaspectratio=true]{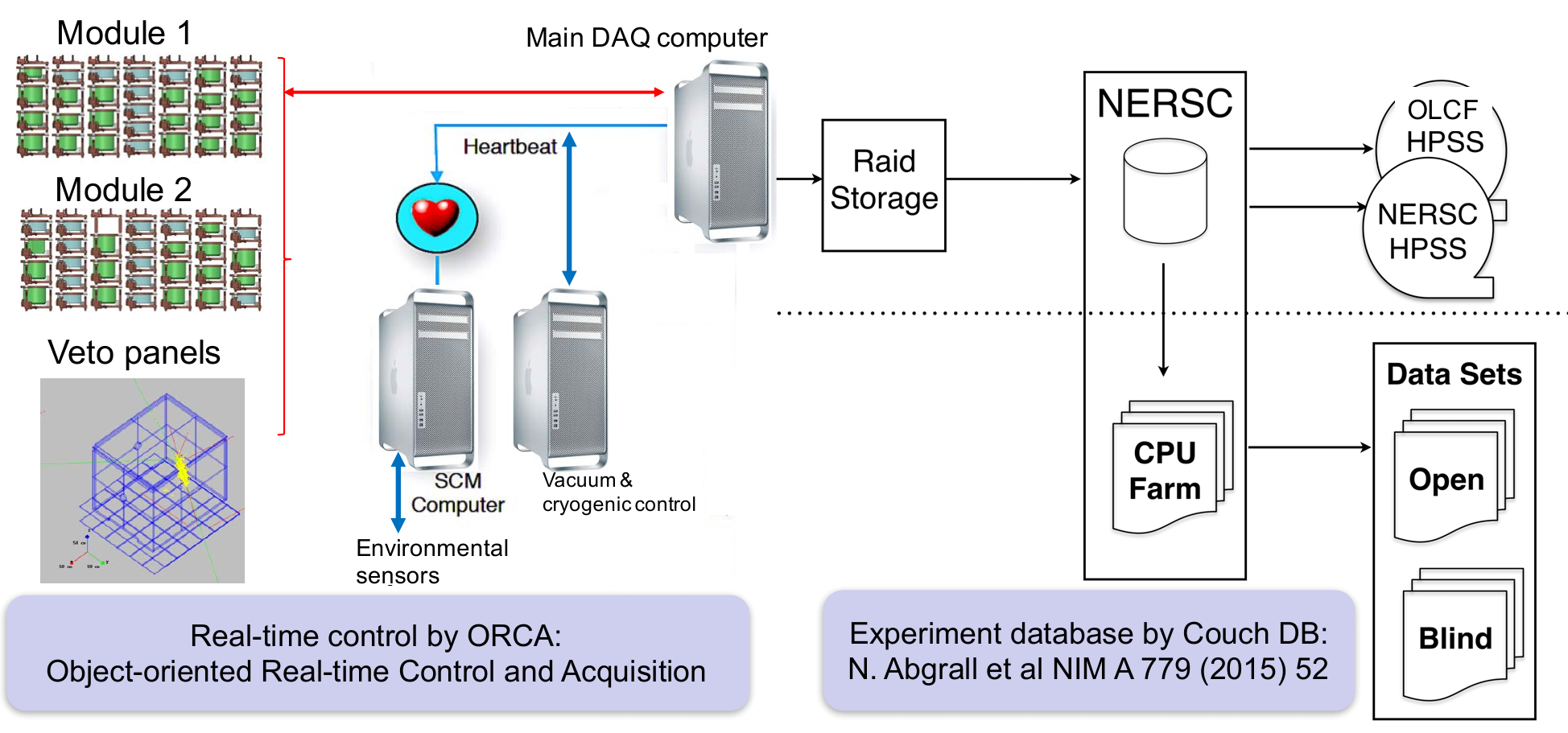}
\caption{High-level diagram of the \MJD\ DAQ systems and data processing. Acronyms are explained in the text. "CPU Farm" refers to the computing resources used at NERSC (see Sec.~\ref{se:underground_data_prod}).}
\label{fig:DAQ}
\end{figure}

The \DEM\ used the Object-oriented Real-time Control and Acquisition (ORCA) software package for data acquisition and control~\cite{howe04, orcaweb, orcagit} running on a dedicated Mac Pro workstation as the primary DAQ computer (Fig.~\ref{fig:DAQ}). 
ORCA is designed for general purpose control of a variety of DAQ and slow controls systems.
It is developed upon the Apple Cocoa framework and optimized for Apple Mac OS X. ORCA treats DAQ hardware as self-contained objects that can be dynamically combined into complete DAQ systems, and therefore it is highly modular. 
ORCA supports run-time configuration, run control, data exploration, custom alarm conditions, and email notifications to specified on-call experts. 
ORCA can also interface with the Apache CouchDB and is capable of replicating local databases to remote sites for near-time monitoring and stored history of the conditions.

Data from the DAQ modules were saved in the native ORCA binary format and transferred to a RAID storage system underground (see Sec.~\ref{se:underground_data_prod}). 
This served as a buffer to allow continuous data-collection even if the network connection to off-site storage was lost.  
The main DAQ computer also provided monitoring data, such as detector rates and baselines, to CouchDB databases at remote sites (Sec.~\ref{se:monitoring}).
An onsite analysis workstation was equipped with ORCA and \DEM\ analysis software in order to continuously monitor the main data stream and to automatically perform more sophisticated analyses at the end of each run to ensure that any issues with data quality would be caught in a timely manner.

After the initial commissioning phases of the \DEM, the DAQ system maintained $>99$\% live time over the course of operation of the enriched Ge detectors.

\section{Slow Controls and Monitoring}
\label{sec:slow_controls}

The underground location of the \DEM\ provided challenges for the control and monitoring of experimental systems.
Power interruptions could occur, sometimes without warning. 
Network connectivity could also be lost. 
Underground personnel access was restricted, and even during normal operations the experiment routinely operated for up to four days at a time with no access.
This placed additional requirements on the monitoring and control systems to operate reliably and maintain the safety of the detector array and other components. 

\subsection{Outline of Slow Controls and Monitoring Systems}
\label{sec:SCMOutline}

The \DEM\ slow controls and monitoring system was controlled and monitored by ORCA and ran on a single Mac workstation called the ``SCM computer'' (Fig.~\ref{fig:DAQ}). 
The workstation tracked the following environmental sensors and controllers:
\begin{itemize}
\item Temperature and relative humidity in the different rooms and inside the shield. The latter levels were important for stable operation of readout electronics. 
\item Particle counts to track cleanliness.
\item Radon levels from a Durridge-RAD7 module.
\item Differential pressure between laboratory and the common corridor to ensure positive pressure and monitor personnel access.
\item Seismic data.
\item LN levels in alcove dewars and boil-off gas generating dewars.
\item LN boiloff flow rates to the shield and glovebox. 
\item Most electronics were powered via remote power managers that allowed hard power cycles. Most of the power managers were controlled from this workstation.
\item The UPS status.
\end{itemize}

A sophisticated system of interlocks was deployed to protect experimental systems, as described in section~\ref{se:cryovac_SCM}. 
For example, the loss of power to the lab or connection to the cryogenic and vacuum systems control computer would trigger the main DAQ computer to ramp down HV for the HPGe detector array.

\subsection{Remote Monitoring}
\label{se:monitoring}

\begin{figure}[ht]
\centering
\includegraphics[width=0.95\textwidth,keepaspectratio=true]{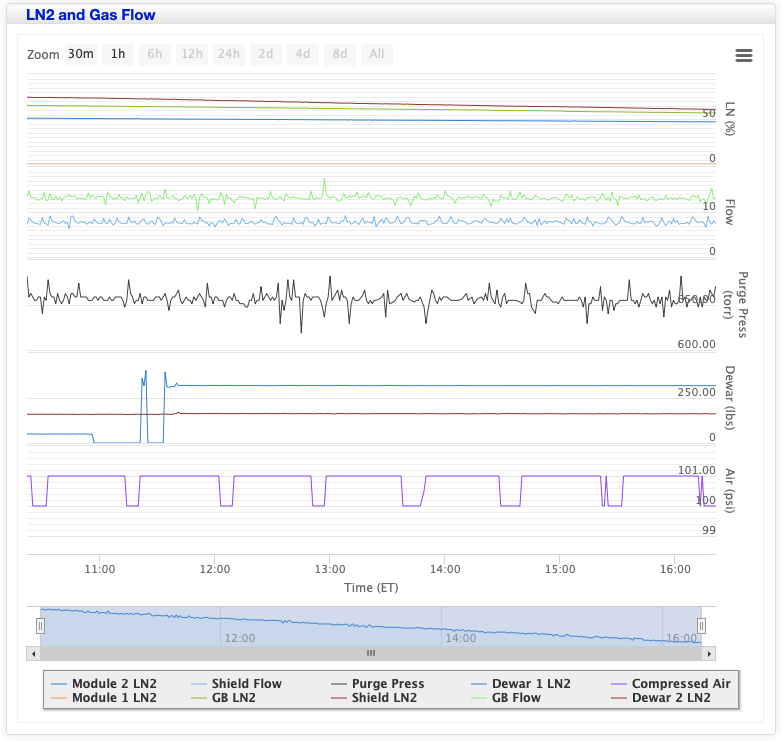}
\caption{A screenshot, taken on 11/7/22, of a part of the passive web-based monitoring tool developed for the \DEM . Descriptions of the different traces are provided at the bottom of the figure.}
\label{fig:feresa_screenshot}
\end{figure}

All underground workstations had remote screen-sharing, which was the primary method of control, reserved for use by experts. 
All slow control data was stored in a database (see section~\ref{se:feresadb}), and a passive, web-based monitoring tool was developed that displayed both real-time and historical data for environmental parameters (Fig.~\ref{fig:feresa_screenshot}). 
In addition to the environmental conditions described above, the web tool displayed detector rates, detector baseline, HV module settings, and HV power supply currents provided by the main DAQ computer. 

 The collaboration instituted remote DAQ shifts to monitor the DAQ performance, slow controls, and environmental parameters several times a day, which included data quality, detector rates, the status of subsystems, the usage of LN, the laboratory environment, the inter-system communications and more. 
 In addition, the DAQ shift personnel checked the run database, which was fully integrated with data-processing and analysis software (see section~\ref{se:run_database}). 
 
 Each of the different control computers running ORCA could contact experts via automated text message and email in case an issue occurred that needed immediate attention.

\section{Databases}
\label{se:databases}
Assembling and operating the \MJD\ was a complex process that required accurate record keeping. 
The collaboration developed and deployed several databases to record parts-fabrication histories, detector cosmic-ray exposure, environmental conditions, and analysis parameters. 
These databases were based on CouchDB, which is a document database with a RESTful API, containing Javascript Object Notation (JSON) documents,  indexed with a single key, with a variety of record identification and retrieval mechanisms.
These databases are described in this section, with updates provided to previous publications. 

\subsection{Parts Tracking Database}
\label{se:parts_tracking}

The collaboration used a database to record the history of all the parts installed inside the shield that were nearest to the detectors. 
The motivation was to ensure that all steps were followed during fabrication and also allow tracing of potential contamination to the origin. 
Importantly, it also tracked the cosmic-ray exposure of EFCu components during shipping, surface machining, and surface storage.
This database was called the \MJ\  Parts Tracking Database (PTDB) and is  described in Ref.~\cite{abgr15}. 
We provide a brief description and update here. 

The PTDB contained records of parts, assemblies of parts, and processes applied to the parts.
Processes include transportation, storage, cleaning by etching and washing, parylene coating, and machining.
The PTDB also recorded relationships between parts, mostly through a parent-child relationship in which a parent material links to child parts, and assemblies, in which parts are used to construct a subsystem of the detector system.   
For example, a sheet of EFCu will be machined into plates that are children of the parent sheet.  
Rods, cut from machined plates, are recorded as children of the plates, with their machining history.
The rod can then become part of a detector unit assembly.
  
The PTDB automatically generated a unique part number for each new part and assembly when it was entered.  
To distinguish similar parts, each part added to the database was laser-engraved with the generated part number, except for the smallest parts where it was impractical.

The database was served by a CouchDB  server.   
The user interface was implemented as a Javascript application document in the database, served by the CouchDB server.

The Javascript application was developed using Bootstrap and Backbone.js Javascript environments.  
The PTDB database contained 15,198 records, of which 9,709 were parts records and 1,135 were assemblies.  
The remaining records were history and processing records.

\subsection{Enriched Ge Tracking Database}
\label{se:enrge_tracking}

Tracking the history of enriched germanium detector material was important in understanding cosmogenic radioisotopes production.
As mentioned earlier, exposure to cosmic-rays at the surface produced \gese\ and \cosixty\ in enriched germanium.  
While the \cosixty\ was removed by the zone refinement, \gese\ remained and built up throughout the processing and reprocessing of the enriched germanium.  
Accurate estimation of cosmogenic isotope production during construction required careful tracking of the enriched germanium through production and its surface exposure.
A separate database was used for this, based on the same user interface as the PTDB, but revised for the unique materials and relationships in producing the detectors, as described in (\ref{sec:geprocessing_and_fabrication}) and Ref.~\cite{abgr18}.
The most important function of the enriched germanium tracking database was to record the altitude-dependent surface level exposure of the Ge to cosmic-rays and the mixing of the different batches and pieces of germanium in the final detectors.
Enriched germanium was stored, as much as possible, in the shielded transport container, Cherokee Caverns in Oak Ridge or finally at SURF in the Davis Campus.  
Enriched germanium was exposed during reduction of the oxide to metal, the zone refinement of the metal into pure metal suitable for use by AMETEK-ORTEC, and during the next level of zone refinement by AMETEK-ORTEC, crystal pulling, detector production, and transportation. 
All of these steps were recorded in the database. 

The enriched germanium tracking database was implemented in the same CouchDB server as the PTDB.  The user interface was also based on the same Javascript environment as the PTDB. 
Appendix~\ref{se:egtd_appendix} shows a sample record and demonstrates the complexity of the history records. 
The database has 2778 records for 35 enriched detectors.
This included 112 records of germanium oxide material, 87 reduced bars, and 691 zone refined bars.
The structure of this database and the calculation of the estimated exposure and cosmic-ray activation of the \DEM\ array is the subject of an upcoming publication. 

\subsection{Slow Controls and Monitoring Database}
\label{se:feresadb}

The slow controls (Sec.~\ref{sec:slow_controls}) and the data acquisition (Sec.~\ref{se:DAQ_Software}) processes created their own CouchDB databases.
The set of documents in the slow controls database contained the most recent status updates from each object in the ORCA configuration. The database associated with the data acquisition process contained information like log files, disk and memory usage, posted alarms, etc. All updates were posted either periodically or when content changed.  Each object's records were identified with a statically defined identification name and documents were identified with Universally Unique Identifier (UUIDs).
Additional databases were created in experiment-specific code within ORCA, for example for monitoring of the germanium detector baselines.
 
For both backup and near-time monitoring purposes, each of the databases on the underground DAQ, vacuum and cryogenic controls, slow controls, and near-time analysis machines were continuously replicated to a dedicated server hosted at UNC-Chapel Hill.  
In total, the server managed replication of 22 CouchDB databases with a combined size of 134~GB in 341 million documents.  
A web service used by the collaboration to monitor the status and history of each experimental subsystem was also hosted by this server (Fig.~\ref{fig:feresa_screenshot}).  
The results of CouchDB document requests and views were retrieved using a simple PHP interface to CouchDB, and Javascript utilities were used to display the data in an interactive web interface. 
This web interface was used by the collaboration to monitor the experimental operating conditions on a regular basis and provided only passive monitoring capabilities.
Sub-system experts were notified by the collaboration's shift personnel who monitor the web interface if issues arose that were not captured by the DAQ automated alarms.
 
\subsection{Run Database}
\label{se:run_database}
The \MJD\ generated approximately 80,000 data files for detector calibration, commissioning, and physics data runs. The large number of data files presented a challenge, since keeping track of run properties was difficult with traditional capabilities of computer operating system directories. 
Instead, a combination of Unix bash, Python, and C++ applications initiated data transfers, executed analysis programs, and stored the processing status in a CouchDB database.
The database provided a method to categorize and access the runs in terms of many attributes such as run type (calibration, test, background data), run quality  and access (open, blind) and to create run lists for analysis. 
The ability to store and display a wide variety of objects in the database --- such as text log files and images as attachments within individual run records --- allowed the database to be used by the collaboration for daily operational checks as well as archiving the run information. Storing processing logs in the database allowed for quality checking of mass-reprocessing tasks.  

The \DEM\ run database implementation allowed access to the database in three ways: via a web browser, via operating system scripts, and through C++ analysis program utilities. Collaboration members viewed the data file status through a web browser on a daily basis to keep track of file transfers, data processing, and data quality. 
Additionally, the database was used by automated programs (cron-jobs) that query the database via operating system level scripts.   
For example, automated production scripts checked the database for files that needed processing and updated the file status and quality fields when the analysis programs completed.

\subsection{Database of Analysis Parameters}
\label{sec:database}

For the offline processing of \MJ\ data, a database was developed to record and provide parameters to the analysis codes.  
These included peak fitting, detector thresholds, analysis cut values, detector masses, and liquid nitrogen fill times.
The latter created periods of increased electronic noise in the detectors, which were removed during analysis.  
The database had ASCII record text to simplify verification and was implemented on CouchDB.
JSON documents represented each instance of a parameter with a single object of parameter data and an additional object containing provenance metadata. 
Each record contained data for one kind of parameter (such as energy calibration constants), including values, uncertainties, and covariances. 
By limiting the scope of each record, the amount of returned data was limited to only what was required.   

Database access was via a custom C++ API, linking the ROOT based analysis and the CouchDB API.
Our custom API provided for creation and maintenance of JSON documents, and manipulation of particular analysis parameters to create, insert, and extract parameters from database documents. 
The software depended heavily on CouchDB views to locate groups of records according to provenance values, libcurl to interact with the RESTful API, and JSON parsing using the TABREE software from the KATRIN collaboration~\cite{Aker_2021}.

This database consisted of approximately 270,000 records, which contained 122,000 energy calibrations, waveform parameters, data quality parameters, and related data for about 80,000 runs.

\section{Data Production}
\label{se:data_prod}
This section provides an overview of the data-taking phases of the \DEM\ and how the data were staged and processed underground and at surface computing facilities.

\subsection{Underground Data Production}
\label{se:underground_data_prod}

The data acquisition system (DAQ) was set up to run continuously and determined the underground-to-surface data bandwidth at SURF.   
Continuous running, without stopping of the digitizers, maximized the livetime of the detector array.
The DAQ wrote every event from all detectors to one file,
including XML configuration data for the acquisition system and each ADC.  
The data files were closed when either 1~hour of acquisition had elapsed or the file 
was 2~gigabytes (GB) in  size.  
One hour was chosen to minimize data losses should file corruption or other errors occur.
The choice of 2~GB kept files to a manageable size. 
The DAQ operation had two primary modes, background running  that produced a file every hour, and weekly source calibrations that produced a 2~GB file every 6~to 15~minutes.
At the end of the \DEM 's operations, the calibration source rate had reduced significantly, which resulted in the lower acquisition rates. 
Other modes were used for troubleshooting the DAQ system and during commissioning.
These data were not used for physics analyses. 
Data files were written to a 900~GB SSD disk to minimize file writing overhead in the data acquisition system.  
To further improve the manageability of the data files, the data transfer task computed a MD5 checksum on each file after it was written, compressing it with gzip, and then computing a checksum on the gzipped file.  
In subsequent file transfers, the checksum traveled with the data file to permit validation of the data.

\begin{figure}
\includegraphics[width=0.9\textwidth]{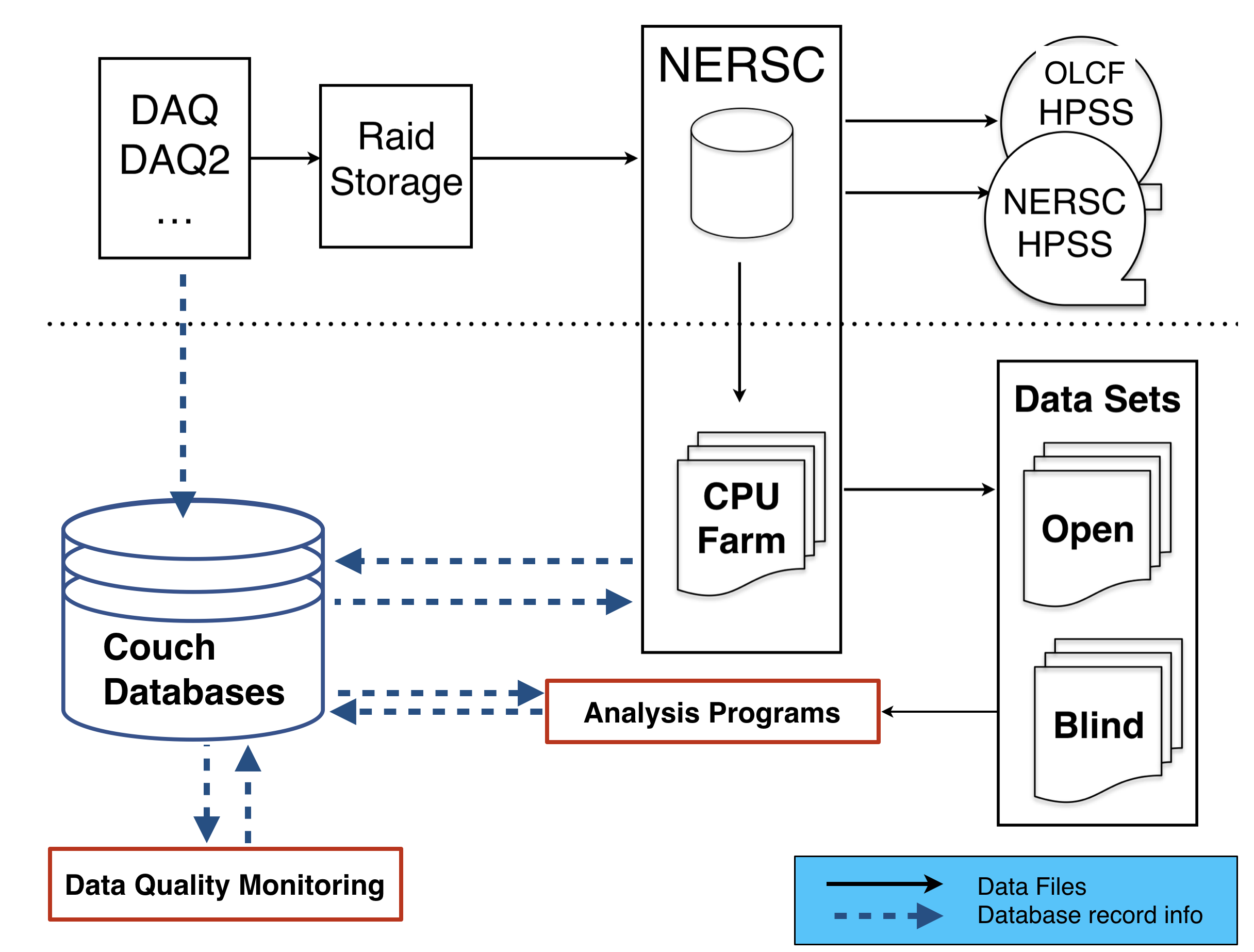}
\caption{High level overview of data processing. The abbreviations are described in the text. To ensure data security, the processes above the dotted line were carried out by a restricted user account. }
\label{fig:data_prod}
\end{figure}

Data files were saved so that there would always be at least two copies of any file.  
The files were saved temporarily underground and permanently at the National Energy Research Scientific Computing Center (NERSC)~\cite{nersc}, where most of the \DEM 's production data analysis occurred (Fig.~\ref{fig:data_prod}).
An additional copy of the data was stored at Oak Ridge Leadership Computing Facility (OLCF).
The underground file store was a 44~terabyte (TB) RAID6 array, used only for file storage and data transfer.
At 10 minute intervals, a crontask on the DAQ computer checked new gzipped data files and copies, using \verb+rsync+, for any that had not been transferred, along with their MD5 signatures.  
Crontasks on the RAID array copied any new data files to NERSC.

Under normal operation, the DAQ wrote about 1~TB of compressed data every two months.  
When the DAQ disk became 75\% full, a semi-automatic process was run to purge the DAQ SSD disk of files.  
The automatic task was built of scripts that compared lists of raw data files on the DAQ computer, the RAID6 array and at NERSC, including tests of the MD5 checksums on each machine. 
The task created a list of files that were safely on all three disks, and thus were available for deletion from the DAQ disk.
At that point, an experienced operator checked the lists and activated a script to purge the disk. 
Data at both NERSC and OLCF were stored on High Performance Storage Systems (HPSS). 
The experiment used the Parallel Distributed Systems Facility (PDSF) and the Cori and Perlmutter supercomputers at NERSC, referred to ``CPU Farm" in figures, for most of its data processing. 

\subsection{File Blinding}
\label{se:file_blinding}

\begin{figure}[ht!]
\centering
\includegraphics[width=0.7\textwidth]{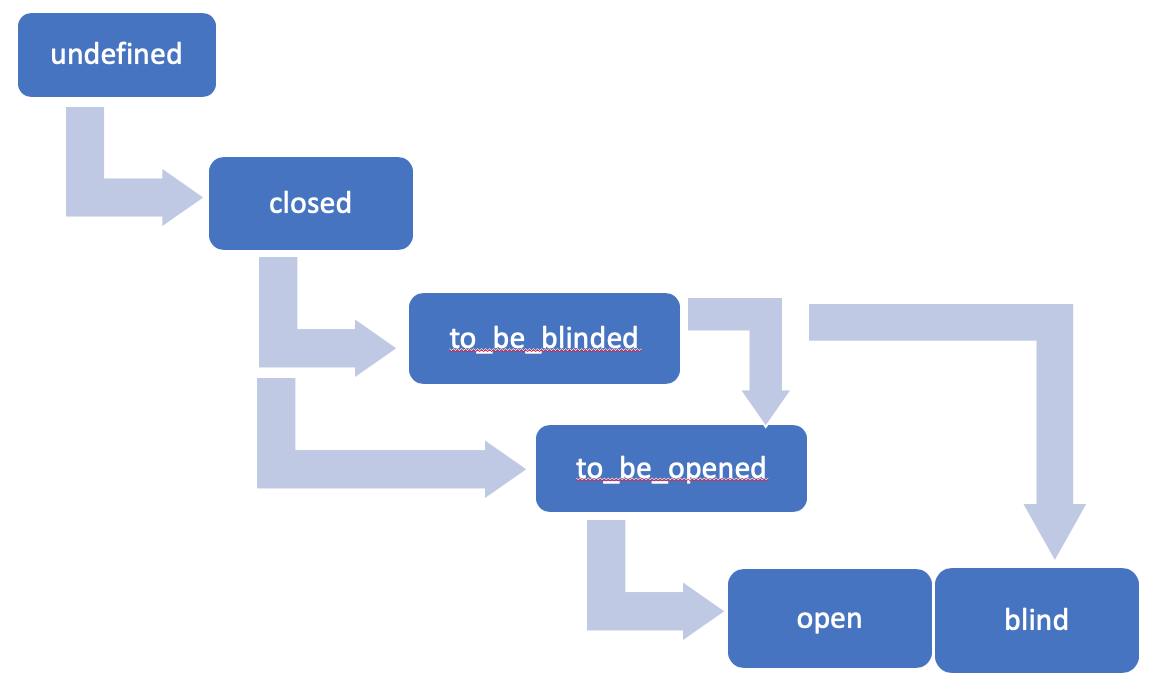}
\caption{Status of file access throughout blinding steps.}
\label{fig:blindness_marking}
\end{figure}

The run database (\S~\ref{se:run_database}) was used to account for the \BBz -decay runs that were covered by the statistical blind analysis approach adopted by \MJ ~\cite{klein-roodman}. 
The \MJD ~blindness scheme called for cycles of 31~hours of open data followed by a period of 93~hours of blinded data to ensure a 25\% open/blind ratio. 
File access control lists were used to limit the analysis to the small subset of ``open" data files using the logic shown in Fig.~\ref{fig:blindness_marking}.
When a raw file arrived on the NERSC disks, its access was changed from undefined to closed.  
Only a very restricted set of users had access to these files.  
During production the raw data files were processed and analysis parameters, such as energy, were calculated.  
During production of the data, the access value of the run files was changed to either ``to\_be\_blinded" or ``to\_be\_opened" based on the run data type provided by ORCA.  
If the run data type corresponded to a \BBz -decay run, then it was marked as ``to\_be\_blinded" while all other data types were set to ``to\_be\_opened".

Files to be ``blinded" were set to either ``blind" or ``to\_be\_opened" according to the 31/93 (open/blind) ratio, and ``to\_be\_opened" raw files had their permissions changed to read access and their access value set to ``open".

Open data were used by the collaboration to check the health of the system and develop analysis cuts. 
These cuts were then applied in a staged approach to the blinded data during the unblinding process used for published results.

\subsection{The \DEM's Data Sets}
\label{sec:data sets}

Data acquired by the \MJD~were divided into data sets characterized by significant differences in the experimental configuration.
Minor changes to the experimental configuration or data acquisition (DAQ) within a data set (DS) were distinguished by subranges denoted by a letter following the data set number.

Dataset-0 (DS0) began on 26~July 2015 with the Module~1 array completely assembled and installed in the Pb shield.  
In place of Module~2, a Pb and outer copper blank monolith was mated to the shield (Sec.~\ref{passive_shield}). 
The inner EFCu shield was not yet installed, and the polyethylene panels on the sides of the shielding structure were not yet in place to allow access to the modules.
Additionally, low-background cryostat seals were still under investigation, so higher activity o-rings were used to seal the cryostats (Sec.~\ref{se:LB_seals}).
As a result, DS0 was a higher background data set that was excluded in the low background configuration used in the \BBz -decay search~\cite{arnq22}. 
However, DS0 served as a complete test of the integrated electronics and supporting subsystems.

Dataset-1 (DS1) began on 31~December~2015, still with Module~1 only, following installation of the EFCu inner shielding.
The waveforms were recorded in a 20~$\mu$s acquisition window at the full sampling rate with the window divided evenly into the pre- and post- trigger regions.
DS-1 was also the start of blindness. 
Examination of data in DS0 and DS1 resulted in the observation of a potential background due to the unexpected response of the HPGe detectors to $\alpha$ particles incident on their passivated surfaces~\cite{Arnquist_2022}.  
Due to the slow collection of charge from such events, data in DS2, beginning on 24 May 2016, were acquired in a different digitizer mode, where each recorded sample in the region 4~$\mu$s after the rising edge was the pre-summed value of four subsequent 10~ns samples, increasing the total acquisition window to 38.2~$\mu$s.

Beginning in August 2016, the \DEM\ began acquiring data with both modules simultaneously deployed in the shield.  
Modules~1 and~2 were recorded with independent DAQ systems in DS3 and DS4, respectively.
With the modules operating on asynchronous DAQ systems, coincidence measurements between modules were not possible.
During DS3 and DS4, the digitizers were reverted to the configuration of DS1 without pre-summing and the polyethylene shielding remained incomplete.

At the start of DS5a, 13 October 2016, the data streams for the modules were merged into a single DAQ system.  
Due to active construction of the remaining polyethylene shielding and optimization of the grounding scheme, DS5a had excess electronics noise which was significantly reduced before the start of DS5b on 27 January 2017.
DS5b and later data sets were acquired in the final shielding configuration.
DS5c began on 17 March 2017 in the same configuration as DS5b, except for implementation of the data blindness scheme.  
DS6a then spanned 11 May 2017 to 16 April 2018 in the same configuration as DS5c except for the re-enabling of the post-rising-edge waveform pre-summing.
Data up to this point were used for the collaboration's second data release~\cite{alvi19b}. 
The detectors operated in the same configuration until the end of 2019 (DS6b), when additional ICPC detectors were installed and the cable and connector upgrade, described in Sec.~\ref{se:CC_upgrades}, was performed.
The \DEM\ subsequently operated with one module (DS7) and then a second module (DS8), before it was decommissioned.

\begin{figure}[ht!]
\centering
\includegraphics[width=0.99\textwidth]{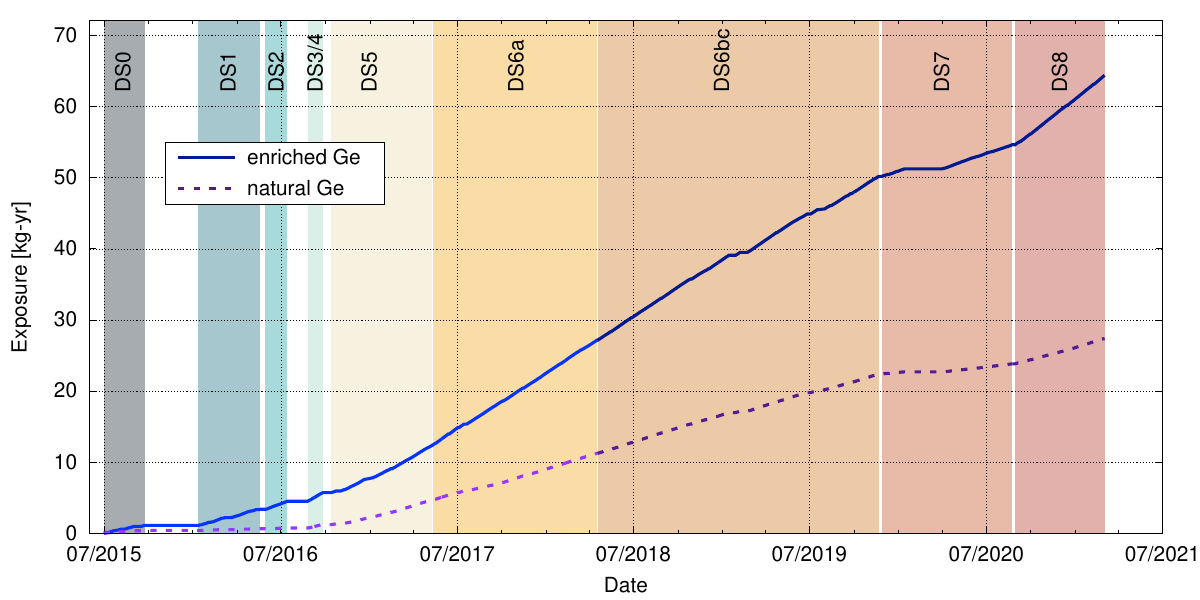}
\caption{The accumulated 65~kg~yr active enriched exposure of the \MJ\ \DEM\ for enriched and natural detectors as a function of time. Also shown are the different datasets that correspond to different configurations. See text for details.}
\label{fig:acc_exposure}
\end{figure}

Fig.~\ref{fig:acc_exposure} shows the accumulated exposure of the \MJ\ \DEM .

\section{Conclusions}
\label{sec:conclusions}

The \MJD\ was an array of  high-purity germanium detectors that operated inside an ultra-low radioactivity shield at the 4850-foot~level of the Sanford Underground Science Laboratory from 2015-2022. 
It demonstrated backgrounds low enough to justify building a tonne-scale experiment to search for the neutrinoless double-beta (\BBz -decay  of \gess\  and published a rich set of searches for physics beyond the Standard Model of particle physics.
Members of the \MJD\ collaboration have joined with members of the GERDA~\cite{agos17} collaboration to pursue a tonne-scale \gess\ experiment called the Large Enriched Germanium Experiment for Neutrinoless $\beta\beta$ Decay (LEGEND)~\cite{abgr17b}. 
The results and experience from the \DEM\ have fed directly into the design of LEGEND. 
In this paper we provided an update of the \DEM , focusing on its construction, commissioning, operation, and physics results. 


\section*{Acknowledgments}

The {\sc Majorana} Collaboration would like to thank the following individuals
for their contributions or assistance in this work: 
Randy Daniels,
John Dunham,
Zhenghao Fu,
Aobo Li,
Jaret Heise,
Gary Holman,
Mark Horten, 
Randy Hughes,
Jonathan Leon,
Patrick Mulkey,
Eric Olivas,
Harry Salazar,
Philip Thompson,
Khang Ton,
Tom Trancynger, 
Cliff Tysor,
Chris Westerfeldt,
and
Doug Will.
This material is based upon work supported by the U.S.~Department of Energy, Office of Science, Office of Nuclear Physics under contract / award numbers DE-AC02-05CH11231, DE-AC05-00OR22725, DE-AC05-76RL0130, DE-FG02-97ER41020, DE-FG02-97ER41033, DE-FG02-97ER41041, DE-SC0012612, DE-SC0014445, DE-SC0017594, DE-SC0018060, DE-SC0022339, and LANLEM77/LANLEM78. We acknowledge support from the Particle Astrophysics Program and Nuclear Physics Program of the National Science Foundation through grant numbers MRI-0923142, PHY-1003399, PHY-1102292, PHY-1206314, PHY-1614611, PHY-13407204, PHY-1812409, PHY-1812356, PHY-2111140, and PHY-2209530. We gratefully acknowledge the support of the Laboratory Directed Research \& Development (LDRD) program at Lawrence Berkeley National Laboratory for this work. We gratefully acknowledge the support of the U.S.~Department of Energy through the Los Alamos National Laboratory LDRD Program, the Oak Ridge National Laboratory LDRD Program, and the Pacific Northwest National Laboratory LDRD Program for this work.  We gratefully acknowledge the support of the South Dakota Board of Regents Competitive Research Grant. 
We acknowledge the support of the Natural Sciences and Engineering Research Council of Canada, funding reference number SAPIN-2017-00023, and from the Canada Foundation for Innovation John R.~Evans Leaders Fund.  
We acknowledge support from the 2020/2021 L'Or\'eal-UNESCO for Women in Science Programme.
This research used resources provided by the Oak Ridge Leadership Computing Facility at Oak Ridge National Laboratory and by the National Energy Research Scientific Computing Center, a U.S.~Department of Energy Office of Science User Facility. We thank our hosts and colleagues at the Sanford Underground Research Facility for their support.

\begin{appendices}

\section{Isotopic Abundance Measurement Uncertainties}
\label{se:abundances}
This appendix  concerns the estimation of the true but unknown abundances $\hat{f}_i$ of isotopic species in a pure sample based on direct measurements of those abundances, e.g.~by mass spectroscopy. We consider here the case where independent measurements $f_i \pm \sigma_i$ are made of each isotope in the sample, where the measurement statistics are taken to be Gaussian. The method can be easily extended to cases in which the measurements are not independent. With additional effort, the method can be extended to cases in which the statistics are non-Gaussian, or in which not all isotopes are measured.

\subsection{Derivation}
We begin by maximizing the likelihood function, which for Gaussian statistics is equivalent to minimization of the $\chi^2$ statistic. However, here we do not simply have the sum of the squared deviations of the the $f_i$ from the $\hat{f}_i$, because the $\hat{f}_i$ are not independent: they must sum to 1. To incorporate this constraint, we would like to multiply the likelihood by something like the Dirac delta function $\delta(1-\Sigma_i \hat{f}_i)$, so that the likelihood function is non-zero only when the constraint is satisfied, but we would like to avoid the infinite value $\delta(0)$; in fact, we would like the extra factor to leave the likelihood function unchanged when the constraint is satisfied. To achieve this, we can approximate the delta function by a Gaussian of width $\epsilon$ and take the limit $\epsilon \rightarrow 0$, but divide by the value of the Gaussian when its argument is zero. Maximization of such an extended likelihood function is equivalent to minimizing the extended $\chi^2$
\begin{equation}
\chi^2 = \left( \sum_{i=1}^N \frac{(f_i - \hat{f}_i)^2}{\sigma_i^2} \right) + \lim_{\epsilon \rightarrow 0} \frac{(1 - \sum_i \hat{f}_i)^2}{\epsilon^2}.
\label{eq:chi2}
\end{equation}
The best-fit values for the $\hat{f}_i$ are obtained by minimizing $\chi^2$ with respect to their variation for generic $\epsilon$, and then taking $\epsilon \rightarrow 0$ at the end. 

The minimization for general $\epsilon$ can be done algebraically. Start by taking the derivative of $\chi^2$ with respect to some $\hat{f}_i$, and setting the result to zero:
\begin{equation}
\frac{f_i - \hat{f}_i}{\sigma_i^2} = -\frac{1-\sum_j \hat{f}_j}{\epsilon^2},
\label{eq:dchi}
\end{equation}
i.e.~at the minimum of $\chi^2$ the $\hat{f}_i$ are shifted from their constraint-free measured values of $f_i$ by an amount proportional to the deviation of the sum of the $\hat{f}_i$ from 1. We can estimate this deviation by summing Eq.~\ref{eq:dchi} for all $i$, and solving for the deviation $1-\sum_i \hat{f}_i$, relabeling summation indices as needed:
\begin{equation}
1-\sum_j \hat{f}_j = \frac{1-\sum_j f_j}{\left(1 + \frac{\sum_j \sigma_j^2}{\epsilon^2}\right)}
\end{equation}
where now no $\hat{f}_i$ appears on the right hand side. Plugging this into Eq.~\ref{eq:dchi} and solving for $\hat{f}_i$ gives
\begin{equation}
\hat{f}_i = f_i + \frac{\sigma_i^2  \left(1- \sum_j f_j\right)}{\epsilon^2 + \sum_j \sigma_j^2}.
\label{eq:fhat}
\end{equation}
We now take the limit $\epsilon \rightarrow 0$. The result is:
\begin{equation}
\hat{f}_i =\frac{\frac{f_i}{\sigma_i^2 }  +  \frac{\overline{f}_i} {\overline{\sigma}_i^2} } { \frac{1}{\sigma_i^2 }  +  \frac{1} {\overline{\sigma}_i^2} }
\label{eq:wtdmean}
\end{equation}
where $\overline{f}_i \equiv 1- \sum_{j \ne i} f_j$ and $\overline{\sigma}_i^2 \equiv \sum_{j \ne i} \sigma_j^2$. We see that the best estimate for $\hat{f}_i$ is given by the weighted mean of the two measurements $f_i \pm \sigma_i$ and the independent measurement $\overline{f}_i \pm \overline{\sigma}_i$ furnished by the sum of the other $f_j$. As can be seen by inspection of either Eq.~\ref{eq:fhat} or \ref{eq:wtdmean}, the best-fit values obey the constraint $\sum_i \hat{f}_i = 1$, even if $\sum_i f_i \ne 1$.

The uncertainty in each of the $\hat{f}_i$ can be computed by standard error propagation. Still assuming that the $f_i$ are all measured independently and are uncorrelated, these uncertainties are characterized by the variances
\begin{equation}
\hat{\sigma}_i^2 = \sigma_i^2 \left(1 - \frac{ \sigma_i^2 }{ \Sigma_j \sigma_j^2}\right).
\label{eq:var}
\end{equation}
The covariance between $\hat{f}_i$ and $\hat{f}_j$ for $i \ne j$ can similarly be computed:
\begin{equation}
\hat{\sigma}_{ij}^2 = - \frac{\sigma_i^2 \sigma_j^2}{ \Sigma_k \sigma_k^2} ~~~ (i \ne j).
\label{eq:cov}
\end{equation}
The covariances are negative, as is required by the constraint $\sum_i \hat{f}_i = 1$. 

Equations~\ref{eq:var} and \ref{eq:cov} can be used to evaluate uncertainties in functions of the $\hat{f}_i$. One of key parameter of interest for example for $0\nu\beta\beta$ experiments is the molar weight:
\begin{equation}
M = \sum_i \hat{f}_i M_i,
\end{equation}
where the $M_i$ are the isotopic molar weights. Assuming the $M_i$ to have negligible uncertainty, the variance in $M$ is given by:
\begin{equation}
\sigma_M^2 = \sum_i M_i^2 \hat{\sigma}_i^2 + 2 \sum_{i \ne j} M_i M_j \hat{\sigma}_{ij}^2.
\end{equation}

\subsection{Example} 
\begin{figure}[htbp]
   \centering
   \includegraphics[width=0.8\textwidth]{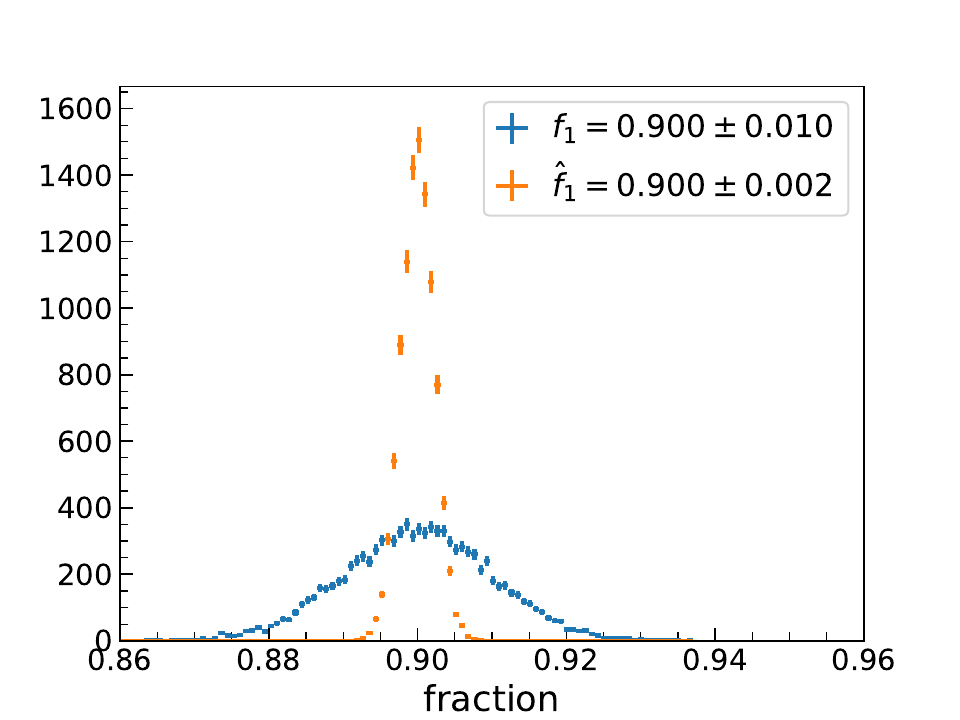}
   \caption{Monte Carlo ($N = 10000$) of measurements of the abundance of species 1 for the example material described in the text. The sampled values of $f_1$ have a wide variance due to their relatively large uncertainty, $\sigma_1$. However the estimate $\hat{f}_1$ for each trial is determined much more precisely due to estimate $\bar{f}_1$ furnished by the more certain measurements of $f_2$ and $f_3$ (not plotted).}
   \label{fig:fs}
\end{figure}

As an example implementation of this method, consider a material containing three atomic species with (unknown) true abundances $f_1 = 0.9$, $f_2 = 0.09$, and $f_3 = 0.01$. A measurement of these abundances is made with independent uncertainties $\sigma_1 = 0.01$, $\sigma_2 = 0.002$, and $\sigma_3 = 0.001$. A typical measured value for $f_1$ will deviate from its true value at the percent level due its relatively large uncertainty, as shown in Fig.~\ref{fig:fs}. However, since $\sigma_2$ and $\sigma_3$ are both much smaller than $\sigma_1$, the estimated value $\hat{f}_1$ will be much closer to the true value. Using Eq.~\ref{eq:var} we estimate an expected uncertainty of $\hat{\sigma}_1 = 0.0022$.

\subsection{Extensions}

The above treatment considered independent, single measurements $f_i$ with Gaussian statistics. If there are multiple measurements of each species, or if the measurements are correlated or the measurement statistics are non-Gaussian, the first term in the $\chi^2$ function (Eq.~\ref{eq:chi2}) requires modification. We conjecture without proof that the likelihood function in these more complex cases must also be maximized by forming (potentially correlated) weighted averages of each $\tilde{f}_i$ with its corresponding $\overline{\tilde{f}}_i$, where the tilde represents the best estimate of each of the $\hat{f}_i$ based on the given measurements ignoring the constraint $\sum_i \hat{f}_i = 1$. This must be the case, since $\tilde{f}_i$ and $\overline{\tilde{f}}_i$, however they are measured, combine to give the best available information on $\hat{f}_i$. The variances and covariances of the $\hat{f}_i$ should then be derivable again from the equivalent of Eq.~\ref{eq:wtdmean} using standard error propagation, incorporating the individual variances and any covariances of the $\tilde{f}_i$. We note, for example, that performing a second iteration of the computation using Eqs.~\ref{eq:wtdmean}-\ref{eq:cov} as inputs leaves the results unchanged.

We now consider the case of not all isotopes in a sample being measured. If all but one species is measured, the best fit is obtained by setting $\hat{f}_i = f_i$ and setting the unknown abundance to achieve the constraint $\sum_i \hat{f}_i = 1$. If more than one species is unmeasured and conjectured to be non-zero, the same procedure can be applied to their summed abundance. The $\hat{f}_i$ are uncorrelated in this case.

As a final note, the method described here does not guard against the unphysical situation $\hat{f}_i < 0$ or $\hat{f}_i > 1$, which by inspection of Eq.~\ref{eq:fhat} will occur whenever $\frac{f_j}{\sigma_j^2} <  \frac{\sum_i f_i -1}{\sum_i \sigma_i^2}$ for some $j$, i.e.~whenever the sum of the $f_i$ exceed one by a statistically significantly larger margin than $f_j$ exceeds zero. In such situations the $\chi^2$ must be minimized respecting these physical boundaries, and numerical methods likely required.

\pagebreak
\section{Axon' cable specifications}
\label{se:axon_specs}

The \DEM\ relied on low mass, low-background cables manufactured by Axon' SAS for wiring inside the cryostat. The specification of these cables are provided in this appendix. 

\begin{table}[h!]
\centering
\begin{tabular}{ |p{7cm}||p{7cm}|  }
 \hline \hline
 \multicolumn{2}{|c|}{Axon' Signal Cable Properties} \\
 \hline
 Property & Value \\
 \hline
 Diameter   & 0.4 mm\\
 Characteristic Impedance & 50 $\Omega$\\
 Capacitance per unit length & 87 pF/m\\
 Mass per unit length & 0.4 g/m\\
 Conductor Material & OFHC Cu\\
 Conductor Gauge & AWG 40\\
 Conductor Diameter & 0.076 mm\\
 Dielectric Material & FEP\\
 Dielectric Diameter & 0.254 mm\\
 Shield Material & OFHC Cu\\
 Shield Gauge & AWG 50\\
 Number of Shield Strands & approximately 30\\
 Jacket Material & FEP\\
 \hline
\end{tabular}
\caption{Table of Axon' signal cable properties.}
\label{table:signal}
\end{table}

\begin{table}[h!]
\centering
\begin{tabular}{ |p{7cm}||p{7cm}|  }
 \hline
 \multicolumn{2}{|c|}{Axon' HV Cable Properties} \\
 \hline
 Property & Value \\
 \hline
 Diameter   & 1.2 mm\\
 Maximum Voltage Rating & 5 kV DC\\
 Mass per unit length & 3 g/m\\
 Conductor Material & OFHC Cu\\
 Conductor Gauge & AWG 34\\
 Conductor Diameter & 0.152 mm\\
 Dielectric Material & FEP\\
 Dielectric Diameter & 0.618 mm\\
 Shield Material & OFHC Cu\\
 Shield Gauge & AWG 50\\
 Jacket Material & FEP\\
 \hline
\end{tabular}
\caption{Table of Axon' HV cable properties.}
\label{table:hv}
\end{table}

\pagebreak

\section{Enriched Germanium Tracking Database Record}
\label{se:egtd_appendix}

\begin{figure*}
\includegraphics[scale=0.4]{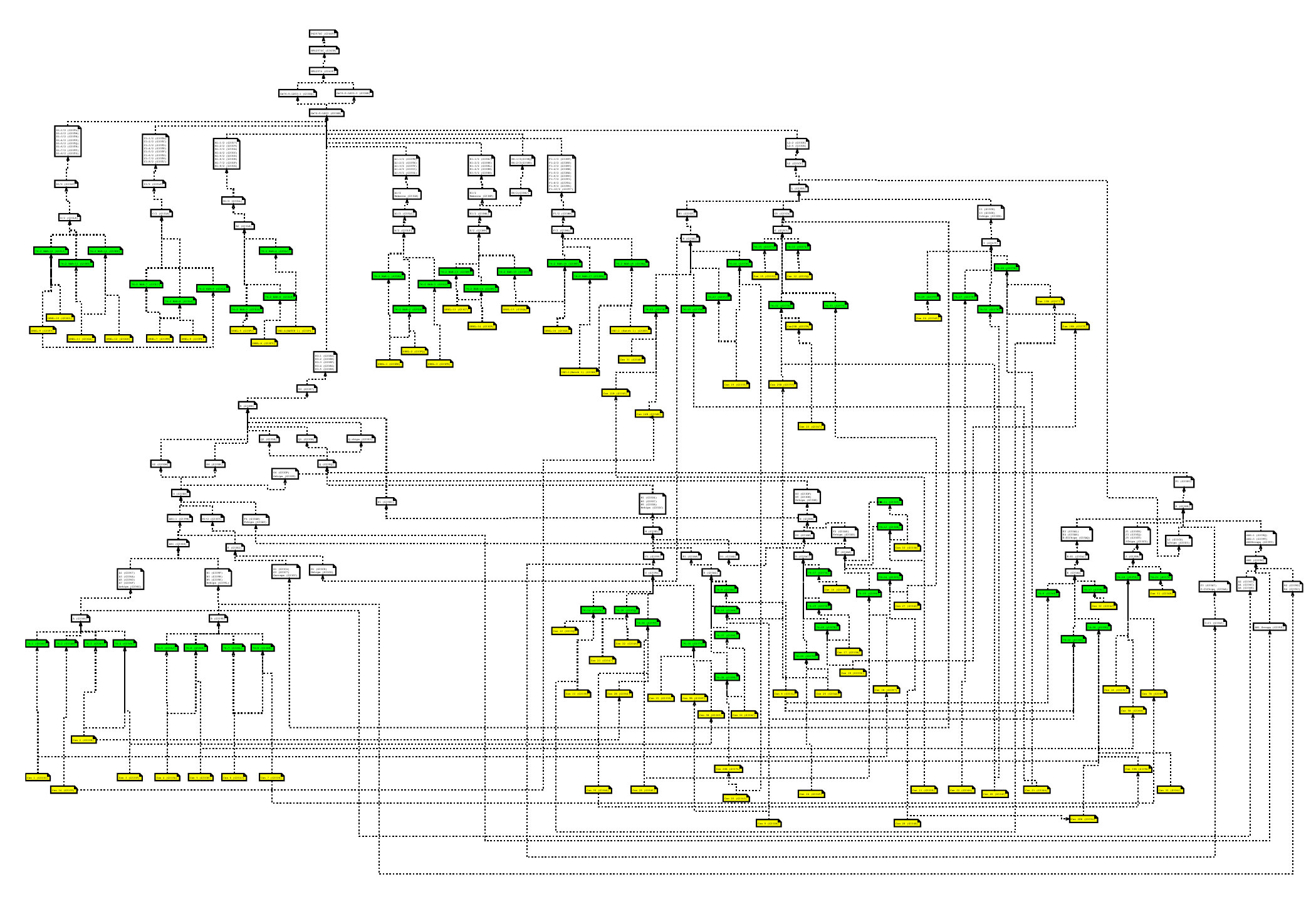}
\caption{Network of enriched Ge used in production of detector P42574C.  Oxide is yellow and reduced bars are green. zone refined bars, crystals, and detectors are white.  The flow of production is from the bottom up.  The detector is the topmost white rectangle.}
\label{fig:enrGe_network}
\end{figure*}

 Here we present more details on the information stored in the enriched germanium tracking database.
 Records in the database contained the following information about germanium lots: 
\begin{itemize}
\item A serial number, generated by the database
\item The form, one of oxide, reduced bar, zone bar, crystal, or detector 
\item Mass
\item Date of creation and date of database entry 
\item Identifying names used in processing
\end{itemize}
It also contained a history, composed of records of processing, transportation and storage.
Here is a sample of the JSON record for a early detector:
\begin{verbatim}
{
  "_id": "G34EA",
  "_rev": "6-c725c4d2ce08211297390a6e1d65cb80",
  "required_forms": {
   "serial_number": "G34EA",
   "record_type": "form_record",
   "content_type": "Detector Blank",
   "esi_identifier": "P42748A",
   "ge_creation_date-time": {"year": "14","month": "3",
                             "day": "24","utctime": ""},
   "db_creation_date-time": {"year": 2016,"month": 8,
                             "day": 2,"utctime": "1551"},
   "mass": {"net": "1002.1"},
   "comment": ""
  },
  "history": [
    {"id": "tf3333KV"},
    {"id": "st3334A6"},
    {"id": "tr333366"}
  ]
}
\end{verbatim}
These history records describe the cutting of the raw crystal that yielded the detector (tf3333KV), the storage of the final detector underground (st3334A6) and the transport of the detector to SURF (tr333366).
To illustrate the potential complexity of the network of enriched germanium, Fig. (\ref{fig:enrGe_network}) is such a network for one of the less complex detectors. 
The diagram does not reflect the time delays in some of the paths, nor the exposure times.

\pagebreak
\section{Equipment and Software Providers}
\label{se:vendors_appendix}

In this appendix we provide information about vendors that provided unique or custom equipment for the \MJD . We also list information about external software packages used. 

\subsection{Equipment Providers}

\noindent 
Axon' Cable SAS, Montmirail, France. Low background and low mass cables.  \url{https://www.axon-cable.com}

\noindent 
CAEN SpA, Viareggio, Italy. Data acquisition equipment.    \url{https://www.caen.it}

\noindent 
Eckert \& Ziegler Analytics, Inc, Berlin, Germany. Radioactive sources.  \url{https://www.ezag.com/home/},

\noindent 
Glenair, Glendale, CA, USA. Electrical connectors.    \url{https://www.glenair.com}

\noindent 
Hovair Systems, Seattle, WA, USA. Air bearing transporter for shield monoliths.  \url{https://www.hovair.com}

\noindent 
ISEG HV, Radeberg, Germany. HV power supplies.    \url{https://iseg-hv.com/en}

\noindent
KME Group, Firenze, Italy. Copper plate stock supplier for outer shield copper. \url{https://www.kme.com/} 

\noindent 
Mirion Technologies (Canberra BNLS) NV, Zellik, Belgium. HPGe detector supplier. \url{https://www.canberra.com/cbns/default.html}

\noindent 
Moxtek, Inc., Orem, Utah, USA. Low noise field-effect transistors. \url{https://www.moxtek.com}

\noindent 
ORTEC/AMETEK, Oak Ridge, Tennessee, USA. HPGe detector supplier.    \url{https://www.ortec-online.com}

\noindent 
Schleuniger Group, Thun, Switzerland. Cabke stripper for Axon' cables.      \url{https://www.schleuniger.com/en/}

\noindent 
Schneider Electric, USA. Uninterruptible power supply (UPS) supplier for underground lab. \url{https://www.se.com/us/en/}

\noindent 
Southern Copper \& Supplies Company Inc, Pelham, AL, USA. Supplier of copper for outer copper shield.    \url{https://southerncopper.com}

\noindent 
Starrett,  Athol MA, USA. Provided metrology instrument for measuring HPGe detector dimensions.   \url{https://www.starrett.com}

\noindent 
Sullivan Metals Inc., Holly Springs, MS, USA. Supplier of lead bricks for lead shield.  \url{http://www.sullivanmetalinc.com}

\noindent 
Xcelite (owned by Cooper Industries) Houston, TX. Crimping tool supplier. \url{https://www.tequipment.net/Xcelite/MIC3020BL/Crimpers/}

\subsection{Software Providers}
The following is a list of external software used by the collaboration: 

\noindent 
Backbone.js \url{http://backbonejs.org}

\noindent 
Bootstrap \url{http://getbootstrap.com}

\noindent
CouchDB, The Apache Software Foundation, \url{http://couchdb.apache.org}

\noindent 
\uppercase{D}ublin \uppercase{C}ore \uppercase{M}etadata \uppercase{E}lement \uppercase{S}et, \uppercase{V}ersion 1.1: \uppercase{R}eference \uppercase{D}escription, DCMI Usage Board, 
\url{https://www.dublincore.org/specifications/dublin-core/dces/2012-06-14/}

 \noindent 
 Ecma International, The \uppercase{JSON} \uppercase{D}ata \uppercase{I}nterchange \uppercase{S}yntax,   \url{https://www.ecma-international.org/publications/standards/Ecma-404.htm}

\noindent 
libcurl - the multiprotocol file transfer library, Daniel Stenberg, \url{https://curl.haxx.se/libcurl/}

\noindent 
Mathematica, {V}ersion 13.1, Wolfram Research{,} Inc.
\url{https://www.wolfram.com/mathematica}

\noindent 
REpresentational \uppercase{S}tate \uppercase{T}ransfer (\uppercase{REST}), or \uppercase{REST}ful \url{https://restfulapi.net}

\end{appendices}

\bibliography{main}

\end{document}